\documentclass[twocolumn,showpacs,aps,prd,amsmath,amssymb,nobibnotes,floatfix]{revtex4-1}
\newcommand{\beq}{\begin{equation}}
\newcommand{\eeq}{\end{equation}}
\newcommand{\beqn}{\begin{eqnarray}}
\newcommand{\eeqn}{\end{eqnarray}}

\newcommand{\apjl}{Astrophys.\ J.\ Lett.}

\newcommand{\mnras}{Mon. Not. Roy. Astron. Soc.}
\newcommand{\ve}[1]{\mbox{\boldmath $#1$}}

\newcommand{\llabel}[1]{\label{#1}}              
\newcommand{\labeq}[2]{ \begin{equation} \llabel{#1}
{#2}
\end{equation}}

\newcommand{\espl}{ \end{split} }
\newcommand{\bspl}{ \begin{split} }
\newcommand{\F}{\mathcal F}
\newcommand{\B}{\mathcal B}
\newcommand{\E}{\mathcal E}
\newcommand{\J}{\mathcal J}
\newcommand{\bS}{\bar{S}}

\newcommand{\sg}{\sqrt{\gamma}\,}
\newcommand{\tem}{T_{\rm EM}}
\newcommand{\Lie}{\mathcal L}
\newcommand{\p}{\partial}
\newcommand{\ga}{\stackrel{>}{ _{\sim}}}

\usepackage{wasysym}
\usepackage{graphicx}
\usepackage{epsf}
\usepackage{graphics,epsfig,placeins,subfigure,wrapfig}

\begin{document}
\title{A new scheme for matching general relativistic ideal magnetohydrodynamics \\ to its force-free limit}

\author{Vasileios Paschalidis}
\author{Stuart~L.~Shapiro}
\altaffiliation{Also at Department of Astronomy and NCSA, University of
  Illinois at Urbana-Champaign, Urbana, IL 61801}
\affiliation{Department of Physics, University of Illinois at
  Urbana-Champaign, Urbana, IL 61801}

\begin{abstract}

We present a new computational method for smoothly matching general
relativistic ideal magnetohydrodynamics (MHD) to its force-free
limit. The method is based on a flux-conservative formalism for MHD
and its force-free limit, and a vector potential formulation for the
induction equation to maintain the zero divergence constraint for the
magnetic field. The force-free formulation we adopt evolves the
magnetic field and the Poynting vector, instead of the magnetic and
electric fields. We show that our force-free code passes a robust
suite of tests, performed both in 1D flat spacetime and in 3D black
hole spacetimes. We also demonstrate that our matching technique
successfully reproduces the aligned rotator force-free solution. Our
new techniques are suitable for studying electromagnetic effects and
predicting electromagnetic signals arising in many different curved
spacetime scenarios. For example, we can treat spinning neutron stars,
either in isolation or in compact binaries, that have MHD interiors
and force-free magnetospheres.

\end{abstract}


\maketitle

\section{Introduction}

In recent years there has been a strong interest in identifying
electromagnetic (EM) counterparts to loud gravitational wave (GW)
events. Apart from the intrinsic information that EM waves carry about
the source, EM signals will also help localize the source on the
sky. Knowledge of the precise location of the source on the sky
eliminates degeneracies and results in improved parameter estimation
from GWs~\cite{Nissanke:2012dj}.

In addition to being strong sources of GWs, compact binaries, such as
binary black hole--neutron stars (BHNSs), and binary neutron
star--neutron stars (NSNSs) are also promising sources of
``precursor'' and ``aftermath'' EM signals. Here, precursor
(aftermath) means before (after) merger has taken place.
For example, BHNS or NSNS mergers may provide the central engine that
powers a short-hard gamma-ray burst. Moreover, during merger
neutron-rich matter can be ejected that can shine as a ``kilonova''
due to the decay of r-process elements
\cite{1998sese.conf..729R,Rosswog:1998hy,Li:1998bw,Kulkarni:2005jw,Metzger:2010sy,Goriely:2011vg,Hotokezaka:2012ze,Kyutoku:2012fv,Rosswog:2013kqa,Grossman:2013lqa,Kyutoku:2013wxa,Bauswein:2013yna}.
While some studies have been performed in Newtonian gravitation or the
conformal flatness approximation of general relativity (GR), only a
fully GR calculation can reliably determine the amount of ejected mass
and its distribution, as well as the GW signature.

Equation of state effects, mass ejection, effects of cooling and
finite temperature, as well as waveforms from the inspiral and merger
of BHNSs and NSNSs, have been computed in full GR via hydrodynamic
simulations (see e.g.
\cite{UIUC_BHNS__BH_SPIN_PAPER,Duez:2009yy,Shibata:2009cn,Foucart:2010eq,Kyutoku:2011vz,2011ApJ...737L...5S,Lackey:2011vz,ShibataBHNSreview,Foucart:2012vn,2012PhRvD..85l4009E,Lovelace:2013vma,Foucart:2013psa}
for BHNSs and
\cite{2002PThPh.107..265S,Shibata:2003ga,Shibata:2006nm,Anderson:2007kz,2008PhRvD..78h4033B,2009CQGra..26k4005B,Kiuchi:2010ze,2012PhRvD..85j4030B,FaberNSNSreview,Sekiguchi:2011zd,Paschalidis:2012ff,Bauswein:2013jpa}
for NSNSs), and magnetohydrodynamic (MHD) simulations (see e.g.
\cite{Chawla:2010sw,UIUC_MAGNETIZED_BHNS_PAPER1,UIUC_MAGNETIZED_BHNS_PAPER2}
for BHNS mergers and
\cite{2008PhRvL.100s1101A,Liu:2008xy,2011PhRvD..83d4014G,Rezzolla:2011da}
for NSNSs). In all of these earlier simulations of magnetized neutron
stars, the magnetic field was confined within the stellar interior,
partly due to the inability of existing ideal MHD schemes to deal with
magnetic fields exterior to the star where the matter magnetization can
become very high.

However, (spinning) neutron stars are believed to be endowed with
dipole magnetic fields extending into the exterior, which comprises a
force-free magnetosphere~\cite{Goldreich:1969sb}.  Thus, toward the
end of a BHNS or NSNS inspiral electromagnetic interactions can give
rise to detectable EM pre-merger signals
\cite{Hansen:2000am,McWilliams:2011zi,Lyutikov:2011tq,Piro:2012rq,Lai:2012qe,D'Orazio:2013kgo},
e.g. either via establishing a unipolar inductor DC circuit
\cite{UI1969,McL2011}, via magnetospheric interactions
\cite{Hansen:2000am} or via emission of magnetic dipole
radiation~\cite{Ioka:2000yb}. As these mechanisms operate in strongly
curved, dynamical spacetimes, numerical relativity simulations are
necessary to reliably determine the amount of EM output. Modelling
these effects requires to first order either a GR resistive MHD
computational scheme
(e.g. \cite{Bucciantini:2012sm,Dionysopoulou:2012zv,Palenzuela:2012my})
or a scheme that matches the ideal MHD interior of the NS to the
exterior force-free magnetosphere, such as those presented in
\cite{Lehner:2011aa,Paschalidis:2013jsa}. Simulations in GR attempting
to model these effects are still in their infancy.  Only recently
have simulations begun to explore the viability of these mechanisms
and calculate the total EM output (see
\cite{Palenzuela:2013hu,Palenzuela:2013kra} for NSNSs and
\cite{Paschalidis:2013jsa} for BHNSs).

In this paper we present the details and tests of our GR force-free
electrodynamics formalism and new code, and our new scheme for
matching ideal MHD to its force-free limit. This code has already been
used and briefly described in \cite{Paschalidis:2013jsa}. We
demonstrate the robustness of our new force-free code in a series of
1D flat spacetime and 3D black-hole spacetime tests, and we test our
new matching scheme by reproducing the force-free aligned rotator
solution for a rotating magnetized star
\cite{Goldreich:1969sb,Contopoulos:1999ga,Komissarov:2005xc,McKinney:2006sd,Spitkovsky:2006np}.

The paper is structured as follows. In Sec.~\ref{sec:conventions} we
discuss the general spacetime and EM field conventions. In
Sec.~\ref{sec:basic_eqns} we review the standard formulation of
force-free electrodynamics, discuss some subtleties arising in
so-called electrovacuum solutions, and derive for the first time some
new identities emerging in this formulation. We also present the
force-free formulation we adopt, and derive several new useful
identities arising in this formulation. In
Sec.~\ref{sec:numerical} we present our methods for numerically
evolving the GR force-free electrodynamics (GRFFE) equations and
matching them to ideal MHD stellar interiors.  Sec.~\ref{tests}
reviews the tests we adopt to demonstrate the robustness of our new
code,  as well as the results from our simulations. We conclude in
Sec.~\ref{sec:summaryandfuturework} with a summary and discussion of
future work.

\section{3+1 Decomposition and General Conventions}
\label{sec:conventions}

In this section we describe the general conventions we use in our
MHD/Force-Free formalism. Throughout we use geometrized units, setting
$c=1=G$. Latin indices denote spatial
components (1--3) and Greek indices denote spacetime components
(0--3). The signature of the spacetime metric is (-+++).

\subsection{3+1 spacetime decomposition}

We use a 3+1 decomposition of spacetime in which
the line element becomes (see e.g. \cite{BSBook})
\beq
  ds^2 = -\alpha^2 dt^2
+ \gamma_{ij} (dx^i + \beta^i dt) (dx^j + \beta^j dt),
\eeq
 where $\gamma_{ij}$ is the induced three-metric in 3D spatial
hypersurfaces of constant time $t$, $\alpha$ is the lapse function and $\beta^i$ the 
shift vector. The full (4D) spacetime metric $g_{\mu \nu}$
is related to the three-metric $\gamma_{\mu \nu}$ by $\gamma_{\mu \nu}
= g_{\mu \nu} + n_{\mu} n_{\nu}$, where
\labeq{normalvector}{
n^{\mu} =(1/\alpha,-\beta^i/\alpha)
}
is the future-directed, timelike unit vector normal to 3D spatial
hypersurfaces.

\subsection{Maxwell 's equations and electromagnetic stress tensor}

The basic equations of ideal GRMHD and their implementation in a 3+1
spacetime decomposition has been treated in a number of papers (see
e.g.~\cite{bs03,Duez:2005sf,Etienne:2010ui,UIUCEMGAUGEPAPER}) and
textbooks (e.g.~\cite{BSBook}), but we review them here to set the stage
for our applications below.

The Faraday tensor $F^{\mu \nu}$ can be decomposed into the 3+1 form
\beq F^{\mu \nu} = n^\mu E^\nu - n^\nu E^\mu - \epsilon^{\mu \nu
  \alpha \beta} B_\alpha n_\beta , 
\eeq 
where $\epsilon^{\mu \nu \alpha \beta}$ is the
Levi-Civita tensor. The electric and magnetic fields measured by 
normal observers are defined as
\beqn E^\mu &=& n_\nu F^{\mu \nu} \label{def:Emu} \\ B^\mu &=&
\frac{1}{2} \epsilon^{\mu \nu \alpha \beta} n_\nu F_{\beta \alpha} =
n_\nu {}^*F^{\nu \mu} ,
\label{def:Bmu}
\eeqn
where 
\beq
  {}^*F^{\mu \nu} = \frac{1}{2} \epsilon^{\mu \nu \alpha \beta} F_{\alpha \beta} 
\eeq
is the dual of $F^{\mu \nu}$. Note that $n_\mu E^\mu = n_\mu B^\mu=0$. Hence both 
$E^\mu$ and $B^\mu$ are purely spatial. It is convenient to introduce the following variables:
\labeq{}{
  \begin{split}
  \F^{\mu \nu} \equiv &\ \frac{F^{\mu \nu}}{\sqrt{4\pi}} \ \ \ , \ \ \ 
  {}^*\F^{\mu \nu} \equiv \frac{{}^*F^{\mu \nu}}{\sqrt{4\pi}} \ \ \ , \\ 
  \E^\mu \equiv &\ \frac{E^\mu}{\sqrt{4\pi}} \ \ \ , \ \ \
  \B^\mu \equiv \frac{B^\mu}{\sqrt{4\pi}} \ \ \ , \ \ \ 
  \J^\mu \equiv \sqrt{4\pi}\, j^\mu .
  \end{split}
}
Here $j^\mu$ is the 4-current density. 

With these new definitions, 
\labeq{dec:Fab}{
  \F^{\mu \nu} = n^\mu \E^\nu - n^\nu \E^\mu - \epsilon^{\mu \nu \alpha \beta} \B_\alpha n_\beta,
}
and
\beq
  {}^*\F^{\mu \nu} = -n^\mu \B^\nu + n^\nu \B^\mu - \E_\alpha n_\beta \epsilon^{\mu \nu \alpha \beta} .
\label{dec:Fsab}
\eeq 
Straightforward calculations yield 
\beq 
  \F^{\mu \nu} \F_{\mu \nu} = 2 (\B^2-\E^2) \ \ \ \mbox{and} \ \ \ 
  {}^*\F^{\mu \nu} \F_{\mu \nu} = 4 \E_\mu \B^\mu ,
\label{eq:dotproducts}
\eeq 
where $\B^2 = \B_\mu \B^\mu = \B_i \B^i$ and $\E^2 = \E_\mu \E^\mu = \E_i \E^i$.

Maxwell's equations can be expressed in terms of the new variables as 
\beq
  \nabla_\mu \F^{\mu \nu} = -\J^\nu \ \ \ , \ \ \ \nabla_{[\alpha} \F_{\beta \gamma]} = 0 .
\label{eq:maxwell}
\eeq
It follows from the antisymmetric property of $\F_{\mu \nu}$ that $\nabla_{[\alpha} \F_{\beta \gamma]} = 0$ 
can be written as 
\beq
  \nabla_\alpha \F_{\beta \gamma} + \nabla_\beta \F_{\gamma \alpha} + \nabla_\gamma \F_{\alpha \beta} = 0.
\label{eq:maxwellabc}
\eeq
In addition, 
\labeq{}{
\begin{split}
  0 = & \frac{1}{2} \epsilon^{\mu \alpha \beta \gamma} \nabla_{[\alpha} \F_{\beta \gamma]} 
 = \frac{1}{2} \epsilon^{\mu \alpha \beta \gamma} \nabla_\alpha \F_{\beta \gamma} \\ 
 = &\ \nabla_\alpha \left( \frac{1}{2} \epsilon^{\mu \alpha \beta \gamma} \F_{\beta \gamma} \right)  
 = \nabla_\nu {}^*\F^{\mu \nu} .
\end{split}
}
Hence $\nabla_{[\alpha} \F_{\beta \gamma]} = 0$ is equivalent to 
\labeq{nablastarFmunu}{
  \nabla_\nu {}^*\F^{\mu \nu} = 0 .
}
The stress-energy tensor associated with the EM field is 
\beq
  \tem^{\mu \nu} = \F^\mu{}_\lambda \F^{\nu \lambda} 
- \frac{1}{4} g^{\mu \nu} \F^{\lambda \sigma} \F_{\lambda \sigma} ,
\label{def:Tem}
\eeq
where $g_{\mu \nu}$ is the spacetime metric. Straightforward calculation yields 
\labeq{eq:Tem31}{
\begin{split}
  \tem^{\mu \nu} = &\ \frac{\B^2+\E^2}{2} (\gamma^{\mu \nu} + n^\mu n^\nu) 
 - (\B^\mu \B^\nu + \E^\mu \E^\nu) \\
 &\ - n_\alpha \E_\beta \B_\lambda (n^\mu \epsilon^{\nu \alpha \beta \lambda} 
 + n^\nu \epsilon^{\mu \alpha \beta \lambda}) ,
\end{split}
}
where $\gamma_{\mu \nu} = g_{\mu \nu} + n_\mu n_\nu$ is the spatial metric on 3D
hypersurfaces of constant time. 
It follows from Eq.~(\ref{def:Tem}) and Maxwell's equations that  
\beq
  \nabla_\nu \tem^{\mu \nu} = -\F^{\mu \nu} \J_\nu .
\label{eq:divTem}
\eeq	
The Poynting vector is defined as 
\beq
  S^\mu = -n_\nu \tem^{\mu \nu} = \frac{\B^2+\E^2}{2} n^\mu 
- \epsilon^{\mu \nu \alpha \beta} n_\nu \E_\alpha \B_\beta .
\label{def:Poynting}
\eeq
It follows that 
\beq
  \B_\mu S^\mu = 0 .
\eeq
In the flat spacetime limit ($g_{\mu \nu} = \eta_{\mu \nu}$) we obtain the
familiar results
\beq
  S^0 = \frac{\B^2+\E^2}{2} \ \ \ , \ \ \ S^i = \epsilon^{ijk} \E_j \B_k .
\label{eq:Poynting_SR}
\eeq

\subsection{Ideal MHD Condition}
\label{sec:MHD}

The ideal MHD condition is 
\beq
  u_\mu \F^{\mu \nu} = 0 ,
\label{mhd_cond}
\eeq
where $u^\mu$ is a unit timelike vector ($u_\mu u^\mu = -1$) equal to the plasma 4-velocity
in ideal MHD, and may be regarded as the plasma 4-velocity in the force-free limit. 
Contracting Eq.~(\ref{mhd_cond}) with $n_\nu$ and using $\E^\mu = n_\nu \F^{\mu \nu}$ yields 
$u_\mu \E^\mu = 0 $.

Comparing Eq.~(\ref{mhd_cond}) with Eq. (\ref{def:Emu}), 
one may interpret the ideal MHD condition as the vanishing electric field measured 
by an observer with four-velocity $u^\mu$. These observers include the one comoving
with the plasma as well as others boosted with respect to this observer
in a direction parallel to the B-field (i.e., $u_\perp^\mu = u_{\perp,\rm comoving}^{\mu}$). 
The magnetic field measured by
such an observer is 
\beq 
  b^\mu = u_\nu {}^*\F^{\nu \mu} .
\label{def:bmu}
\eeq
The Faraday and electromagnetic stress tensors can be decomposed by $u^\mu$ and $b^\mu$ by analogy to the decomposition
with $n^\mu, B^{\mu}, E^\mu$ presented in the previous section, i.e., 
\beqn
\F^{\mu \nu} &=& \epsilon^{\mu \nu \alpha \beta} u_\alpha b_\beta \label{mhd:Fab} \\ 
{}^*\F^{\mu \nu} &=& b^\mu u^\nu - u^\mu b^\nu \label{mhd:Fsab} \\ 
\F^{\mu \nu} \F_{\mu \nu} &=& 2b^2 \label{mhd:FdotF} \\ 
{}^*\F^{\mu \nu} \F_{\mu \nu} &=& 0 \label{mhd:FdotFs} \\ 
\tem^{\mu \nu} &=& b^2 u^\mu u^\nu + \frac{b^2}{2} g^{\mu \nu} - b^\mu b^\nu . 
\label{mhd:Tem}
\eeqn
Eqs.~(\ref{mhd:FdotFs}) and (\ref{eq:dotproducts}) yield
\beq
  \E_\mu \B^\mu = 0 .
\eeq
and combining Eq.~(\ref{mhd:Fsab}) with $\B^\nu = n_\mu {}^*\F^{\mu \nu}$ yields 
\beq
  \B^\nu = u^\nu n_\alpha b^\alpha - b^\nu n_\alpha u^\alpha .
\eeq
It is straightforward to show that \cite{Duez:2005sf}
\beq
  b^\mu = \frac{P^\mu{}_\nu \B^\nu}{-n_\alpha u^\alpha} = \frac{P^\mu{}_\nu \B^\nu}{\gamma_v} ,
\label{eq:bmufromBmu}
\eeq
where
\beq
  P^\mu{}_\nu = \delta^\mu{}_\nu + u^\mu u_\nu . 
\eeq
is the projection tensor, $\gamma_v=-n_\alpha u^\alpha=\alpha u^0$ is the Lorentz factor corresponding
to the relative velocity of $u^\mu$ with respect to a normal observer $n^\mu$. It follows from Eq.~(\ref{eq:bmufromBmu}) that 
\beq
  b^2 = b^\mu b_\mu = \frac{P_{\mu \nu} \B^\mu \B^\nu}{\gamma_v^2} 
= \frac{\B^2 + (u_\mu \B^\mu)^2}{\gamma_v^2} .
\eeq
Hence $b^2$ is positive-definite, and $b^2=0$ if and only if $\B^\mu=0$, which also 
implies $b^\mu=0$ and $\F^{\mu \nu} =0$ from Eqs.~(\ref{eq:bmufromBmu}) 
and (\ref{mhd:Fab}). By use of Eqs.~(\ref{mhd:FdotF}), 
(\ref{eq:dotproducts}) and the condition $b^2\geq 0$ we have
\beq
  \F^{\mu \nu} \F_{\mu \nu} \geq 0 \ \ \ \mbox{and} \ \ \ \B^2 \geq \E^2 .
\eeq
The equality holds if and only if $\F^{\mu \nu}=0$ or, equivalently, $\B^\mu=\E^\mu=0$. 
Therefore, the ideal MHD condition forbids the (vacuum EM wave) solution $B^2=E^2$ with $B^2 > 0$.

\section{Force-Free Electrodynamics (FFE)} 
\label{sec:basic_eqns}

In this section we present the FFE conditions and briefly review the two most
popular formulations of FFE. The first one uses the electric and
magnetic fields as the fundamental dynamical variables \cite{k04}, and
the second one replaces the electric field by the Poynting vector
\cite{Komissarov:2002my,m06}.  We include derivations of several key
equations in order to clarify subtle points, correct typos in the
literature, and to present the basis of our approach.

\subsection{FFE conditions}
\label{sec:ffeaxi}

The force-free conditions are \cite{bz77,Komissarov:2002my}
\beqn
  \F^{\mu \nu} \J_\nu &=& 0 , \label{ffe:FdotJ} \\
  {}^*\F^{\mu \nu} \F_{\mu \nu} &=& 0 , \label{ffe:FdotFs} \\ 
  \F^{\mu \nu} \F_{\mu \nu} & > & 0 . \label{ffe:FdotF} 
\eeqn
The above conditions can be regarded as axioms of FFE (in addition to
the Maxwell and Einstein equations). Physically, these conditions are
expected to apply when the magnetic fields dominate over the inertia
of the matter \cite{Goldreich:1969sb,bz77}.

In terms of the 3+1 variables, Eqs.~(\ref{ffe:FdotJ})--(\ref{ffe:FdotF}) become 
\beqn
  \rho \E^i + \epsilon^{ijk} J_j \B_k &=& 0 , \label{ffe:FdotJ31} \\ 
  \E_i \B^i &=& 0 , \label{ffe:FdotFs31} \\
  \B^2 & > & \E^2 . \label{ffe:FdotF31}
\eeqn
These can be regarded as the FFE axioms in terms of $\E$ and $\B$ fields,  
where $\epsilon^{ijk}=n_\mu \epsilon^{\mu ijk}$ is the Levi-Civita tensor associated 
with the spatial metric $\gamma_{ij}$, and the 4-current density has been decomposed 
into the 3+1 form
\beq
  \J^\mu = \rho n^\mu + J^\mu
\eeq
with 
\beq
  \rho = -n_\mu \J^\mu \ \ \ , \ \ \ 
  J^\mu = \gamma^\mu{}_\nu \J^\nu .
\eeq
Contracting Eq.~(\ref{ffe:FdotJ}) with $n_\mu$ and using $\E^\mu = n_\nu \F^{\mu \nu}$ gives 
\beq
  \J_\mu \E^\mu = \J_i \E^i = 0. 
\label{ffe:JdotE}
\eeq

The conditions~(\ref{ffe:FdotFs}) and (\ref{ffe:FdotF}) are properties
of the ideal MHD condition, and as it was first shown in \cite{m06},
the ideal MHD condition is contained in the force-free conditions. In
particular, it can be shown that if the conditions (\ref{ffe:FdotFs}) and
(\ref{ffe:FdotF}) are satisfied, there exists a one-parameter family
of timelike unit vectors $\{U^\mu \}$ so that $u_\nu \F^{\mu \nu}=0$
for any $u^\mu \in \{U^\mu \}$. This one-parameter family of unit
timelike vectors is given by 
\beq u^\mu_L =
\sqrt{\frac{\B^2}{\B^2(1-L^2) - \E^2} } \left( n^\mu -
\frac{\epsilon^{\mu \beta \gamma \delta} n_\beta \E_\gamma
  \B_\delta}{\B^2} + L \frac{\B^\mu}{\B} \right)
\label{eq:umuLtext}
\eeq
where the $L$ parameter is restricted by
\beq
  |L| < \sqrt{ \frac{\B^2-\E^2}{\B^2}} .
\label{eq:Lrestricttext}
\eeq
In Appendix~\ref{timelikevectors_exist} we present a proof of Eqs. \eqref{eq:umuLtext}, 
\eqref{eq:Lrestricttext} using standard 3+1 notation.

As was pointed out in \cite{m06} in this family of unit timelike
vectors, the one that has the \textit{minimum} Lorentz factor is given
by $L=0$, i.e.\ $u^\mu$ is orthogonal to $\B^\mu$. The corresponding
$u^\mu$ is 
\beq u_{(m)}^\mu = \sqrt{\frac{\B^2}{\B^2 - \E^2} } \left(
n^\mu - \frac{\epsilon^{\mu \beta \gamma \delta} n_\beta \E_\gamma
  \B_\delta}{\B^2} \right) ,
\label{eq:ummu}
\eeq
or 
\beqn
  u_{(m)}^0 &=& \frac{1}{\alpha} \sqrt{\frac{\B^2}{\B^2 - \E^2} } \label{eq:um0} \\
  v_{(m)}^i &=& \frac{u_{(m)}^i}{u_{(m)}^0} = \alpha \frac{\epsilon^{ijk} \E_j \B_k}{\B^2} - \beta^i 
= \alpha \frac{\gamma^{ij} S_j}{\B^2} - \beta^j .
\label{eq:vmi}
\eeqn
In the flat spacetime limit, $u_{(m)}^\mu$ reduces to 
\beq
  u_{(m)}^0 = \gamma_v = \sqrt{\frac{\B^2}{\B^2 - \E^2} } \ \ \ , \ \ \ 
  v^i_{(m)} = \frac{u_{(m)}^i}{u_{(m)}^0} = \frac{\epsilon^{ijk} \E_j \B_k}{\B^2} .
\eeq
\\
The three-velocity $v^i_{(m)}$ appearing in this last equation is also known as the drift velocity.

Finally, by use of Eqs.~(\ref{ffe:FdotJ}) and (\ref{eq:divTem}), we obtain 

\beq
  \nabla_\nu \tem^{\mu \nu} = 0 .
\label{ffe:divTem}
\eeq
Hence, FFE can be regarded as a limiting case of the MHD in which the
plasma has negligible inertia. It is this property that motivates
our scheme for matching ideal MHD to its force-free limit, which 
we present in Sec.~\ref{machingMHDFFE}.

\subsection{On the $\ve{\E_i \B^i=0}$ Condition}
\label{sec:EBcondition}

In some literature (e.g.~\cite{k04,k11}), it is claimed that Eq.~(\ref{ffe:FdotFs31}) follows 
from Eq.~(\ref{ffe:FdotJ31}). We argue that this is not true. 

Taking a dot product of Eq.~(\ref{ffe:FdotJ31}) with $\B^i$ gives
$\rho \E_i \B^i=0$, while taking the cross product of
Eq.~(\ref{ffe:FdotJ31}) with $\E^i$ and using Eq. \eqref{ffe:JdotE} gives $J^k
(\E_i \B^i)=0$. Hence, from a mathematical point of view $\E_i \B^i=0$
follows only if $\J^\mu \neq 0$. Hence, the condition $\E_i B^i=0$
can be violated in regions where $\J^\mu=0$, if one uses only
Eqs. \eqref{ffe:FdotJ31} and \eqref{ffe:FdotF31} as the FFE conditions. One
simple example is the initial data $\E^i = \E_0^i/\sg$ and
$\B^i=\B_0^i/\sg$ with $\E_0^i$ and $\B_0^i$ being constant vectors
and $\E_{0i} \B_0^i \neq 0$ and $|\B_o^i| > |\E_0^i|$. Clearly the
initial data satisfy the Maxwell constraints $D_i \E^i = \rho$ and
$D_i \B^i =0$ for $\J^\mu=0$, as well as the remaining force-free 
constraints \eqref{ffe:FdotJ31} and \eqref{ffe:FdotF31}. Hence they are 
valid EM initial data but not valid force-free initial data. Moreover, 
Eq.~(\ref{ffe:FdotJ31}) holds while Eq.~(\ref{ffe:FdotFs31}) does not. 

The situation $\J^\mu=0$ and $T^{\mu \nu}=\tem^{\mu \nu}$ everywhere
in the spacetime is known as the \textit{electrovacuum}. In the electrovacuum, both
$\E_i \B^i=0$ and $\B^2 > \E^2$ conditions can be violated.  Examples
of electrovacuum solutions that are not force-free include the
Kerr-Newmann black holes ($\E^2>\B^2$ and $\E_i \B^i \neq 0$), and Wald's
electrovacuum solution in rotating black holes ($\E_i \B^i \neq
0$)~\cite{w74}.

One may therefore choose to replace the condition~(\ref{ffe:FdotFs}) by
$\J^\mu \neq 0$. However, doing this will exclude some of the
electrovacuum solutions that are also force-free under the
condition~(\ref{ffe:FdotFs}). One example is Wald's electrovacuum solution
in Schwarzschild spacetime, which has been used to test GRFFE
codes (see~\cite{k04} and Sec.~\ref{wald} below) or even a nonrotating 
star with a dipole magnetic field. Therefore, we suggest that
the condition~(\ref{ffe:FdotFs}) should be kept in favor of $\J^\mu \neq
0$. One may also define FFE as a limiting case of ideal MHD, as was
done in~\cite{k02}. In that case, the condition~(\ref{ffe:FdotFs}) is
inherited from the ideal MHD conditions. The advantage of the
axiomatic approach we adopt is the ability to formulate FFE without
reference to the 4-velocity $u^\mu$ [see also in~\cite{k02}, where 
the ideal MHD condition $u_\mu \F^{\mu \nu}=0$ is replaced by
(\ref{ffe:FdotFs}) and (\ref{ffe:FdotF})].

While it may come as a surprise that there exist electrovacuum
solutions (no matter present) that are also FFE solutions (tenuous
plasma present) this is not a contradiction.  As a model of physical
reality, force-free electrodynamics applies to cases where a
highly-conducting tenuous plasma is involved. Hence, physically,
force-free environments cannot be the same as an electrovacuum
environment.  However, mathematically, any electrovacuum solution
satisfying the force-free conditions \eqref{ffe:FdotJ} -
\eqref{ffe:FdotF}, will also be a force-free solution. For example an
electrovacuum solution in which $\E^i=0$ and $\B^i \neq 0$, is
simultaneously a force-free solution. 

\subsection{Evolution Equations for $\ve{\E}$ and $\ve{\B}$}
\label{sec:formalism1}

Perhaps the most popular formulation of FFE uses the $\ve{\E}$ and
$\ve{\B}$ fields as dynamical variables. As shown
in~\cite{1982MNRAS.198..339T,bs03}, without any assumption of MHD the
general Maxwell equations (\ref{eq:maxwell}) can be brought into the
3+1 form:
\beqn
D_i \E^i & = & \rho \label{divE} \\
\partial_t \E^i & = & \epsilon^{ijk} D_j ( \alpha \B_k)
        - \alpha J^i + \alpha K \E^i + \Lie_{\bf \beta} \E^i
        \label{Edot} \\
D_i \B^i & = & 0 \label{divB} \\
\partial_t \B^i & = & - \epsilon^{ijk} D_j (\alpha \E_k) + \alpha K \B^i
        + \Lie_{\bf \beta} \B^i  \label{Bdot} ,
\eeqn
where $D_i$ is the covariant derivative associated with the spatial metric $\gamma_{ij}$,
$K=K^i{}_i$ is the trace of the extrinsic curvature, and $\Lie_{\bf \beta}$ is the Lie
derivative along the shift vector $\beta^i$. 

The general set of Maxwell Eqs. \eqref{divE}-\eqref{Bdot}, coupled to the
general fluid equations for the matter [$\nabla_{\mu} (T_{\rm matter}^{\mu\nu} + T_{\rm EM}^{\mu\nu}) = 0$],
reduce to the equations of ideal MHD (e.g., Eqs. (5.168) - (5.175) in \cite{BSBook})
whenever collision timescales are sufficiently short for the plasma
to behave as an isotropic fluid and the magnetic Reynolds number is
sufficiently large that resistivity can be ignored.

To apply Eqs. \eqref{divE}-\eqref{Bdot} for FFE, an 
expression for the 3-current density $J^i$ is needed. It is useful to decompose $J^i$ 
into a component perpendicular to $\B^i$ and a component parallel to $\B^i$: 
\beq
  J^i = J_\perp^i + J_\parallel \frac{\B^i}{\B^2} ,
\label{eq:Jidecomp}
\eeq
with 
\beq
  J_\parallel \equiv \B_i J^i = \B_\mu \J^\mu \ \ \ \mbox{and} \ \ \ 
  J_\perp^i \equiv J^i - (\B_k J^k) \frac{\B^i}{\B^2} .
\label{def:Jpara_Jperp}
\eeq
The perpendicular component (after contracting Eq.~(\ref{ffe:FdotJ31}) 
with $\epsilon_i{}^{lm} \B_l$ and taking the cross product with $\B^i$) is given by 
\beq
  J_\perp^i = \rho \frac{\epsilon^{ijk} \E_j \B_k}{\B^2} 
  = (D_m \E^m) \frac{\epsilon^{ijk} \E_j \B_k}{\B^2} .
\label{eq:Jperpi}
\eeq
The parallel component can be determined by demanding that the evolution equations 
preserve the constraint $C_{EB}=\E_i \B^i=0$, i.e.\ $\partial_t C_{EB}=0$ (see e.g. \cite{m06,k11}), 
eventually giving
\beq
  J_\parallel = \epsilon^{ijk} (\B_i D_j \B_k - \E_i D_j \E_k) - 2 \E^i \B^j K_{ij} .
\label{eq:Jpara}
\eeq
Combining the results yields 

\labeq{ffe:Ji}{
\begin{split}
  J^r = &\ (D_m \E^m) \frac{\epsilon^{rjk} \E_j \B_k}{\B^2} +  \\
 & \ \frac{\epsilon^{ijk} (\B_i D_j \B_k - \E_i D_j \E_k) - 2 \E^i \B^j K_{ij}}{\B^2} \B^r.
\end{split}
}
Equation~(\ref{ffe:Ji}) is known as the Ohm's law in dissipationless GRFFE. 
In the flat spacetime limit, it reduces to the well-know expression (see, e.g.~\cite{g99}) 
\beq
  \ve{J} = \frac{(\ve{\E}\times \ve{\B}) (\ve{\nabla}\cdot \ve{\E}) + \ve{\B} [ \ve{\B}\cdot 
(\ve{\nabla}\times \ve{\B}) - \ve{\E} \cdot (\ve{\nabla} \times \ve{\E})]}{\B^2} .
\eeq
Note that since $K_{ab} = -\nabla_{(a} n_{b)} - n_{(a} a_{b)}$ ($a_b = n^c \nabla_c n_b$), 
$-2\E^i \B^j K_{ij} = \E^\mu \B^\nu (n_{\mu;\nu} + n_{\nu;\mu})$. Hence Eq.~(\ref{eq:Jpara}) 
agrees with the expression in Eq.~(25) of~\cite{m06}. However, the $-2\E^i \B^j K_{ij}$ term is 
missing in Eq.~(80) of~\cite{k11}.


While the constraint $C_{EB}\equiv \E_i \B^i =0$ is preserved due to the FFE Ohm's law~(\ref{eq:Jpara}), the constraint 
$C_{dB}\equiv D_i \B^i=0$ is preserved by Eq.~(\ref{Bdot}). To see this, consider 
\[
  \partial_t (\sg C_{dB}) = \partial_t [ \partial_i (\sg \B^i) ] 
= \partial_i (\B^i \partial_t \sg + \sg \partial_t \B^i) , 
\]
where $\gamma$ is the determinant of the spatial metric $\gamma_{ij}$. It follows 
from the Arnowitt-Deser-Misner equations [see e.g. Eq. (2.136) in \cite{BSBook}] that 
\beq
  \partial_t \sg = \sg (-\alpha K + D_i \beta^i) .
\eeq
Using the identities 
\beq
  D_i [\epsilon^{ijk} D_j(\alpha E_k)] = \epsilon^{ijk} D_i D_j(\alpha E_k) = 0 , 
\eeq
and, for any spatial vector $w^i$, 
\beq
  D_{[i} D_{j]} w^i = {}^{(3)}R_{ij}{}^i{}_k w^k = {}^{(3)}R_{jk} w^k ,
\eeq
where ${}^{(3)}R_{ijkl}$ is the Riemann tensor associated with $\gamma_{ij}$,
we find after some algebra
%
%
\beqn
  \partial_t (\sg C_{dB})
&=& \partial_j (\sg C_{dB} \beta^j).
\label{eq:CdBdot}
\eeqn
%
Equation~(\ref{eq:CdBdot}) shows that if $C_{dB}=0$ initially, the evolution equations preserve 
the constraint.

Equations~(\ref{Edot}) and (\ref{Bdot}), combined with
Eq.~(\ref{ffe:Ji}) are the evolution equations for $\E^i$ and $\B^i$,
which are subject to the three constraints $C_{dB}=D_i \B^i=0$,
$C_{EB}=\E_i \B^i=0$ and $\B^2 > \E^2$. The first two constraints are
preserved by the evolution equations but not the last one. Violation
of $\B^2 > \E^2$ indicates the breakdown of FFE, which typically
occurs in current sheets. Mathematically, if violation occurs the
initial value problem for Eqs. \eqref{Edot}, \eqref{Bdot} with
\eqref{ffe:Ji} becomes ill-posed \cite{k02,Pfeiffer:2013wza}.
Moreover, the constraint \eqref{divE} is automatically satisfied, if
one uses it to compute the charge density. Finally,
Eqs. \eqref{eq:maxwell} also imply charge conservation, as it is
straightforward to see that \labeq{}{ \nabla_\mu \J^\mu = 0, } which
can be used as an evolution equation for the charge density.

Perhaps the greatest advantage of using
Eqs. \eqref{divE}-\eqref{Bdot}, is that they are general and can be
used to find both force-free solutions [as long as the current is
given by Eq. \eqref{ffe:Ji}], and electrovacuum solutions (as long as
one sets $\J^\mu = 0$). The disadvantage is that this formulation is
not straightforward to use in conjunction with the well-known
constrained transport methods for preserving the Maxwell constraints
see e.g. \cite{Tóth2000605,1988ApJ...332..659E,Etienne:2010ui}. Thus,
common numerical implementations of this formulation usually resort to
divergence cleaning methods to maintain the Maxwell constraints (see
e.g.  \cite{Dedner2002645,Palenzuela:2010xn}).

\subsection{Evolution Equations for $\ve{S_i}$ and $\ve{\B^i}$}
\label{sec:formalism2}

Instead of evolving $\ve{\E}$ and $\ve{\B}$, Refs.~\cite{k02} and \cite{m06} 
suggest the adoption of $S_i$ and $\B^i$ as dynamical variables. These
are the fundamental variables we adopt.

It follows from Eq.~(\ref{def:Poynting}) that $S_i = \epsilon_{ijk} \E^j \B^k$. Taking the cross 
product of $S_i$ with $\B^i$ and using $\E_i \B^i=0$ gives 
\beq
  \epsilon^{ijk} \B_j S_k = \epsilon^{ijk} \epsilon_{klm} \B_j \B^m \E^l  =\B^2 \E^i .
\eeq
The condition $\B^2 > \E^2$ guarantees that $\B^2>0$. Hence 
\beq
  \E^i = \frac{\epsilon^{ijk} \B_j S_k}{\B^2} .
\label{eq:E-SB}
\eeq
The above equation can be rewritten 
using the identities $S_\mu = -n_\nu \tem^{\nu}{}_\mu$ and 
$\epsilon^{\mu \nu \alpha} = n_\beta \epsilon^{\beta \mu \nu \alpha}$ as 
\beq
  \E^\alpha =-\frac{\epsilon^{\alpha \beta \gamma \delta} \B_\beta S_\gamma n_\delta}{\B^2} = 
\frac{\epsilon^{\alpha \beta \gamma \delta} \B_\beta T^{\mu}_{\rm EM}{}_\gamma n_\mu n_\delta}{\B^2} ,
\label{eq:E-SBn}
\eeq
(see also Eq.~(10) in~\cite{m06}). Note that the constraint $C_{EB}=\B_i \E^i =0$ is automatically 
satisfied by Eq.~(\ref{eq:E-SB}). Contracting Eq.~(\ref{eq:E-SB}) with $\E_i$ gives 
\labeq{E2S2B2}{
  \E^2 = \frac{\bar{S}^2}{\B^2} - \frac{C_{SB}^2}{\B^4} = \frac{\bar{S}^2}{\B^2}, 
}
where $C_{SB}\equiv \B^i S_i=0$ and 
\labeq{}{
  \bar{S}^2 \equiv \gamma^{ij} S_i S_j .
}

In this formulation the condition $\B^2 > \E^2$ is expressed through Eq. \eqref{E2S2B2} as 
\beq
  \bar{S}^2 < \B^4 .
\label{eq:S2ltB4}
\eeq

If we define the densitized magnetic field 
\beq
  \tilde{\B}^i \equiv \sg \B^i = \sg n_\nu {}^*\F^{\nu i} = \alpha \sg {}^*\F^{i0},
\eeq
the time component of the Maxwell Eq. \eqref{nablastarFmunu} yields the constraint 
equation 
\beq
  \partial_i \tilde{\B}^i = 0 , 
\label{eq:divBeq0}
\eeq
whereas the spatial components of Eq. \eqref{nablastarFmunu} give the equation 
\beq
  \partial_t \tilde{\B}^i + \partial_j (\beta^i \tilde{\B}^j - \beta^j \tilde{\B}^i 
+ \alpha \sg \epsilon^{ijk} \E_k) = 0 ,
\eeq
where Eqs.~(\ref{dec:Fsab}) and \eqref{normalvector} have been used. 
Substituting $\E_k$ using Eq.~(\ref{eq:E-SB}) yields the induction equation 
\beq
  \partial_t \tilde{\B}^i + \partial_j \left( \alpha S_k
\frac{\tilde{\B}^i \gamma^{jk}-\tilde{\B}^j \gamma^{ik}}{\B^2}
+\beta^i \tilde{\B}^j - \beta^j \tilde{\B}^i \right) = 0 .
\label{eq:induction3}
\eeq
Introducing a 3-vector $v^i$ defined as 
\beq
  v^i = \alpha \frac{\gamma^{ij}S_j}{\B^2} - \beta^i .
\label{def:vi}
\eeq
Then the induction equation~(\ref{eq:induction3}) takes the familiar form
\beq
  \partial_t \tilde{\B}^i + \partial_j (v^j \tilde{\B}^i - v^i \tilde{\B}^j) = 0 .
\label{eq:induction}
\eeq
The induction equation clearly preserves the constraint $C_{dB}=0$: 
\beq
  \partial_t (\sg C_{dB}) = \partial_t (\partial_i \tilde{\B}^i) 
= -\partial_i \partial_j (v^j \tilde{\B}^i - v^i \tilde{\B}^j) = 0 .
\eeq

The evolution equation for $S_i$ can be derived from Eq.~(\ref{ffe:divTem}), which gives 
$\nabla_{\nu} \tem^{\nu}{}_i=0$ or 
\beq
  \partial_t \tilde{S}_i + \partial_j (\alpha \sg \tem^j{}_i ) = \frac{1}{2} \alpha \sg \tem^{\mu \nu} 
\partial_i g_{\mu \nu} , 
\label{eq:Sidot}
\eeq
where 
\labeq{densSi}{
  \tilde{S}_i = \sg S_i 
}
is the densitized spatial Poynting vector, and the EM stress-energy tensor can be expressed 
in terms of $\B^i$ and $S_i$ via Eqs.~(\ref{eq:Tem31}), (\ref{def:Poynting}) and the 
first equality of Eq. (\ref{eq:E-SBn}). The quantities $S_0$, $S^\mu$, $\B^0$ and 
$\B_\mu$ can be expressed in terms of $\B^i$ and $S_i$ 
using Eq.~(\ref{def:Poynting}) and $n_\mu \B^\mu=0$ as 
\beq
  \B^0 = 0 \ \ , \ \ \B_0 = \gamma_{ij} \beta^i \B^j \ \ , \ \ 
  \B_i = \gamma_{ij} \B^j , 
\eeq 
and 
\labeq{}{
\begin{split}
  S_0 = &\ -\alpha \frac{\B^2 +\bar{S}^2/\B^2}{2} + \frac{\beta^i S_i}{\alpha}, \\
  S^0 = &\ \frac{\B^2 +\bar{S}^2/\B^2}{2\alpha}, \\
  S^i = &\ -\frac{\B^2 +\bar{S}^2/\B^2}{2\alpha}\beta^i + \gamma^{ij} S_j .
\end{split}
}

Note that the time component of 
Eq.~(\ref{ffe:divTem}) also implies $\nabla_{\nu} \tem^{\nu}{}_0=0$, which gives the energy equation 
\beq
  \partial_t (\sg S_0) + \partial_j (\alpha \sg \tem^j{}_0 ) = \frac{1}{2} \alpha \sg \tem^{\mu \nu}
\partial_t g_{\mu \nu} .
\label{eq:S0dot}
\eeq 
However, Eqs.~(\ref{eq:induction3}) and (\ref{eq:Sidot}) already
provide a complete system of evolution equations for $S_i$ and $B^i$,
which can be used to calculate $\E^i$ using Eq.~(\ref{eq:E-SB}). Thus,
the energy equation~(\ref{eq:S0dot}) is either a constraint or
redundant and it must be able to be derived from
Eqs.~(\ref{eq:divBeq0}), (\ref{eq:induction3}), (\ref{eq:Sidot}) and
(\ref{eq:E-SB}). We show that the energy is indeed redundant and not a
constraint in Appendix~\ref{energy_redundant}.  Finally, the Maxwell
equation $\nabla_\alpha \F^{\nu \alpha}=\J^\nu$ implies that one of
the force-free conditions $\F^{\mu \nu} \J_\nu=0$ is also enforced by
the evolution equations and the constraint $D_i \B^i=0$.

The evolution equations~(\ref{eq:induction3}) and (\ref{eq:Sidot})
consist of a system of 6 coupled partial differential equations for 6
variables $\B^i$ and $S_i$, which contain the same number of equations
as Eqs.~(\ref{Edot}) and (\ref{Bdot}) in \S~\ref{sec:formalism1}. The
system of partial differential equations in \S~\ref{sec:formalism1}
are subject to two constraints: $C_{dB}=D_i \B^i=0$ and $C_{EB}=\E_i
\B^i=0$. In the present system, the constraint $C_{dB}=0$ remains, but
$C_{EB}=0$ is automatically satisfied by Eq.~(\ref{eq:E-SB}). This fact was
also pointed out in \cite{m06}, but, another constraint that arises in
this formulation was ignored: A simple change of variables cannot
change the number of constraints in a dynamical system.  The
constraint that replaces $C_{EB}=0$ in the $\ve{S}$-$\ve{\B}$
formulation of GRFFE is $C_{SB}\equiv\B^i S_i=\B^\mu S_\mu=0$. It can
be shown that the evolution equations (\ref{eq:induction3}) and
(\ref{eq:Sidot}) preserve the constraint $C_{SB}=0$ as long as both
$C_{dB}=0$ and $C_{SB}=0$ initially (see Appendix~\ref{evolCsb}).

As the $\ve{S}$-$\ve{\B}$ formulation is equivalent to the
$\ve{E}$-$\ve{\B}$ formulation of GRFFE, one can use
Eq. (\ref{ffe:Ji}) to compute the 4-current density $\J^\mu = \rho
n^\nu + J^\mu$.  It is possible to prove that in the
$\ve{S}$-$\ve{\B}$ formalism the FFE current density is given by the same
equation as in the $\ve{E}$-$\ve{\B}$ formalism (see Appendix~\ref{FFE_current}).

\section{Numerical Method}
\label{sec:numerical}

Here we summarize the formulation and numerical methods we use to solve the equations 
of GRFFE and our new scheme for matching the ideal MHD to its force-free limit.

\subsection{Evolution scheme for the GRFFE equations}
\label{sec:GREFFEalgorithm}

The greatest advantage of the $\ve{S}$-$\ve{\B}$ formulation is that
it is straightforward to implement numerically, if one has already
developed a GRMHD code. There are at least two more reasons for
adopting the $\ve{S}$-$\ve{\B}$ formulation: a) the evolution
equations for $S_i$ and $\B^i$ are basically the same as their MHD
counterparts. This already hints that the same evolution equations can
be used to match ideal MHD domains to force-free domains. b) The
constrained-transport method can be used to enforce the $D_i \B^i=0$
constraint as in the MHD case. The remaining constraint $S_i \B^i=0$,
which was ignored in \cite{m06}, is algebraic and can be enforced by
replacing $S_i \rightarrow S_i - (S_j \B^j)\B^i/\B^2$ after each
evolution timestep, i.e., in the same way the $\E_i \B^i$ constraint
is enforced in the $\ve{\E}$-$\ve{\B}$ formulation see
e.g. \cite{Spitkovsky:2006np,Palenzuela:2010xn}. See also
\cite{Palenzuela:2010xn,Alic:2012df} for other alternatives for
enforcing the $\E_i \B^i$ constraint. So, to transform a GRMHD
high-resolution shock capturing code to a force-free code all one has
to do is to remove from the GRMHD code all terms related to the
perfect fluid stress tensor (i.e. the matter is ignored), and add a
new algorithm for the primitives recovery.

To summarize, the complete set of evolution equations are the induction and momentum equations
\beqn
  \p_t \tilde{B}^i + \p_j (v^j \tilde{B}^i - v^i \tilde{B}^j) &=& 0 \label{eq:inductionrecap}  \\ 
  \p_t \tilde{S}_i + \p_j (\alpha \sg \tem^j{}_i) &=& \frac{1}{2} \alpha \sg 
\tem^{\mu \nu} \partial_i g_{\mu \nu} \label{eq:dtSiforcefree},
\eeqn
with 
\beq
  v^i = 4\pi \alpha \frac{\bar{S}^i}{B^2} - \beta^i 
= 4\pi \alpha \frac{\gamma^{ij} \tilde{S}_j}{\sg B^2} - \beta^i ,
\label{eq:vifromcons}
\eeq
where $\bar{S}^i = \gamma^{ij} S_j$. 
Note that the factor $4\pi$ has reappeared since our GRMHD code uses $B^i$ 
instead of $\B^i$. 
The evolution equations can be made to look even more similar to the MHD equations by 
introducing the unit timelike 4-vector $u^\mu$ as 
\labeq{eq:u0ui}{
\begin{split}
  u^0 = &\ \frac{1}{\alpha} \sqrt{ \frac{B^2}{B^2-E^2} } 
= \frac{1}{\alpha} \sqrt{ \frac{\tilde{B}^2}{\tilde{B}^2-16\pi^2 \gamma \tilde{S}^2/\tilde{B}^2}},\\
  u^i = &\ u^0 v^i ,
\end{split}
}
where $\tilde{S}^2=\gamma^{ij} \tilde{S}_i \tilde{S}_j$. 
Note that this is exactly the same as $u^\mu_{(m)}$ in Eq.~(\ref{eq:ummu}) -
the unit timelike 4-vector that satisfies $u_\mu F^{\mu \nu}=0$ 
with the minimum Lorentz factor $\gamma_v = -n_\mu u^\mu = \sqrt{B^2/(B^2-E^2)}$. 
The EM stress-energy tensor is given by Eq.~(\ref{mhd:Tem})
\beq
  \tem^{\mu \nu} = b^2 u^\mu u^\nu + \frac{b^2}{2} g^{\mu \nu} - b^\mu b^\nu ,
\label{eq:Tembmu}
\eeq
where $b^\mu$ can be computed from $B^\mu$ 
and $u^\mu$ using Eq.~(\ref{eq:bmufromBmu}) 
\beq
  b^\mu = \frac{P^\mu{}_\nu B^\nu}{\sqrt{4\pi} \gamma_v} .
\label{eq:bmufromBmu2}
\eeq
 Equations (\ref{eq:inductionrecap})--(\ref{eq:bmufromBmu2}) give the
 complete evolution equations for $B^i$ and $S_i$. We embed this GRFFE
 formulation in the conservative ideal GRMHD, high-resolution shock
 capturing infrastructure we have presented and tested in
 \cite{Duez:2005sf,Etienne:2010ui,UIUCEMGAUGEPAPER}, and in which we
 preserve the $\partial_i \tilde B^i = 0$ constraint via a vector
 potential formulation which is equivalent to the standard
 staggered-mesh constrained-transport scheme in uniform-resolution
 grids~\cite{Etienne:2010ui,UIUCEMGAUGEPAPER}. To close the system we
 choose the generalized Lorenz gauge we developed and used in
 \cite{Farris:2012ux,UIUC_MAGNETIZED_BHNS_PAPER2,Paschalidis:2013jsa}.

The evolution (``conservative'') variables are $\tilde{B}^i$ and $\tilde{S}_i$. The 
``primitive'' variables are $B^i$ and $v^i$, as in the MHD case. Reconstructions 
are done on the primitive variables. 

The inversion from 
conservative to primitive variables is trivial in GRFFE: $B^i = \tilde{B}^i/\sg$ 
and $v^i$ from Eq.~(\ref{eq:vifromcons}). The electric field $E^i$ is not needed 
for evolution but may be computed from Eq.~(\ref{eq:E-SB})
\beq
  E^i = 4\pi \frac{\epsilon^{ijk} B_j S_k}{B^2} .
\eeq

Inversion fails whenever the condition $B^2 > E^2$ 
is violated as it leads to superluminal velocity (i.e.\ $\gamma_v$ becomes purely imaginary). 
Thus, the condition for the primitive inversion to yield a physical solution is 
\beq
  \tilde{S}^2 < \frac{\tilde{B}^4}{16\pi^2 \gamma} .
\label{eq:stineq}
\eeq
It should be noted that the inequality~(\ref{eq:stineq}) should be checked after removing 
the component of $\tilde{S}$ along the magnetic field, i.e.\ imposing 
the constraint $\tilde{B}^i \tilde{S}_i=0$ by the procedure 
$\tilde{S}_i \rightarrow \tilde{S}_i - (\tilde{S}_j \tilde{B}^j) \tilde{B}^i/\tilde{B}^2$. 
It is straightforward to show that removing the $\tilde{B}^i$ component from $\tilde{S}_i$ always 
leads to smaller $\tilde{S}^2$. The inequality~(\ref{eq:stineq}) may be imposed by specifying 
a maximum Lorentz factor $\gamma_{\rm max}$ and requiring that 
$\gamma_v = \alpha u^0 \leq \gamma_{\rm max}$. It follows from Eq.~(\ref{eq:u0ui}) that 
the condition $\gamma_v \leq \gamma_{\rm max}$ is equivalent to 
\beq
  \tilde{S}^2 \leq (1-\gamma_{\rm max}^{-2}) \frac{\tilde{B}^4}{16\pi^2 \gamma} .
\eeq
Define a factor 
\beq
  f \equiv \sqrt{ (1-\gamma_{\rm max}^{-2}) \frac{\tilde{B}^4}{16\pi^2 \gamma \tilde{S}^2}} .
\eeq
The inequality~(\ref{eq:stineq}) can be imposed by setting 
\beq
  \tilde{S}_i \rightarrow \tilde{S}_i \min(1,f) .
\eeq
Imposing the condition $B^2>E^2$ when the FFE is supposed to break
down (as in e.g. a current sheet) is effectively to add artificial dissipation to the fields and
remove energy immediately to bring the fields back to the FFE
regime. We typically set $\gamma_{\rm max} = 2000$. In addition, as
was proposed in \cite{m06} in current sheets we null the inflow
velocity normal to the current sheet, i.e., if $\tilde n^i$ is the
normal to the current sheet we set
\labeq{}{
\tilde n_i v^i = 0,
}
within an infinitesimal region above and below the current sheets
covered by four zones.  For a discussion motivating this approach and
of its possible shortcomings we refer the interested reader to
\cite{m06}.

\subsection{Matching ideal MHD to its force-free limit}
\label{machingMHDFFE}

Force-free magnetospheres appear in many occasions in astrophysical
environments, e.g., including neutron stars. The interior of a NS is
highly conducting and the assumption of perfect conductivity is
well-justified. As a result ideal MHD applies to the NS
interior. However, existing high-resolution shock capturing MHD
schemes cannot deal with high magnetizations and as a result they
cannot typically deal with magnetic fields exterior to the highly
conducting matter. On the other hand NSs are typically endowed with a
force-free magnetosphere and since force-free electrodynamics can be
regarded as the limit of ideal MHD in which the magnetic fields
dominate the inertia of the matter, there must exist ways of making
this transition from the ideal MHD interior to the force-free
exterior. Such a scheme for matching ideal GRMHD to its force-free
limit was first proposed in \cite{Lehner:2011aa} using the
$\ve{\E}$-$\ve{\B}$ formulation, but the implementation required the
introduction of new variables and coding of additional evolution
equations, as well as prescribing a penalty function based on the
rest-mass density for transitioning from the interior to the exterior.

Our scheme for matching ideal MHD interiors to force-free exteriors
utilizes the fact that the magnetic field is frozen-in and is simply
advected with the fluid for sufficiently weak magnetic fields. So, we
propose that the frozen-in condition be enforced in the dense interior
of the star and the surface values for the B-field and the Poynting
vector then provide the boundary conditions for the exterior FFE
evolution using the $\ve{S}$-$\ve{\B}$ formalism we outlined in
Sec. \ref{sec:formalism2}.

We point out here that our matching scheme does not allow for any
back-reaction of the exterior magnetic field onto the interior matter. This
back-reaction potentially may become important in a thin layer near
the surface of a star. However, resistive MHD studies of NSs, which
include magnetic field back-reaction, indicate that neglecting it leads
only to small errors \cite{Palenzuela:2012my}.

\subsubsection{Matching when the fluid rest-mass density and four velocity are given}
\label{easy_matching}

First we will consider the case where we are evolving the EM field of
a star with a well-defined surface, and that both the interior fluid
four-velocity $u_\mu$ and the rest-mass density distribution $\rho_0$ are
known and given for all times (e.g. a stationary rotating star with a
weak interior field). Physically, the rest-mass density in a
force-free magnetosphere \textit{cannot} be zero.  However, the
equations of FFE \textit{ignore} the existence of matter, and for
\textit{numerical} purposes we can safely set the rest-mass density
exterior to the star equal to zero. Therefore, in our algorithm the
stellar surface is defined as the 2-surface where the rest-mass
density transitions from $\rho_0^{(\rm num)}=\rho_0 \neq 0$ to
$\rho_0^{(\rm num)} = 0$, i.e., the \textit{numerical}
magnetosphere has zero density.

\begin{itemize}
 
\item Interior to the star $\rho_0^{(\rm num)} \neq 0$, and the
  frozen-in condition is enforced.  
  We evolve the induction equation for the A-field
  \cite{UIUCEMGAUGEPAPER}
\labeq{dtAi}{
\partial_t A_i = \epsilon_{ijk}v^j B^k - \partial_i(\alpha \Phi - \beta^j A_j)
}
with any convenient EM gauge choice to determine the scalar potential
$\Phi$, setting the three-velocity $v^j$ equal to the given fluid velocity. In
the continuum limit this truly enforces the frozen-in condition, while in
the discrete limit small deviations from the frozen-in condition are
expected. These converge away with increasing resolution. Given the
A-field, we then determine the B-field, and compute the E-field
using the ideal MHD condition \eqref{eq:udotF=0} where again the fluid
3-velocity is used. We set the interior $\ve{\tilde S}^{\rm (in)}$ equal to
\labeq{Siinter}{ \tilde S_i^{\rm (in)} = -\sqrt{\gamma}T_{\mu\nu}n^\mu
  \gamma^\nu{}_i, 
}
where $T_{\mu\nu} = T^{(\rm matter)}_{\mu\nu} + T^{(\rm
  EM)}_{\mu\nu}$ 
%
%
where $T^{(\rm EM)}_{\mu\nu}$ is given by Eq. \eqref{eq:Tembmu}.
Notice that as we approach the stellar surface the matter inertia
contribution becomes subdominant: $T_{\mu\nu} \approx T^{(\rm
  EM)}_{\mu\nu}$, and Eq. \eqref{Siinter} smoothly becomes the
densitized Poynting vector $\tilde S_i$ of Eq. \eqref{densSi} [see
  also Eq. \eqref{def:Poynting}].  This approach provides valid
boundary conditions for $\tilde S_i$ and $\tilde B^i$ for the exterior
force-free evolution.

\item Exterior to the star, $\rho_0^{(\rm num)} = 0$, and the
  force-free limit applies. In the exterior we again evolve both the
  induction equation \eqref{dtAi} and the Poynting vector
  \eqref{eq:dtSiforcefree}, only now the 3-velocity is given by
  Eq.  \eqref{eq:vifromcons}, and the evolution methods are those
  described in Sec.~\ref{sec:GREFFEalgorithm}.
 
Note that the same EM gauge has to be used in the interior and exterior
to ensure that the magnetic field will smoothly join from the ideal-MHD regime to its force-free
limit. 

\end{itemize}

The method we have just described applies to cases where we can treat
the numerical magnetosphere as if it has no matter. We have used this
method successfully in \cite{Paschalidis:2013jsa} where we studied
BHNS magnetospheres. In this paper, we demonstrate the validity of our
approach by reproducing the aligned rotator solution in
Sec. \ref{aligned_rotator}. In all these cases a dynamical GRMHD
evolution of the matter is redundant, because the fluid four-velocity
is known, and an unambiguous definition of the stellar surface is
possible. This method is ideally suited for studying the dependence on
the orbital separation of the total EM output generated from compact
binaries endowed with force-free magnetospheres. This study can be
performed by using a sequence of quasiequilibrium initial data for the
fluid and the spacetime and running simulations similar to those we
presented in \cite{Paschalidis:2013jsa}, but at multiple orbital
separations.  Moreover, the approach we described in
\cite{Paschalidis:2013jsa} is ideal for preparing relaxed EM initial
data for binary inspiral simulations, i.e., a dynamical evolution in
full GR. Important studies (such as those in
\cite{Palenzuela:2013hu,Palenzuela:2013kra}), could be enhanced by
adopting relaxed initial exterior EM fields, thereby avoiding the
initial transient behavior associated with unrelaxed fields.

In cases where a dynamical evolution is required, such as merging
binary BHNSs or NSNSs, our scheme is also applicable,
but with some modifications.

\subsubsection{Matching when the fluid rest-mass density and four velocity is determined dynamically through an evolution}

Now we will consider the case where we require a dynamical evolution of a
star. Here neither is its surface sharply defined (because most
high-resolution-schock-capturing schemes require a tenuous atmosphere
and because a dynamical evolution will cause the stellar surface to
oscillate) nor do we know a priori the fluid four-velocity.

First, the stellar surface must be defined. We propose that the ratio
$\rho_0/b^2$ be used to determine the transition from the dense MHD interior
to the tenuous force-free exterior: this ratio indicates how dominant
the magnetic field is with respect to the inertia of the matter. For
example, for an ideal gas the condition for EM dominance
$T_{\mu\nu}^{(\rm matter)} \ll T_{\mu\nu}^{(\rm EM)}$, generally
implies $\rho_0/b^2 \ll 1$ near the stellar surface where $P\ll
\rho_0 \leq \rho$. 

Therefore, if $ \rho_0/b^2\lesssim \mbox{few} \%$, then the
environment is practically force-free and the exterior velocity should
be recovered using Eq. \eqref{eq:vifromcons}. The remaining MHD
primitive variable $\rho_0$ can be recovered given the exterior
4-velocity, magnetic field and Poynting vector setting a floor value
to prevent it from becoming too small. If $ \rho_0/b^2\gtrsim
\mbox{few} \%$, then the environment is sufficiently dense and the
primitives recovery can be performed the usual way, e.g. see
\cite{UIUC_MAGNETIZED_BHNS_PAPER1}. This scheme has not been fully
implemented, yet and we will report on it in the near future.


\begin{figure*}
  \centering
    \subfigure{\includegraphics[trim =0cm 6.5cm 0cm 6.5cm,clip=True,width=0.8\textwidth]{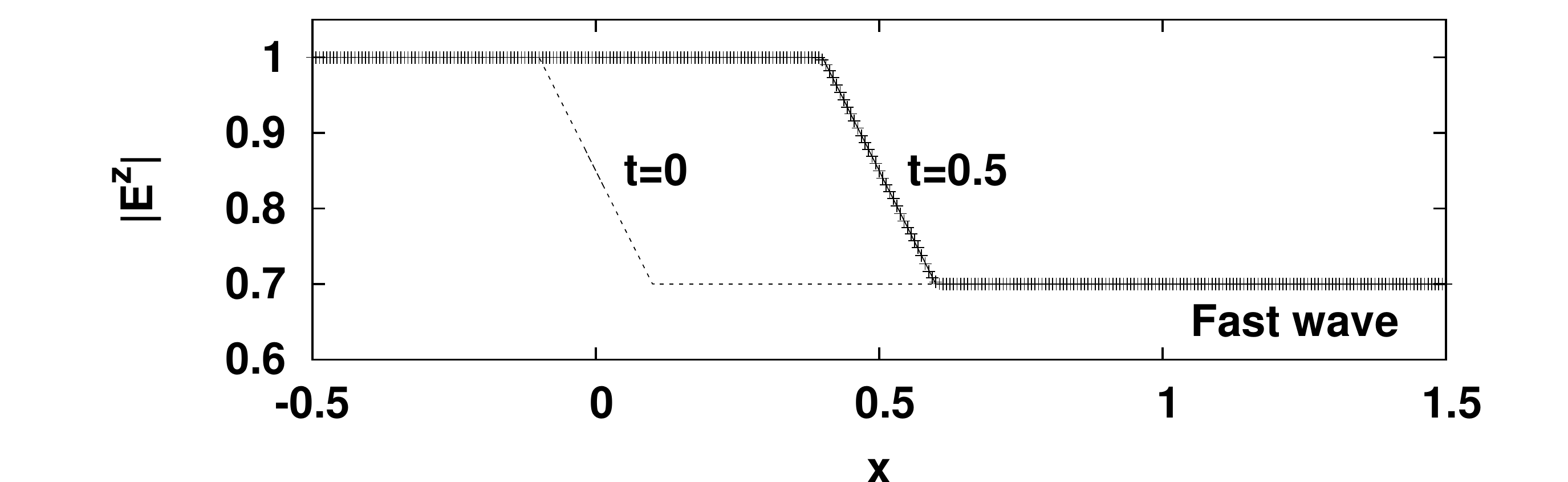}}
    \subfigure{\includegraphics[trim =0cm 6.5cm 0cm 6.5cm,clip=True,width=0.8\textwidth]{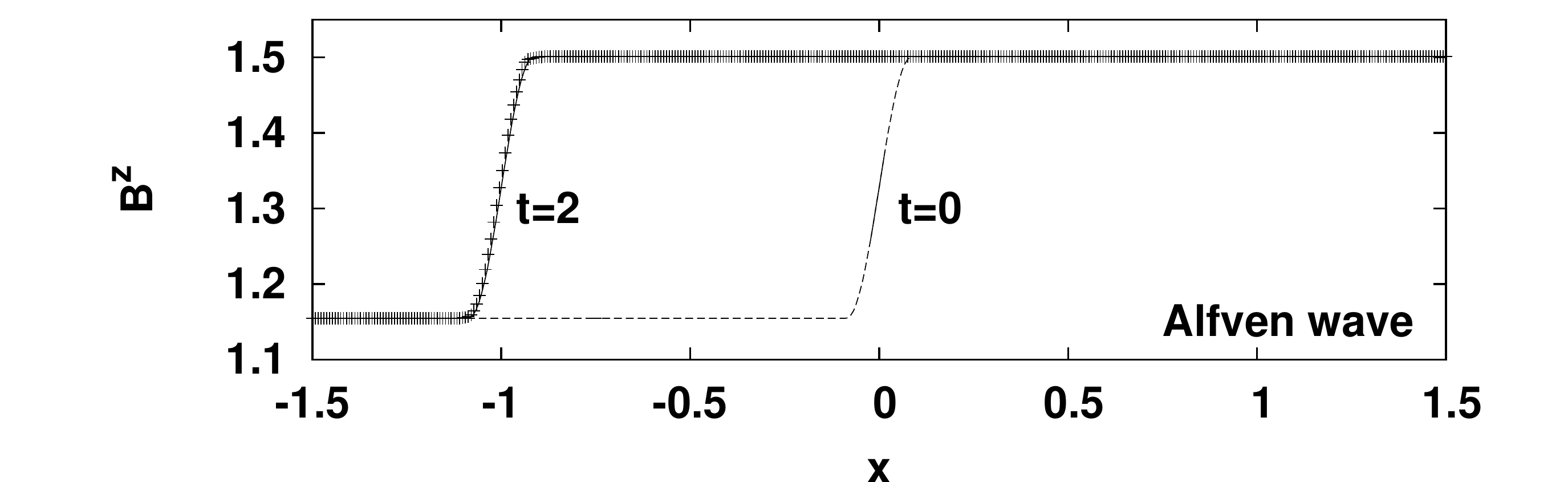}}
    \subfigure{\includegraphics[trim =0cm 6.5cm 0cm 6.5cm,clip=True,width=0.8\textwidth]{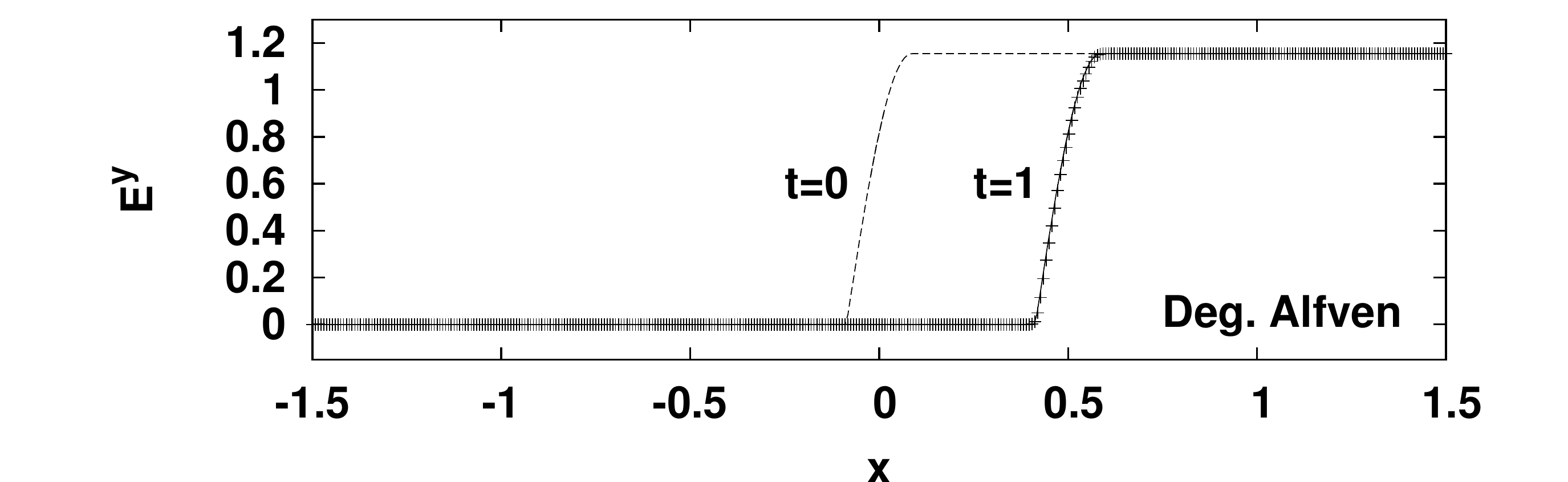}}
    \subfigure{\includegraphics[trim =0cm 6.5cm 0cm 6.5cm,clip=True,width=0.8\textwidth]{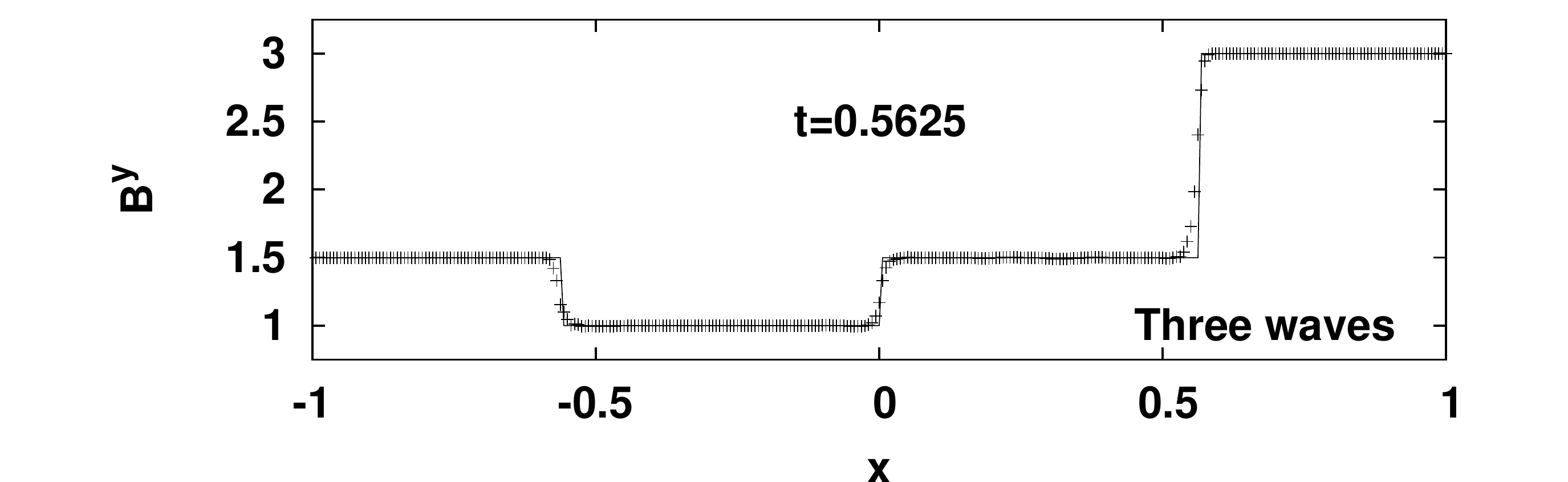}}
    \subfigure{\includegraphics[trim =0cm 6.5cm 0cm 6.5cm,clip=True,width=0.8\textwidth]{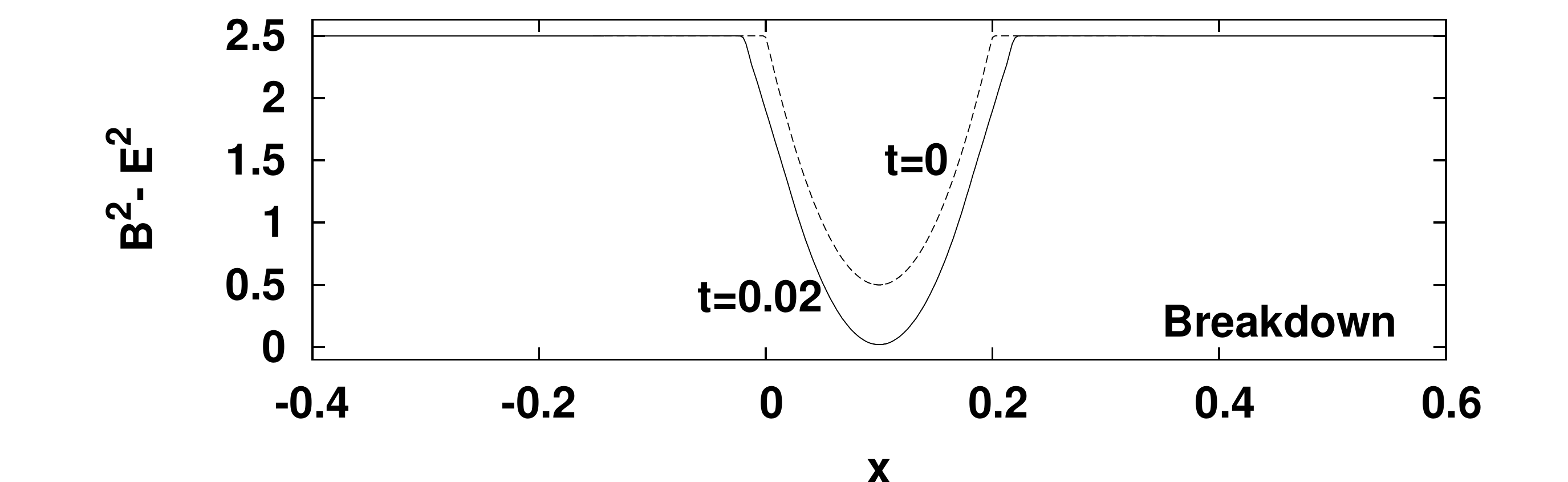}}
      \caption{Results from 1D force-free tests: dashed lines indicate
        initial data, solid lines the analytic solution (at the
        indicated time) and crosses the numerical solution, except for
        the force-free breakdown test (bottom), where the solid line
        indicates the numerical solution.
        \label{1Dtests}
      }
\end{figure*}

\section{Code Test Problems}
\label{tests}

In this section we test our new methods for evolving the GRFFE
equations and for matching ideal GRMHD to its force-free limit.  We test
our force-free implementation with a robust suite of standard 1D solutions in Minkowski
spacetime and 3D solutions in BH spacetimes, and finally we test our
new matching method by reproducing the aligned rotator solution.

For the tests shown in this section, the GRFFE equations are evolved
by a high-resolution shock-capturing technique that employs the
PPM~\cite{PPM} reconstruction, coupled to the Harten, Lax, and van
Leer approximate Riemann solver~\cite{HLL}. 

\subsection{One-Dimensional Tests in Minkowski Spacetime}

These 1D tests are based on those considered in~\cite{k02,k04}.  We
now present the grid setup, initial data for the vector potential,
and, for comparison and completion the magnetic and electric field
initial data. We do so in part to correct the initial data presented
in the literature or to use slightly modified values. All these tests
are evolved using the generalized Lorenz gauge and on uniformly-spaced
spatial grids using three resolutions. A standard Runge-Kutta 4th
order time integration scheme is employed with the Courant factor set
equal to 0.5. Results from these simulations are shown in
Fig.~\ref{1Dtests}, where it is demonstrated that our code reproduces
the exact solutions. All these plots show results from our ``medium
resolution'' runs.

\subsubsection{Fast wave} 
\label{fast_wave}

The initial configuration is defined by~\footnote{Note that~\cite{k02,k04} give the initial data for $\B^i$ 
and $\E^i$. However, since the FFE equations are invariant if $\B^i$ and $\E^i$ are multiplied 
by a constant factor, initial data with $B^i$ and $E^i$ having the same values as the ones 
with $\B^i$ and $\E^i$ are equally valid and the subsequent evolution will be exactly the same 
as the old set of initial data after multiplying an appropriate factor. Therefore, the initial data 
listed in this note are not multiplied by the factor $\sqrt{4\pi}$.} 
\labeq{}{
 \begin{split}
 B^x(0,x)= &\ 1.0, \\
 B^y(0,x) = &\ \left \{ \begin{array}{lll} 1.0 & \mbox{if} & x \leq -0.1 \\ 
						1.0 - 1.5 (x+0.1) & \mbox{if} & -0.1 \leq x \leq 0.1 \\
						0.7 & \mbox{if} & x \geq 0.1 \end{array} \right. , \\
 B^z(0,x) = &\ 0 , 
\end{split}
}

\beq
  E^x(0,x) = 0.0  \ , \  E^y(0,x) = 0.0  \ , \  E^z(0,x) = -B^y(0,x) .
\eeq
The initial data for $v^i$ can be computed using Eq.~(\ref{def:vi}), which, in Minkowski spacetime,
reduces to
\beq
  \ve{v}=\frac{\ve{E}\times \ve{B}}{B^2} .
\eeq

A vector potential generating these $B^i$ initial data is 
\labeq{}{
  \begin{split}
   A_x = &\ 0,\quad A_y=0,\\
   A_z = &\ y + \left \{ \begin{array}{lll} -x-0.0075 & \mbox{if} & x \leq -0.1 \\
				0.75x^2-0.85x & \mbox{if} & -0.1 \leq x \leq 0.1 \\
				-0.7x - 0.0075 & \mbox{if} & x \geq 0.1 \end{array} \right. .
\end{split}
}

The fast wave travels to the right with speed $\mu=1$. Hence 
the solution at time $t$ is given by 
\beq
  Q(t,x) = Q(0,x-t) ,
\eeq
where $Q$ denotes $B^i$, $E^i$, or $v^i$.

We perform this test in a domain $x\in [-0.5,1.5]$ using low, medium
and high resolutions covering the domain with 160, 320, 640 zones, respectively.

\subsubsection{Alfv\'en wave}
\label{alfven_wave}

The initial data at the wave frame are 

\labeq{}{
  \begin{split}
  B'^{x'}(x') = &\ 1.0,\ B'^y(x') = 1.0, \\
  B'^z(x') = &\ \left \{ \begin{array}{lll} 1.0 & \mbox{if} & x' \leq -0.1 \\
				1.0+0.15 f(x') & \mbox{if} & -0.1 \leq x' \leq 0.1 \\
				1.3 & \mbox{if} & x' \geq 0.1 \end{array} \right. ,
  \end{split}
}
where $f(x)=1+\sin (5\pi x)$.

\beq
  E'^{x'}(x') = -B'^z(0,x') \ \ , \ \ E'^y(x') = 0.0 \ \ , \ \ E'^z(x') = 1.0  .
\eeq
 The above data are taken from~\cite{k04}. The initial data in 
the grid frame are given by simple Lorentz boost

\labeq{eq:Bboost}{
  \begin{split}
  B^x(0,x) = &\ B'^{x'}(\gamma_\mu x) , \\
  B^y(0,x) = &\ \gamma_\mu [ B'^y(\gamma_\mu x) - \mu E'^z(\gamma_\mu x) ] , \\ 
  B^z(0,x) = &\ \gamma_\mu [ B'^z(\gamma_\mu x) + \mu E'^y(\gamma_\mu x) ] , 
\end{split}
}
\labeq{eq:Eboost}{
  \begin{split}
  E^x(0,x) = &\ E'^{x'}(\gamma_\mu x) , \\ 
  E^y(0,x) = &\ \gamma_\mu [ E'^y(\gamma_\mu x) + \mu B'^z(\gamma_\mu x) ] ,\\ 
  E^z(0,x) = &\ \gamma_\mu [ E'^z(\gamma_\mu x) - \mu B'^y(\gamma_\mu x) ],
\end{split}
}
where $\mu$ is the wave speed relative to the grid frame and $\gamma_\mu = (1-\mu^2)^{-1/2}$. 
Note that the Lorentz contraction 
$x'=\gamma_\mu x$ has been taken into account in the above transformation.
The value of $\mu$ can be anything 
between $-1$ and 1, and is set to $-0.5$ for this test. 
A vector potential that generates the initial $B^i$ is 
\labeq{}{
  \begin{split}
  A_x = &\ 0,\\
  A_y = &\ \left \{ \begin{array}{lll} \gamma_\mu x - 0.015 & \mbox{if} & x \leq -0.1/\gamma_\mu \\ 
1.15 \gamma_\mu x - 0.03g(x) & \mbox{if} & -0.1/\gamma_\mu \leq x \leq 0.1/\gamma_\mu \\ 
  		1.3 \gamma_\mu x - 0.015 & \mbox{if} & x \geq 0.1/\gamma_\mu \end{array} \right. , \\ 
  A_z = &\ y - \gamma_\mu (1-\mu)x ,
\end{split}
}
where $g(x) = \cos (5\pi \gamma_\mu x)/\pi$.
The solution at time 
$t$ is given by 
\[
  Q(t,x) = Q(0,x-\mu t) .
\]

We perform this test in a domain $x\in [-1.5,1.5]$ using low, medium
and high resolutions covering the domain with 200, 400, 800 zones.

\subsubsection{Degenerate Alfv\'en wave}
\label{deg_alfven_wave}

The initial data in the wave frame are 

\labeq{}{
  \begin{split}
  \ve{E}'(x') = &\ 0,  B'^{x'}(x') = 0, \\ 
  B'^y(x') = &\ 2 \cos \phi ,\\ 
  B'^z(x') = &\ 2 \sin \phi , 
\end{split}
}
where 
\beq
  \phi(x') = \left \{ \begin{array}{lll} 0.0 & \mbox{if} & x' \leq-0.1 \\ 
			2.5\pi (x'+0.1) & \mbox{if} & -0.1 \leq x' \leq 0.1 \\ 
		 	0.5 \pi & \mbox{if} & x' \geq 0.1 \end{array} \right. .
\eeq
The grid frame $B^i$ and $E^i$ can be obtained by Eqs.~(\ref{eq:Bboost}) and (\ref{eq:Eboost})
with arbitrary $\mu \in (-1,1)$. For this test, $\mu$ is set to 0.5.
A vector potential that generates the initial $B^i$ is 
\labeq{}{
\begin{split}
  A_x = &\ 0 , \\ 
  A_y = &\ \left \{ \begin{array}{lll} -0.8/\pi & \mbox{if} & x \leq -0.1/\gamma_\mu \\ 
	-(0.8/\pi)h_1(x) & \mbox{if} & -0.1/\gamma_\mu \leq x \leq 0.1/\gamma_\mu \\ 
			2 (\gamma_\mu x - 0.1) & \mbox{if} & x \geq 0.1/\gamma_\mu \end{array} \right. ,
\end{split}
}
where $h_1(x)= \cos [ 2.5\pi (\gamma_\mu x+0.1) ]$,
\beq
  A_z = \left \{ \begin{array}{lll} -2(\gamma_\mu x + 0.1) & \mbox{if} & x \leq -0.1/\gamma_\mu \\
	-(0.8/\pi) h_2(x) & \mbox{if} & -0.1/\gamma_\mu \leq x \leq 0.1/\gamma_\mu \\ 
			-0.8/\pi & \mbox{if} & x \geq 0.1/\gamma_\mu \end{array} \right. .
\eeq
where $h_2(x)=\sin [ 2.5\pi (\gamma_\mu x+0.1) ]$.

The Alfv\'en speeds are given by (see~\cite{k02}) 
\beq
  \mu_a^{\pm} = \frac{B_z E_y - B_y E_z \pm \sqrt{ B_x^2 (B^2-E^2)}}{B^2} .
\eeq
For the initial data set considered here, $\mu_a^+ = \mu_a^-=\mu$, hence the 
Alfv\'en wave is said to be degenerate. The solution at time $t$ is 
\[
  Q(t,x) = Q(0,x-\mu t) .
\]

We perform this test in a domain $x\in [-1.5,1.5]$ using low, medium
and high resolutions covering the domain with 200, 400, 800 zones.

\subsubsection{Three waves}
\label{three_waves}

For this test, the initial discontinuity at $x=0$ splits into two fast discontinuities 
and a stationary Alfv\'en wave. The initial data are 
\beq
  \ve{B}(0,x) = \left \{ \begin{array}{lll} (1.0,1.5,3.5) & \mbox{if} & x<0 \\
					    (1.0,3.0,3.0) & \mbox{if} & x>0 \end{array} 
\right. , 
\label{eq:3wavesB}
\eeq
\beq
  \ve{E}(0,x) = \left \{ \begin{array}{lll} (-1.0,-0.5,0.5) & \mbox{if} & x<0 \\
                                            (-1.5,2.0,-1.5) & \mbox{if} & x>0 \end{array} 
\right. .
\label{eq:3wavesE}
\eeq
A vector potential that generates the initial $B^i$ is 
\labeq{}{
  \bspl
  A_x = &\ 0 , \\ 
  A_y = &\ 3.5x H(-x) + 3.0x H(x) , \\
  A_z = &\ y - 1.5x H(-x) - 3.0x H(x) ,
  \end{split}
}
where $H$ is the Heaviside step function.

Note that this set of initial data is not the same as that in~\cite{k02}. 
The initial data in~\cite{k02} are not adopted here because they do not 
satisfy the $\ve{E} \cdot \ve{B}=0$ constraint. The initial data 
(\ref{eq:3wavesB}) and (\ref{eq:3wavesE}) are composed of three waves: 
\labeq{}{
  \bspl
  \ve{B}(0,x) = &\ \ve{B_a}(0,x) + \ve{B_+}(0,x) + \ve{B_-}(0,x), \\ 
  \ve{E}(0,x) = &\ \ve{E_a}(0,x) + \ve{E_+}(0,x) + \ve{E_-}(0,x) ,
  \end{split}
}
where 
\labeq{}{
  \begin{split}
  \ve{B_a}(0,x) = &\ \left \{ \begin{array}{lll} (1.0,1.0,2.0) & \mbox{if} & x<0 \\
					(1.0,1.5,2.0) & \mbox{if} & x>0 \end{array} 
\right. , \\ 
  \ve{E_a}(0,x) = &\ \left \{ \begin{array}{lll} (-1.0,1.0,0.0) & \mbox{if} & x<0 \\
					(-1.5,1.0,0.0) & \mbox{if} & x>0 \end{array} 
\right. 
\end{split}
}
corresponding to a stationary Alfv\'en wave, 
\labeq{}{\bspl
  \ve{B_+}(0,x) =&\ \left \{ \begin{array}{lll} (0.0,0.0,0.0) & \mbox{if} & x<0 \\
                                        (0.0,1.5,1.0) & \mbox{if} & x>0 \end{array} 
\right. , \\
  \ve{E_+}(0,x) = &\ \left \{ \begin{array}{lll} (0.0,0.0,0.0) & \mbox{if} & x<0 \\
                                        (0.0,1.0,-1.5) & \mbox{if} & x>0 \end{array} 
\right.
\end{split}
}
corresponding to the right-going fast wave, and 
\labeq{}{
  \bspl
  \ve{B_-}(0,x) = &\ \left \{ \begin{array}{lll} (0.0,0.5,1.5) & \mbox{if} & x<0 \\
                                        (0.0,0.0,0.0) & \mbox{if} & x>0 \end{array}
\right.  , \\
  \ve{E_-}(0,x) = &\ \left \{ \begin{array}{lll} (0.0,-1.5,0.5) & \mbox{if} & x<0 \\
                                        (0.0,0.0,0.0) & \mbox{if} & x>0 \end{array}
\right. \end{split}
}
corresponding to the left-going fast wave.  The solution at $t$ is given by 
\beq
 Q(t,x) = Q_a(0,x) + Q_+(0,x-t) + Q_-(0,x+t) .
\eeq

We perform this test in a domain $x\in [-1.,1.]$ using low, medium and
high resolutions covering the domain with 160, 320, 640 zones.

\subsubsection{FFE breakdown test}
\label{ffe_breakdown}

The initial data are 

\labeq{}{
  \bspl
  \ve{B}(0,x) = &\ \left \{ \begin{array}{lll} (1.0,1.0,1.0) & \mbox{if} & x<0 \\
					  (1.0,z(x),z(x)) & \mbox{if} & 0 < x < 0.2 \\
                                          (1.0,-1.0,-1.0) & \mbox{if} & x>0.2         \end{array} 
\right. , \\
  \ve{E}(0,x) = &\ (0.0,0.5,-0.5)  ,
   \end{split}
}
where $z(x) = -10.0x+1.0$.

A vector potential that generates the initial $B^i$ is 
\labeq{}{
  \bspl
  A_x = &\ 0, \\ 
  A_y = &\ \left \{ \begin{array}{lll} x-0.2 & \mbox{if} & x<0 \\
					  -5.0x^2 + x -0.2 & \mbox{if} & 0 < x < 0.2 \\
                                          -x & \mbox{if} & x>0.2         \end{array} 
\right. , \\ 
   A_z = &\ y - A_y
   \end{split}
}

According to the simulation reported in~\cite{k02}, $B^2 -E^2$
decreases in time and approaches 0 at $t \ga 0.02$, leading to the
breakdown of FFE.  We perform this test in a domain $x\in [-0.4,0.6]$
using low, medium and high resolutions covering the domain with 200,
400, 800 zones. 

In Fig.~\ref{1Dtests} we show the solution obtained with our code. The
only difference between our solution and the one shown in Fig. 5
of~\cite{k02}, is due to the fact that we plot $B^2-E^2$, while
$(B^2-E^2)/B^2$ is plotted in~\cite{k02}. When we plot $(B^2-E^2)/B^2$
our results are in excellent agreement with~\cite{k02}. However, we
prefer to show $B^2-E^2$ as was done in~\cite{m06}, with which our
results also agree. The important aspect of this problem is to
demonstrate that $B^2 - E^2 =0$ occurs at $t\approx 0.02$. As can be
seen in Fig. 1 our code reproduces the solution.



\subsection{Multidimensional, Black-Hole Spacetime Tests}

These tests are based on the 3D BH tests considered in~\cite{k04}, 
only that we perform them here in Cartesian coordinates, corresponding
to shifted Kerr-Schild (KS) coordinates, i.e., the radial coordinate on our
grid is $r = r_{\rm KS} - r_0$, where $r_{\rm KS}$ is the KS radial coordinate
and $r_0$ is a constant by which we shift the coordinate. This choice is
convenient because it excludes the BH singularity from our domain, as $r=0$
corresponds to $r_0$ in KS coordinates. The transformation from shifted KS spherical 
coordinates to Cartesian is done in the usual way. 
 
We now describe these tests, present the grid setup, and the results of 
our simulations which all reproduce the expected solutions.

\subsubsection{Split monopole}
\label{split_monopole}

The split monopole solution is derived from the Blandford-Znajek monopole solution
by inverting the solution in the lower hemisphere. 

The Blandford-Znajek monopole solution is an approximate solution for
small black-hole spin $a_{*} = a/M= J/M^2 \ll 1$.  The derivation can be found
in~\cite{bz77,mg04}. The solution in~\cite{mg04} is given in spherical
Kerr-Schild coordinates and is the one we use to perform the test .

The 4-vector potential is given by (dropping the subscript ``KS'' in r)
\beqn
 A_r   &=& -\frac{aC}{8}\cos\theta \left( 1 + \frac{4M}{r} \right) 
\sqrt{ 1 + \frac{2M}{r}} + O(a_{*}^{3}),
\hfill \label{eq:Ar-bzmono} \\
  A_\phi  &=&  M^2 C [1-\cos \theta + a^2 f(r) \cos \theta \sin^2 \theta] + O(a_{*}^4), \qquad
\label{eq:Aphi-bzmono}  \\ 
  {\cal A}_t &=& -\frac{a}{8M^2} A_\phi + O(a_{*}^3) , \label{eq:At-bzmono}
\eeqn
where $C$ is a constant and $f$ is the radial function given by Eq.~(41) of~\cite{mg04}.
\beqn
  f(r) &=& \frac{r^2(2r-3M)}{8M^3} L \left( \frac{2M}{r}\right)
+ \frac{M^2+3Mr-6r^2}{12M^2}\ln \frac{r}{2M} \cr 
&& + \frac{11}{72} + \frac{M}{3r} + \frac{r}{2M} - \frac{r^2}{2M^2} .
\eeqn
where $L$ is the dilogarithm function defined as 
\beq
  L(x) = {\rm Li}_2(x) + \frac{1}{2} \ln x \ln (1-x)  \ \ \ \mbox{for } 0<x<1
\label{def:Lx01}
\eeq
and ${\rm Li}_2$ is defined as 
\beq
  {\rm Li}_2(x) = -\int_0^1 \frac{\ln(1-tx)}{t} dt = \sum_{k=1}^\infty \frac{x^k}{k^2} .
\label{def:Li2}
\eeq

Note that our Eq.~\eqref{eq:Aphi-bzmono} is not exactly the same as in
~\cite{mg04}; the term $CM^2$ has been added to the original
expression given in~\cite{mg04} to prevent the Cartesian components of
$A_i$ from diverging on the upper $z$-axis.


The magnetic field is given by Eqs.~(47)--(49) of~\cite{mg04}. However, there is 
a factor of $\alpha$ different between their definition of $B^i\equiv {}^*F^{it}$ and the 
$B^i=n_{\nu} {}^*F^{\nu i}=\alpha {}^*F^{it}$ adopted here. So, we have
\labeq{}{
  \bspl
  B^r =&\ \frac{C \alpha M^2}{r^2} + \\
   &\ \frac{C \alpha a^2 M^2}{2r^4} \left[ -2\cos \theta + \left(\frac{r}{M}\right)^2 (1+3 \cos 2\theta) f(r) \right], \\
  B^\theta =&\ - \frac{C \alpha a^2}{r^2} \sin \theta \cos \theta f'(r), \\ 
  B^\phi =&\ -\frac{C \alpha a M}{8r^2} \left( 1 + \frac{4M}{r}\right) . 
\end{split}
}

\begin{figure}
  \centering
    \subfigure{\includegraphics[width=0.23\textwidth]{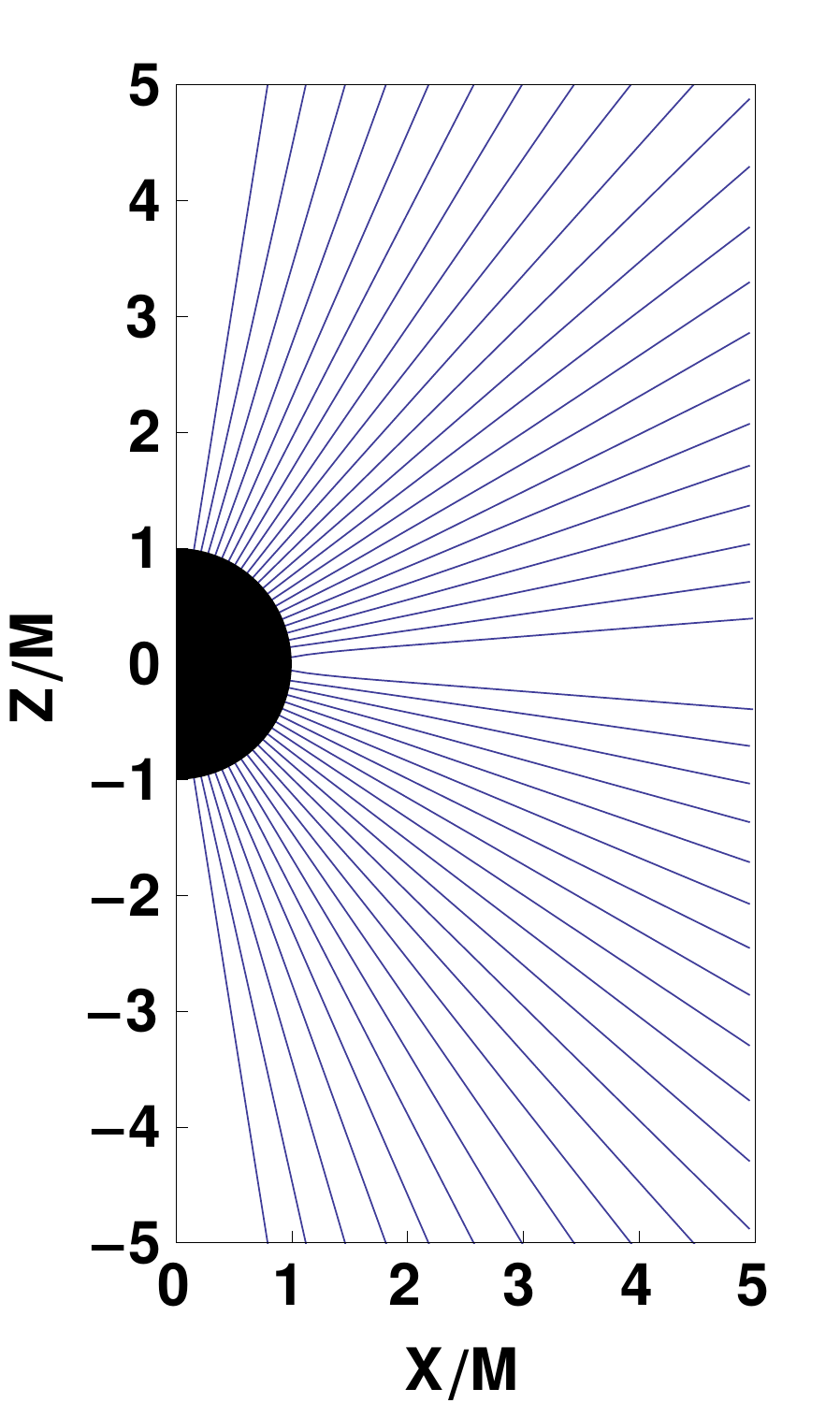}}
    \subfigure{\includegraphics[width=0.23\textwidth]{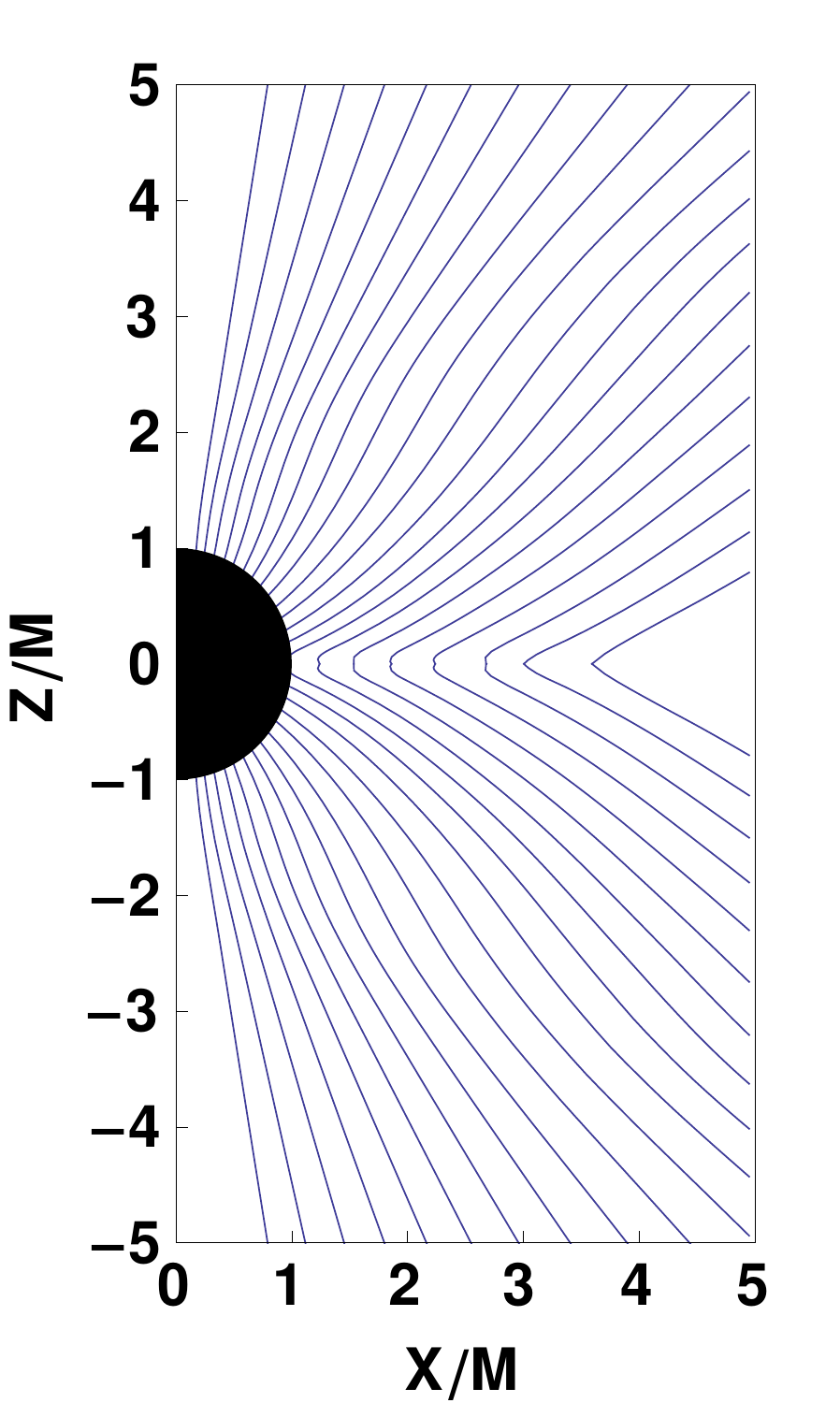}}
      \caption{Left: Split monopole initial poloidal magnetic fields lines in blue
        (black in greyscale). Black shaded area designates the
        BH interior.  Right: Same as in left, but at t = 5M. 
        \label{split_mon}
      }
\end{figure}

The Faraday tensor $F_{\mu \nu}$ is given by Eqs.~(27)--(31) of~\cite{mg04}. They can be used 
to compute the electric field: 
\beq
  E_i = n^{\mu} F_{i\mu} = -\frac{1}{\alpha} F_{ti} + \frac{\beta^j}{\alpha} F_{ji} ,
\eeq
where, for the Kerr-Schild metric in spherical coordinates, 
\labeq{eq:ksabg}{
  \bspl
  \alpha = &\ \left( 1 + \frac{2Mr}{\Sigma} \right)^{-1/2} , 
  \beta^r = \frac{2Mr}{\Sigma + 2Mr}, \beta^\theta=  \beta^\phi=0 , \\
  \sg =&\ \Sigma \sin \theta \sqrt{1 + \frac{2Mr}{\Sigma} } ,
\end{split}
}
and where $\Sigma = r^2+a^2\cos^2 \theta$. Finally, 
\beqn
  E_r &=& -\frac{C a^3}{8\alpha M^3} f'(r) \cos \theta \sin^2 \theta  \\ 
  E_\theta &=& -\frac{Ca}{8\alpha}[\sin \theta + a^2 f(r) \sin \theta (2 \cos^2 \theta-\sin^2 \theta) ] \cr
&&- \beta^r \sg \frac{a C}{8 r^2}\left( 1+\frac{4M}{r}\right) \\ 
E_\phi &=& \frac{\beta^r}{\alpha M} Ca^2 f'(r) \cos \theta \sin^2 \theta .
\eeqn

Note that $f(r) \sim -r^2 \ln r/4$ as $r \rightarrow \infty$, invalidating the 
solution at large $r$ (because $B^2 < E^2$ at sufficiently large $r$). Following 
~\cite{k04,m06} we drop terms involving $f(r)$ and $f'(r)$, making the solution accurate 
only to first order in $a_*$.

To perform the split monopole the constant $C$ is changed to $-C$ in 
the lower hemisphere ($\theta > \pi/2$), in the expressions of $B^i$ and $E^i$. 
The vector potential for the split monopole test then can be written as 
\beqn
  A_r &=& - \frac{Ca}{8} |\cos \theta| \left( 1 + \frac{4M}{r} \right)
\sqrt{ 1 + \frac{2M}{r}} + O(a_{*}^3) \qquad \\
  A_\phi &=& C (1- |\cos \theta|) + O(a_{*}^2) .
\eeqn

As pointed out in \cite{k04}, the split-monopole configuration is
sensible only if there exists a conducting disc in the equatorial
plane of the black hole to sustain it. Otherwise, the equatorial
current sheet cannot be stable -- the magnetic field lines will
reconnect and be pushed away. If one assumes that the equatorial
current sheet is stable, because it is sustained by a disk, then no
reconnection is expected to take place. We can model both scenarios by
turning off and on our resistivity prescription, i.e., the nulling of
the inflow velocity in the current sheet. If we do null the inflow
velocity into the equatorial current sheet, no reconnection takes
place and our solution is in agreement with the solution found in
\cite{m06}, as expected.  The results of this test without the
resistivity prescription are shown in Fig. \ref{split_mon}, and are in
good agreement with the ones obtained in \cite{k04}.

We perform this test setting $a_{*} = 0.1$, and chose $r_0 = 1.0M$, so that
the BH horizon corresponds to $r \approx 0.995M$ in the shifted KS
radial coordinate. We perform this test on a fixed-mesh-refinement
grid hierarchy with 6 levels of refinement setting the outer boundary
at 100M. The half-side length of the refinement levels is $3.125\times
2^{6-n}M, \ \ n=1,2,\ldots, 6$, $n=1$ indicating the coarsest level in
the hierarchy. The resolution of each level is $\Delta x_{\min}\times
2^{6-n}, \ \ n=1,2,\ldots, 6$, where $\Delta x_{\min}$ is the
resolution of the finest level. We use 3 resolutions $\Delta
x_{\min}=M/8$, $\Delta x_{\min}=M/16$, and $\Delta x_{\min}=M/24$. The
Courant factor is set to $0.03125\times 2^{n-1}, \ n=1,\ldots, 3$ and
0.5 for $n=4,5,6$.  We use the generalized Lorenz gauge to run the
test with damping parameter $\xi = 4/M$.  The
solution shown in Fig.~\ref{split_mon} corresponds to our low
resolution run, and the results of all other resolutions are
almost overlapping, indicating that the resolutions used are
sufficiently high.


\subsubsection{The Wald solution}
\label{wald}

The EM field of the solution to Maxwell's equations in the
electrovacuum about a black hole is generated by the 4-vector
potential 
\beq {\cal A}_\mu = \frac{B_0}{2} (\phi_\mu + 2 a t_\mu) ,
\eeq
 where $B_0$ is a constant, $\phi^\mu = (\partial /\partial \phi)^\mu$ and $t^\mu
 = (\partial /\partial t)^\mu$.

In the Schwarzschild black hole case ($a=0$), 
this electrovacuum solution is also a force-free solution, which will be the 
case considered in this test. The 4-vector potential in this case is 
\beq
  {\cal A}_\mu = \frac{B_0}{2} \phi_\mu .
\eeq

The 3-vector potential is $A_i = B_0 \phi_i/2= B_0 g_{\phi i}/2$. In Kerr-Schild 
metric written in spherical coordinates, the only nonvanishing component is 
\labeq{Aphiwald}{
  A_\phi = \frac{B_0}{2} g_{\phi \phi} = \frac{B_0}{2} r^2 \sin^2 \theta .
}
The magnetic field is given by 
\beq
  B^i = \epsilon^{ijk} \partial_j A_k = \frac{[ijk]}{\sqrt{\gamma}} \partial_j A_k ,
\eeq
where $[ijk]$ denotes the antisymmetric permutation symbol. Hence the components of 
$B^i$ in spherical Kerr-Schild coordinates are 
\beqn
  B^r &=& \frac{1}{\sg} \p_\theta A_\phi = B_0 \left( 1+ \frac{2M}{r}\right)^{-1/2} \cos \theta \\ 
  B^\theta &=& -\frac{1}{\sg} \p_r A_\phi = -\frac{B_0}{r} \left( 1+ \frac{2M}{r}\right)^{-1/2} \sin \theta \\ 
  B^\phi &=& 0 \\
  B_r &=&  B_0 \cos \theta \sqrt{1+ \frac{2M}{r}} \\ 
  B_\theta &=& -B_0 r \sin \theta \left( 1+ \frac{2M}{r}\right)^{-1/2} \\ 
  B_\phi &=& 0  \\ 
  B^2 &=& B_0^2 \left( 1 - \frac{2M}{r+2M} \sin^2 \theta \right) .
\eeqn
As $r \rightarrow \infty$, $B^i$ becomes a uniform vector with magnitude $B_0$ and points in the 
$z$-direction. To compute $E^i$, first calculate 
\labeq{Ei}{
  E_i = n^\nu F_{i\nu} = \frac{1}{\alpha} (F_{i0} - \beta^j F_{ij}) .
}
Given that 
$
  F_{i0} = \p_i {\cal A}_0 - \p_t {\cal A}_i =\frac{B_0}{2} \p_i \phi_0 
= \frac{B_0}{2} \p_i g_{t\phi} = 0 $ and $F_{ij}=\p_i A_j - \p_j A_i$. 
The nonzero  $F_{\mu\nu}$ components are 
\labeq{Fmnunuwald}{
  \bspl
  F_{r\phi} = &\ -F_{\phi r} = B_0 r \sin^2 \theta, \\ 
  F_{\theta \phi} = &\ -F_{\phi \theta} = B_0 r^2 \sin \theta \cos \theta .
  \end{split}
}
By use of Eq. \eqref{Fmnunuwald}, Eq. \eqref{Ei}  yields
\beq
  E_r = E_\theta = 0 \ \ \ , \ \ \ 
  E_\phi = 2M B_0 \left( 1+ \frac{2M}{r}\right)^{-1/2} \sin^2 \theta ,
\eeq
and 
\beqn
  E^r &=& E^\theta=0 \ \ \ , \ \ \ 
  E^\phi = \frac{2MB_0}{r^2} \left( 1+ \frac{2M}{r}\right)^{-1/2} \\ 
  E^2 &=& \frac{4M^2 B_0^2 \sin^2 \theta}{r^2} \left( 1+ \frac{2M}{r}\right)^{-1} .
\eeqn

The electric field vanishes as $r\rightarrow \infty$. It is now
straightforward to see that ${\bf E}\cdot {\bf B}=0$ and $E^2 <
B^2$. Hence, this electrovacuum solution is also a force-free solution.

The 3-velocity can be calculated by 
\labeq{vi}{
  v^i = \alpha \frac{\epsilon^{ijk} E_j B_k}{B^2} - \beta^i  
 = \frac{[ijk] E_j B_k}{B_0^2 r \sin \theta (r+2M\cos^2 \theta)} - \beta^i .
}
and we find
\beqn
  v^r &=& -\frac{2M\cos^2 \theta}{r+2M\cos^2 \theta} \\
  v^\theta &=& \frac{M \sin 2\theta}{r (r+2M \cos^2 \theta)} \\ 
  v^\phi &=& 0 .
\eeqn

\begin{figure}
  \centering
    \subfigure{\includegraphics[width=0.23\textwidth]{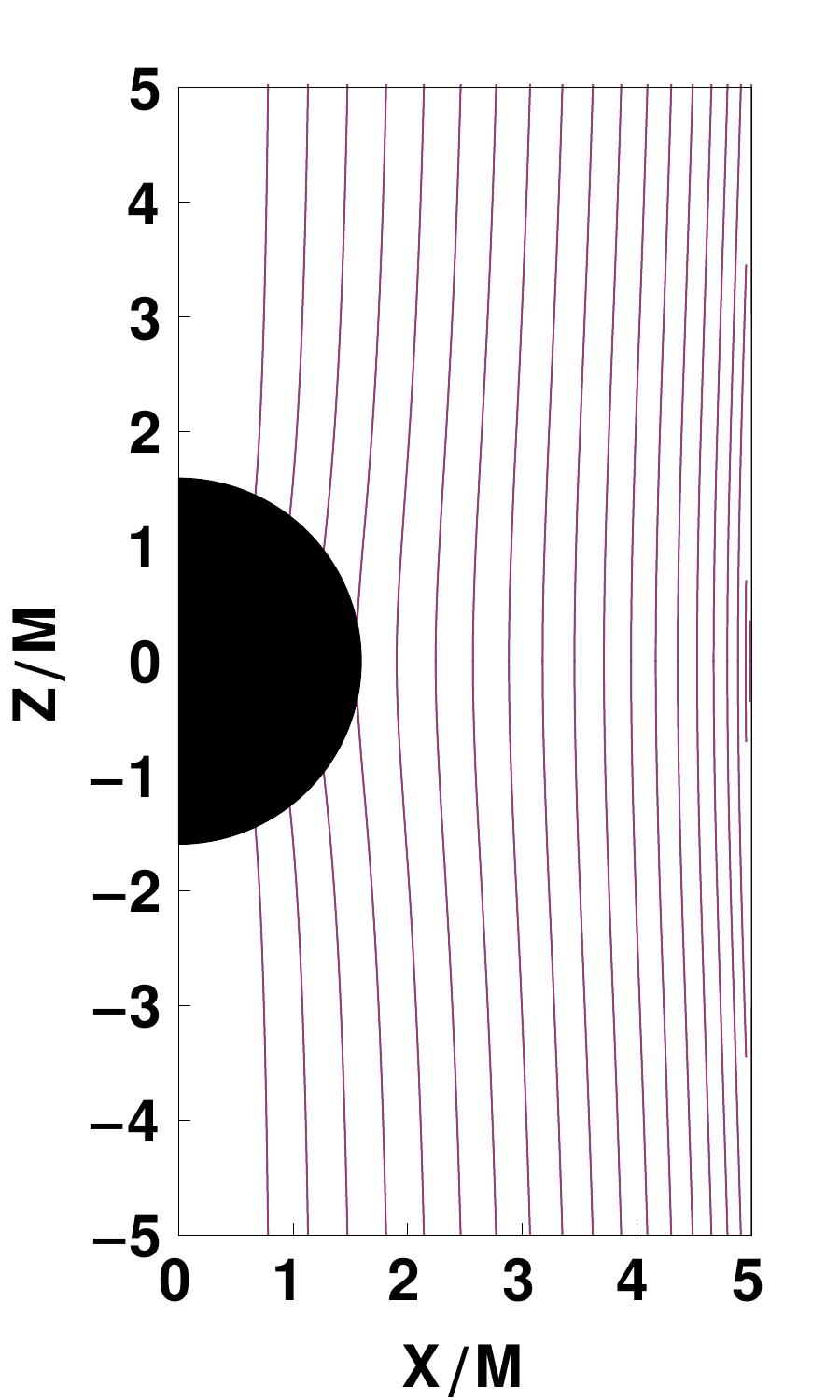}}
      \caption{Poloidal magnetic field lines for the Schwarzchild electrovacuum Wald solution. Poloidal magnetic fields lines at $t=0$M in blue
        (black in greyscale), and at $t=5$M in red (grey in greyscale). Black shaded area designates the
        BH interior. The lines at $t=5$M are overlapping with those at $t=0$M.
        \label{vac_wald}
      }
\end{figure}

For this test we arbitrarily chose $r_0 = 0.4M$, so that the BH horizon corresponds to $r = 1.6M$ in the shifted
KS radial coordinate. We perform this test on the same fixed-mesh-refinement grid hierarchy as the split-monopole
test, using the same 3 resolutions and EM gauge. In Fig.~\ref{vac_wald} we show the Poloidal field lines at $t=0M$ and $t=5M$ for the 
low resolution run - the two overlap and cannot be distinguished by eye. Since this test is the only smooth 3D
exact solution in the testbeds we consider we use it to also show that our code is convergent. Our convergence test
study is presented in section Sec.~\ref{convergence}.

\subsubsection{Magnetospheric Wald Problem}
\label{magnetospheric_wald}

\begin{figure}
  \centering
    \subfigure{\includegraphics[width=0.23\textwidth]{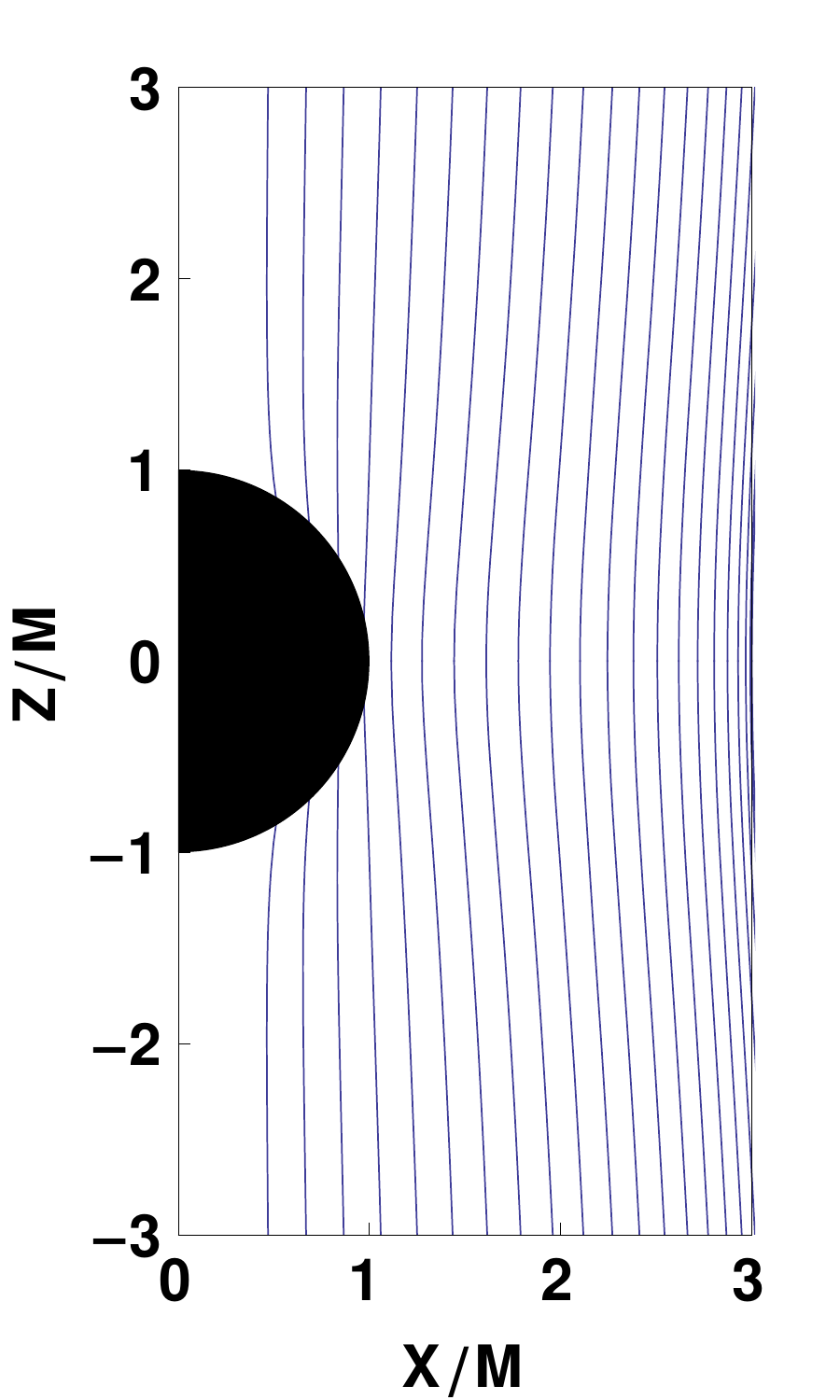}}
    \subfigure{\includegraphics[width=0.23\textwidth]{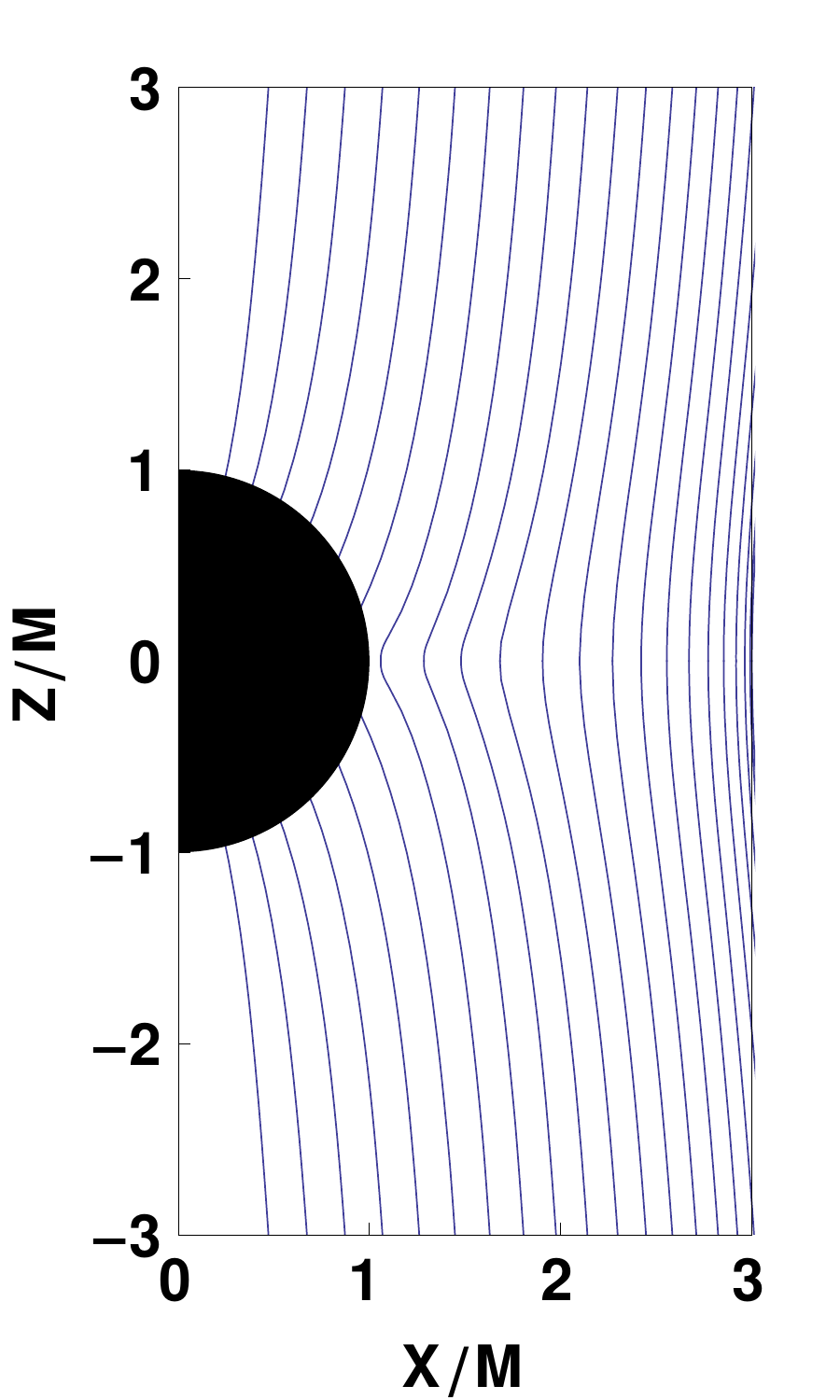}}
      \caption{Left: Magnetospheric Wald initial poloidal magnetic fields lines in blue
        (black in greyscale). Black shaded area designates the
        BH interior.  Right: Same as in left, but at t = 126M. 
        \label{fig:wald}
      }
\end{figure}

This again is a force-free problem. The initial data for the magnetic
field are given by the same spatial vector potential as the Wald's
solution, i.e., \beq A_i = \frac{B_0}{2} (\phi_i + 2 a t_i) =
\frac{B_0}{2} (g_{i\phi} + 2a g_{ti}) .  \eeq However, the electric
field is set to 0 initially, as in~\cite{k04}. Hence $S_i=0$ and
$v^i=-\beta^i$.  There is no analytic solution to this problem. The
evolution of the initial data is expected to reach a steady state
similar to the one reported in~\cite{k04}. Following \cite{k04}, we
perform this test setting $a_*=0.9$. We also set $r_0 = 0.4359M$, so that
the BH horizon lies at $r \approx 1.0M$ on our grid. We perform this
test on the same fixed-mesh-refinement grid hierarchy as the other BH
tests, using the same 3 resolutions and EM gauge. In Fig.~\ref{fig:wald} we plot
the poloidal magnetic field lines at $t=126M$ at which point the
solution has reached steady state and is very similar to that obtained
in \cite{k04}. 

\subsection{Force-free aligned rotator}
\label{aligned_rotator}

Here we reproduce the aligned rotator, force-free solution in flat
spacetime
\cite{Contopoulos:1999ga,Komissarov:2005xc,McKinney:2006sd,Spitkovsky:2006np}.
However, instead of applying the boundary condition on the NS surface,
we use our new matching technique, which we described in
Sec.~\ref{easy_matching}, to evolve the magnetic field both interior
and exterior to the star. In this approach the density profile of the
star can be anything, as the magnetic field does not back-react onto
the matter and an integration of the ideal MHD fluid equations is not
performed. Instead, the density and velocity are evolved simply by
``rotating'' their initial values as described in
~\cite{Paschalidis:2012ff}. The density profile serves only as a proxy
for locating the surface of the star. 
We endow the star with a uniform rotational three-velocity
\labeq{}{
\mathbf{v} = \Omega \mathbf{e}_z \times \mathbf{r},
}
where $\mathbf{e}_z$ is the unit vector in the z-direction, and
$\Omega$ is the stellar angular velocity. As in
\cite{Spitkovsky:2006np} we choose a spherical star and set $\Omega$
such that the theoretically expected location of the light cylinder
radius, $R_{\rm LC}$, is 5 stellar radii away from the stellar center,
i.e., $\Omega = 1/5R_{\rm NS}$. In the exterior, the 3-velocity is set
to 0. The electric field is set according to $\ve{E} =
-\ve{v}\times\ve{B}$ everywhere and the Poynting vector is calculated
using Eq. \eqref{def:Poynting}.

The star and its magnetosphere are endowed with a magnetic field
corresponding to a dipole determined by the toroidal vector potential 
\labeq{}{ A_\phi = \frac{\mu\varpi^2}{r^3},
} 
where $\mu = B_p R_{\rm NS}^3 /2$ is the magnetic dipole moment, the
cylindrical radial coordinate $\varpi^2 = x^2+y^2$, and
$r=\sqrt{x^2+y^2+z^2}$ is the radial coordinate.

We perform the test using 8 levels of refinement and set the outer
boundary at $35.3R_{\rm LC}$.  The length of each refinement box is
$2.94R_{\rm NS}\times 2^{8-n}, n=1,2,\ldots,8$, where $n=8$
corresponds to the finest refinement level.  We use 3 resolutions: the
low, medium and high resolutions cover the stellar radius with 34, 68
and 87 zones, respectively.

\begin{figure}
  \centering
    \subfigure{\includegraphics[width=0.23\textwidth]{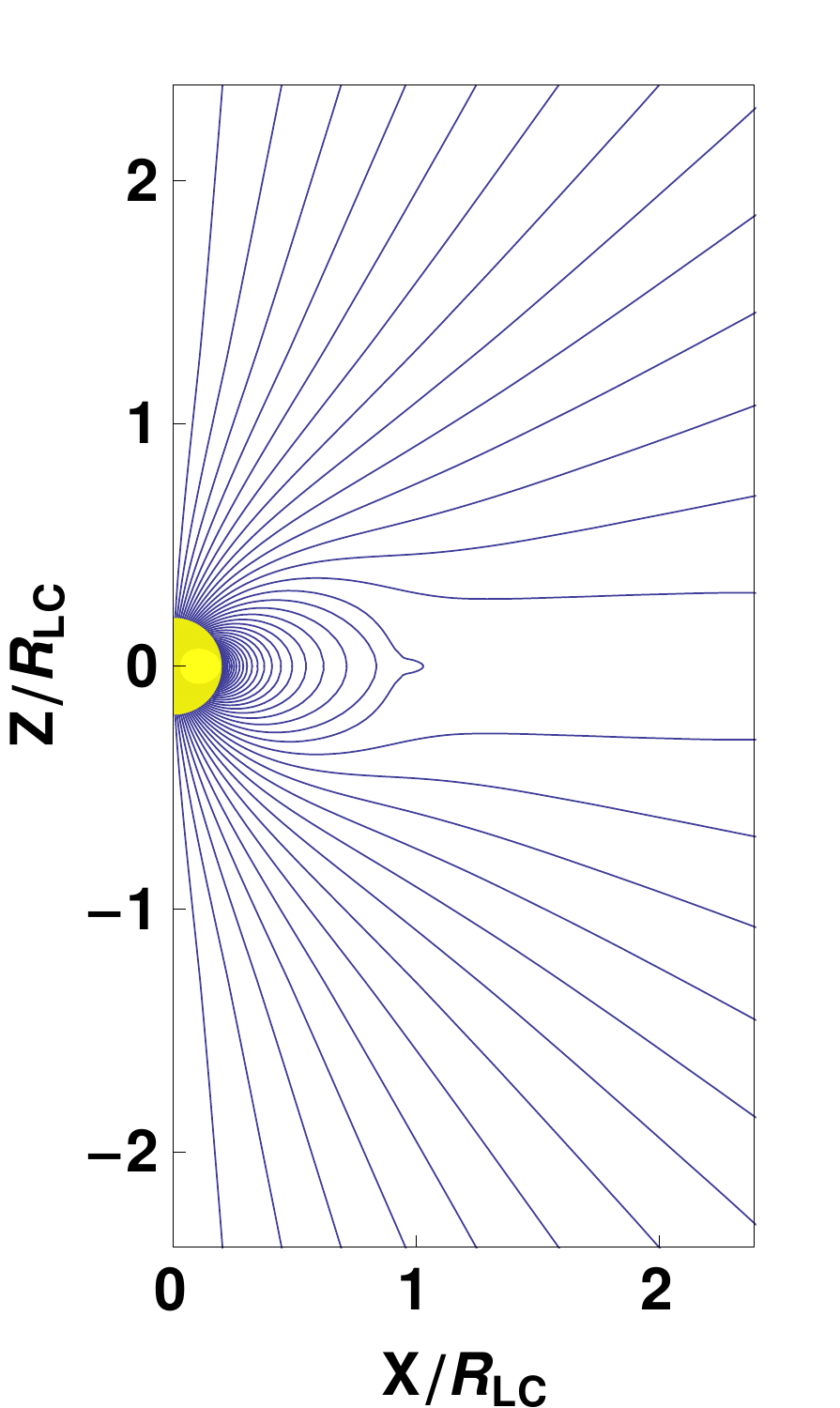}}
    \subfigure{\includegraphics[width=0.246\textwidth]{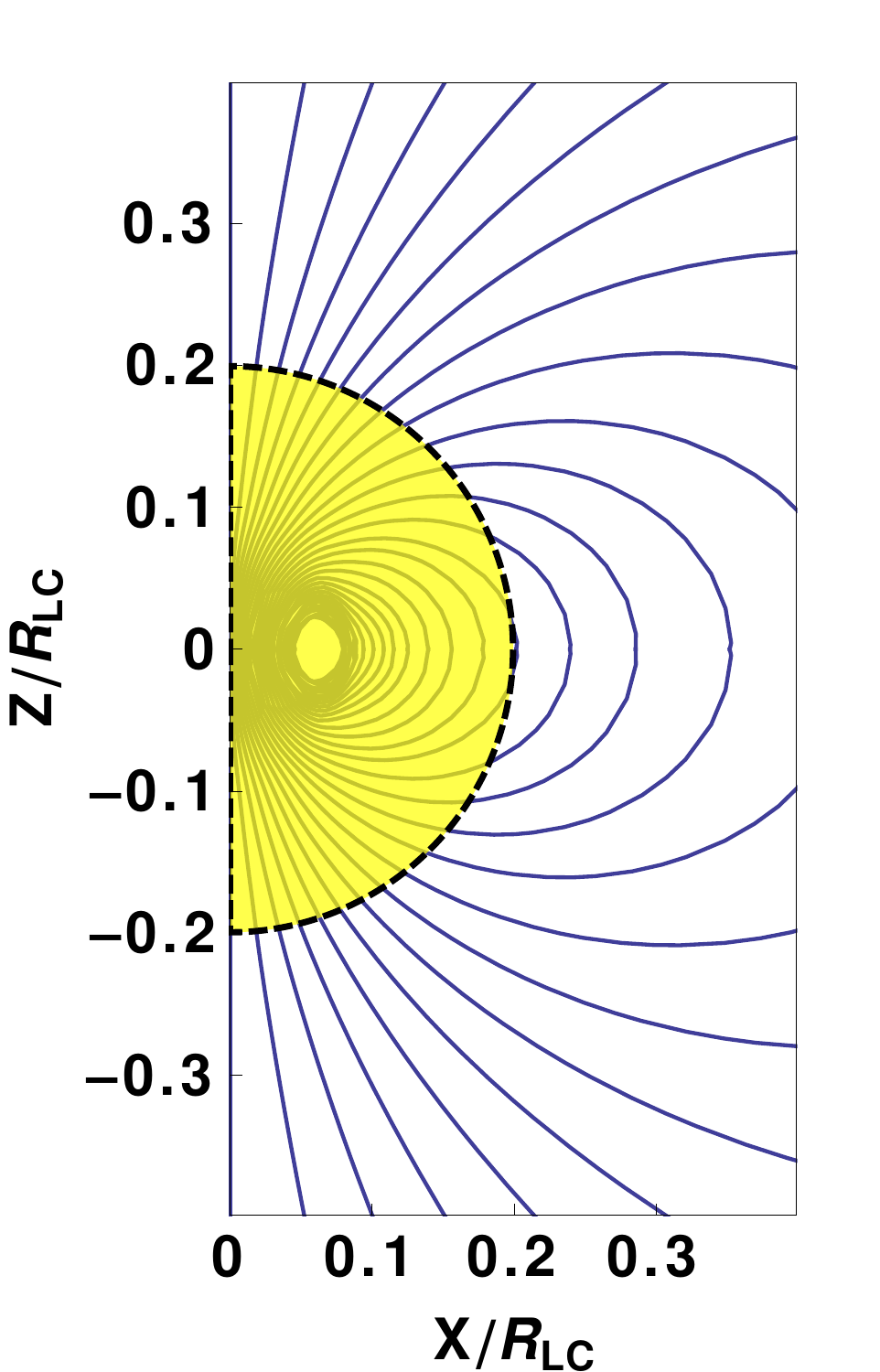}}
      \caption{Left: Far field poloidal magnetic fields lines in blue
        (black in greyscale). Yellow (grey in greyscale) shaded area designates the
        stellar interior.  Right: Same as in left, but zooming in on the
        near zone and showing the interior magnetic field, and how it smoothly
        joins to the exterior one. The time $t=6\pi/\Omega$, at which point the field
        has reached a stationary state.
        \label{rot_star}
      }
\end{figure}

\begin{figure}
  \centering
    \subfigure{\includegraphics[trim =2.cm 4.cm 1.7cm 4.5cm,clip=True,width=0.48\textwidth]{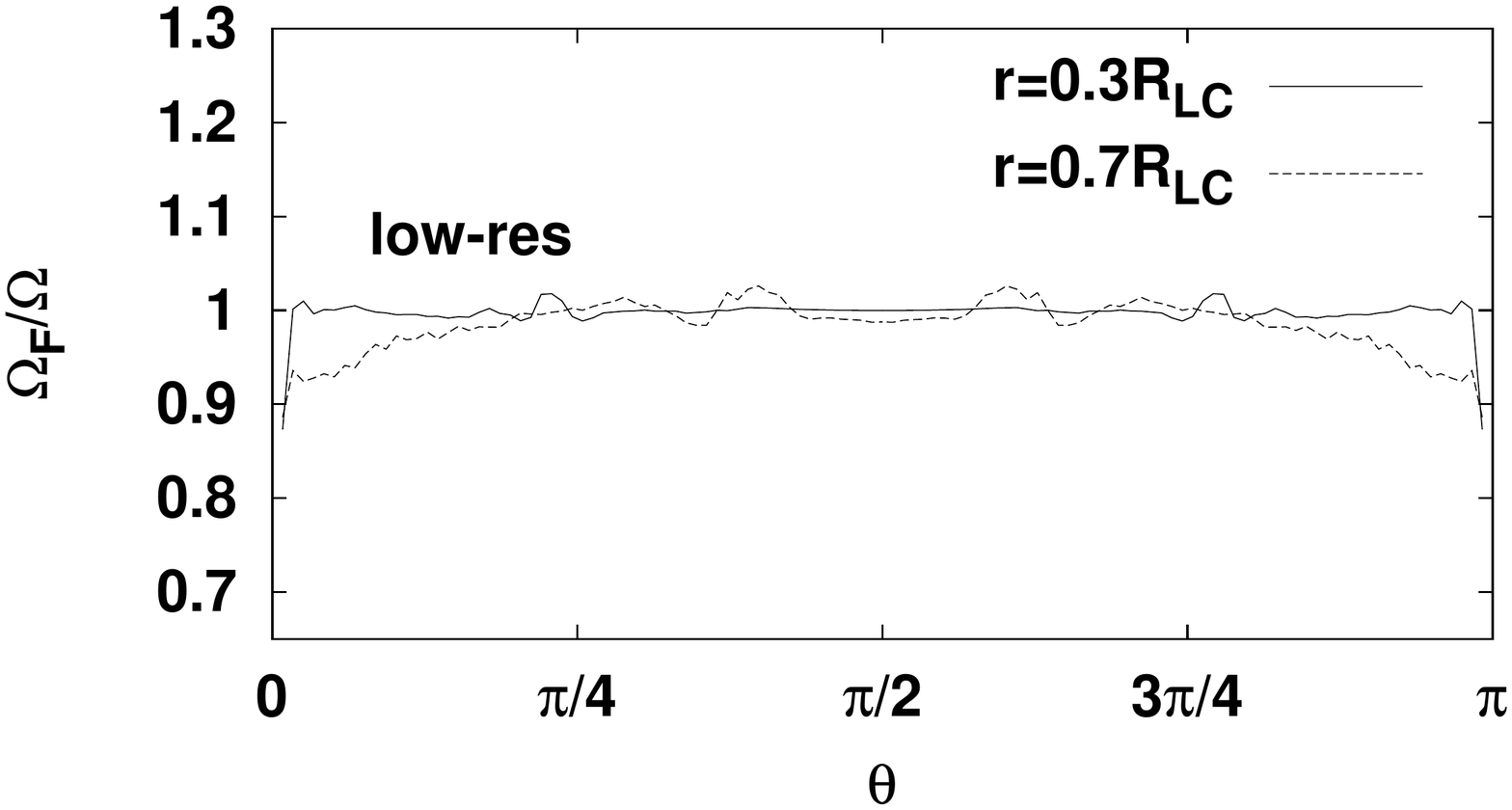}}
    \subfigure{\includegraphics[trim =2.cm 4.cm 1.7cm 4.5cm,clip=True,width=0.48\textwidth]{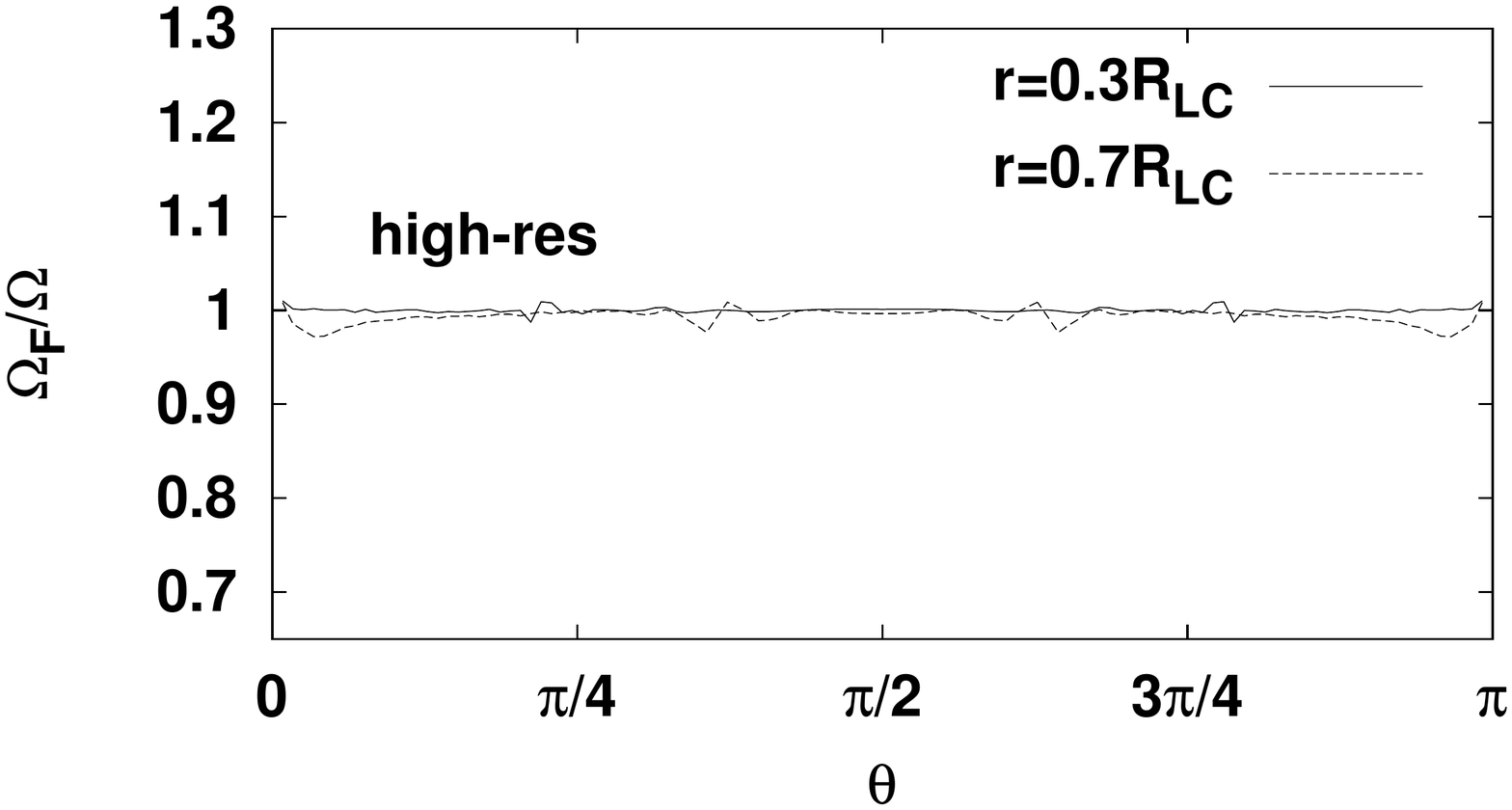}}
      \caption{Angular frequency of the magnetic field lines
        $\Omega_F$ normalized by the spin angular frequency of the
        star $\Omega$ vs the polar angle $\theta$ on the x-z plane for
        $r = 0.3 R_{\rm LC}$ (solid lines) and $r=0.7R_{\rm LC}$
        (dashed lines). The value of this ratio should be unity. Two
        resolutions are shown here: low resolution (top) and high
        resolution (bottom). It is clear that the magnetosphere within
        the light cylinder is in near corotation with the star even
        for the low resolution case.
        \label{OmegF}
      }
\end{figure}

In Fig. \ref{rot_star} we show the poloidal magnetic field lines in the x-z plane, 
where it is clear that our code successfully captures the standard features of the
pulsar magnetosphere: 1) the formation of a Y-point at the expected location of 
the light cylinder, 2) open field lines above the equatorial current sheet and 
beyond the light cylinder, and 3) dipole magnetic field structure within the light cylinder. 
In addition, the evolved interior field remains ``frozen in'' to the rotating matter. 
The right panel of the figure shows the structure of the magnetic field in the interior
and the immediate exterior of the star, demonstrating that our matching technique is 
smooth. 

The expected spin-down luminosity of an aligned rotator is $L = (1\pm
0.05)\mu^2 \Omega^4$ \cite{Spitkovsky:2006np}. We have calculated the
outgoing EM radiation using the Poynting flux and the Penrose scalar
$\phi_2$ (see e.g.~\cite{Paschalidis:2013jsa}), and we find that it
converges to a value within $4\%$ of $\mu^2 \Omega^4$, and hence in
good agreement with previous studies.

We plot the angular frequency of the magnetic field lines in the
exterior \cite{bz77} 
\labeq{}{ \Omega_F(r,\theta) =
  \frac{F_{tr}}{F_{r\phi}} = \frac{F_{t\theta}}{F_{\theta \phi}} , 
}
on the x-z plane at $r = 0.3R_{\rm LC}$ and $r = 0.7R_{\rm LC}$ as a
function of the polar angle $\theta$.  The result after $\sim 3$ periods of
evolution is shown in Fig. \ref{OmegF}
(cf. \cite{Komissarov:2005xc,McKinney:2006sd} who perform axisymmetric
high-resolution simulations). It is clear that the magnetosphere
within the light cylinder corotates with the star and that the higher
the resolution, the closer is the magnetosphere to corotation.

\subsection{Convergence}
\label{convergence}

The Wald vector potential which generates the stationary
magnetic field is itself time independent provided the proper
electromagnetic gauge choice is made. 
 A straightforward calculation demonstrates that
\labeq{ucrossB}{
\begin{split}
\epsilon_{ijk} v^j B^k = &\ \epsilon_{ijk}\bigg(\alpha \frac{\epsilon^{j\ell m}E_\ell B_m}{B^2}-\beta^j\bigg) B^k \\
                       = &\ -\alpha E_i - \epsilon_{ijk}\beta^j B^k \\
                       = &\ -\alpha E_i - \tilde\epsilon_{ijk}\tilde\epsilon^{k\ell m}\beta^j \partial_\ell A_m \\
                       = &\ -\alpha E_i - \beta^j( \partial_i A_j - \partial_j A_i) \\
                       = &\ -\alpha E_i -\beta^j F_{ij} = 0,
\end{split}
} where in the first line we used Eq. \eqref{vi}, in the second line
we used the degeneracy constraint (${\bf E}\cdot {\bf B} = 0$), in the
second and fourth lines we used the property $\epsilon_{ijk}\epsilon^{j\ell m} =
(\delta_i{}^\ell\delta_k{}^m-\delta_i{}^m\delta_k{}^\ell)$ and in the
third line we used the definition $\tilde B^k=\tilde\epsilon^{k\ell
  m}\partial_\ell A_m$. The last equality in the fifth line holds
true because of Eq. \eqref{Ei} and $F_{i0} = 0$. This result
implies that $\partial_t {\bf B} = 0$ from the magnetic induction equation \eqref{eq:induction},
but also has an interesting consequence regarding the typical electromagnetic gauges we use in
our code: The evolution equation for the vector potential is given by Eq. \eqref{dtAi}.
In the original algebraic electromagnetic gauge \cite{Etienne:2010ui} $\alpha \Phi = \beta^j A_j$. 
Hence, the evolution equation \eqref{dtAi} preserves the initial $A_i$ field ($\partial_t A_i = 0$). 

In fact, a straightforward calculation using the Wald vector potential
shows that $\beta^i A_i = 0$, which implies that the right-hand-side
of Eq. \eqref{dtAi} must be
\labeq{walddtAi}{
\partial_t A_i =  - \partial_i(\alpha \Phi)
}

Thus, any electromagnetic gauge condition, for which $\partial_i(\alpha\Phi)=0$ will
preserve the initial vector potential. Thus, the gauge $\alpha\Phi=\mbox{const.}$ also
preserves the initial A-field.

\begin{figure}[t]
  \centering
    \subfigure{\includegraphics[trim =5cm 1.5cm 3cm 1.5cm,clip=True,width=0.44\textwidth]{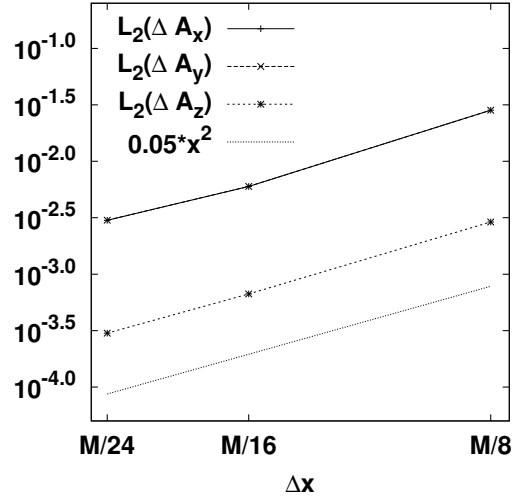}}
      \caption{Convergence test for the time-independence of the
        vector potential $A_i$. The plot shows the $L_2$ norms of the
        difference at $t=5$M between the numerical and analytic solutions in the
        volume contained inside a coordinate sphere of radius $r =
        90M$. The plot demonstrates that the error norms converge to
        zero as $\sim \Delta x^2$, i.e., the order of convergence of
        our code is 2. Note that $L_2(A_x)\simeq 3.52\times 10^4$, so
        that $L_2(\Delta A_x)/L_2(A_x) \sim 10^{-6}$ even for the
        lowest-resolution run.
        \label{vac_wald_conv}
      }
\end{figure}

This is the case in the (generalized Lorenz) gauge \cite{UIUCEMGAUGEPAPER,BriansLatest2012} we have developed, 
as we now demonstrate. The generalized Lorenz gauge is 
\labeq{}{
\partial_t (\sqrt{\gamma}\Phi) + \partial_j(\alpha \sqrt{\gamma}A^j) - \partial_i (\sqrt{\gamma}\beta^i \Phi) = -\xi \alpha\sqrt{\gamma} \Phi,
} 
where $\xi$ is the damping parameter. For the Wald solution $\partial_j(\alpha \sqrt{\gamma}A^j) = 0$, 
hence, 
\labeq{}{
\partial_t (\sqrt{\gamma}\Phi) - \partial_i (\sqrt{\gamma}\beta^i \Phi) = -\xi \alpha\sqrt{\gamma} \Phi.
} 
This means that if $\Phi = 0$ initially, $\partial_t \Phi = 0$. Thus, even the generalized 
Lorenz gauge will preserve the Wald A-field \eqref{Aphiwald}, as long as the initial value for 
$\Phi$ is 0. Of course due to truncation error the right-hand-side of the evolution equation for 
$A_i$ will not be exactly 0, but will converge to 0 at second order which is the accuracy of our
scheme. 

A test of this convergence is shown in Fig. \ref{vac_wald_conv}, where we plot the L2 norm
of the difference between the numerical and analytic solution for the vector potential defined as 
\labeq{}{
L_2(\Delta A_i) = \sqrt{\int (A_i^{\rm num}-A_i^{\rm ex})^2 d^3x}, 
}
where $A_i^{\rm num},\ A_i^{\rm ex}$ designate the numerical and exact solutions, respectively. 
This norm should converge to 0 with increasing resolution. Fig. ~\ref{vac_wald_conv} demonstrates
that our code is second-order convergent.

\section{Summary}
\label{sec:summaryandfuturework}

Neutron stars either in isolation or in compact binaries are likely to
be endowed with a force-free magnetosphere. For inpiralling binaries,
the GWs in the premerger regime can be accompanied by detectable
``precursor'' electromagnetic signals propagating through this
magnetosphere. To study these effects numerical relativity simulations
are necessary and require a scheme that matches the ideal MHD interior
of the NS to the exterior force-free magnetosphere.

Here we present a new method for matching general relativistic ideal
MHD to its force-free limit. We have tested out force-free code using
a series of 1D flat spacetime tests, as well as 3D stationary black
hole tests. We confirmed the validity of our new matching scheme by
reproducing the well-known aligned rotator solution. We demonstrated
the robustness of our algorithms and code and new techniques and we
have shown that for smooth solutions our new code is second-order
convergent.

This new method has already been used in \cite{Paschalidis:2013jsa},
where we presented the first GR simulations of a binary black hole -
neutron star magnetosphere. We plan to use this code to simulate
other complicated dynamical spacetime scenarios involving neutron
stars and their magnetospheres.  In a future paper we also plan to extend
our code to handle dynamical scenarios to extend our study to the
inspiral of compact binaries involving neutron stars.

\acknowledgments 

It is a pleasure to thank Yuk Tung Liu, Zachariah B. Etienne, Roman
Gold, and Milton Ruiz for useful discussions. This paper was supported
in part by NSF Grants PHY-0963136 and PHY-1300903 as well as NASA
Grants NNX11AE11G and NN13AH44G at the University of Illinois at
Urbana-Champaign. VP gratefully acknowledges support from a
Fortner Fellowship at UIUC. This work used the Extreme Science and
Engineering Discovery Environment (XSEDE), which is supported by NSF
grant number OCI-1053575.

\appendix

\section{Existence of a family of timelike vectors in GRFFE satisfying the ideal MHD condition}
\label{timelikevectors_exist}

In this Appendix we demonstrate that the force-free conditions imply
the ideal MHD condition.

{\it Theorem}: If the conditions (\ref{ffe:FdotFs}) and
(\ref{ffe:FdotF}) are satisfied, there exists a one-parameter family
of timelike unit vectors $\{U^\mu \}$ so that $u_\nu \F^{\mu \nu}=0$
for any $u^\mu \in \{U^\mu \}$.

{\it Proof}: The proof can be established by finding the solution of the equation 
$u_\mu \F^{\mu \nu}=0$. If true, it follows that $0=u_\mu n_\nu \F^{\mu \nu}=u_\mu \E^\mu$. 
Substituting Eq.~(\ref{dec:Fab}) for $\F^{\mu \nu}$ in $u_\mu \F^{\mu \nu}=0$ gives 
\beq
  \E^\alpha - v_\beta \B_\gamma n_\delta \epsilon^{\alpha \beta \gamma \delta} = 0 ,
\label{eq:udotF=0}
\eeq
where $v^\mu = u^\mu/\gamma_v$ and $\gamma_v = -n_\mu u^\mu$. This is the GR version 
of the flat spacetime $\ve{E}+\ve{v}\times \ve{B}=0$ equation. We note that our Eq. \eqref{eq:udotF=0}
is different than  Eq.~(14) of~\cite{m06} by a sign. We now prove the assertion by first decomposing $v^\mu$ in terms 
of 4 mutually orthogonal vectors $n^\mu$, $\B^\mu$, $\E^\mu$ and 
$\epsilon^{\mu \nu \alpha \beta} n_\nu \B_\alpha \E_\beta$:
\beq
  v_\beta = G \left( \frac{-\epsilon_{\beta \mu \nu \lambda} n^\mu \E^\nu \B^\lambda}{B^2} \right)
 + H n_\beta + K\left(\frac{\E_\beta}{\E}\right) + L \left( \frac{\B_\beta}{\B}\right) ,
\label{eq:vexp}
\eeq
where $G$, $H$, $K$ and $L$ are coefficients to be determined. It follows from $u_\mu \E^\mu=0$ and 
$v_\mu n^\mu =-1$ that $K=0$ and $H=1$. Substituting Eq. (\ref{eq:vexp}) into Eq. (\ref{eq:udotF=0}) and 
using $\E_\mu \B^\mu=0$ yields $G=1$. It follows from $u_\mu u^\mu=-1$ that $\gamma_v^2 v_\mu v^\mu=-1$ 
or 
\beq
  \gamma_v = \sqrt{ \frac{\B^2}{\B^2(1-L^2) - \E^2} } .
\label{eq:gammav}
\eeq
Straightforward algebra also yields
\beq
  \gamma_v = \sqrt{1 + \gamma^{ij} u_i u_j} .
\eeq
Since $\gamma^{ij}$ is positive-definite, $\gamma_v \geq 1$. Thus, in order for Eq.~(\ref{eq:gammav}) 
to be a valid (real) solution, $L$ has to be restricted by 
\beq
  |L| < \sqrt{ \frac{\B^2-\E^2}{\B^2}} .
\label{eq:Lrestrict}
\eeq
With this restriction, Eq.~(\ref{eq:gammav}) guarantees that $\gamma_v \geq 1$ since $\B^2>\E^2$. 
As a result the desired one-parameter family of unit timelike vectors is given by 
\beq
  u^\mu_L = \sqrt{\frac{\B^2}{\B^2(1-L^2) - \E^2} } \left( n^\mu 
- \frac{\epsilon^{\mu \beta \gamma \delta} n_\beta \E_\gamma \B_\delta}{\B^2} 
+ L \frac{\B^\mu}{\B} \right) 
\label{eq:umuL}
\eeq
with the parameter $L$ taking values in the range given by Eq.~(\ref{eq:Lrestrict}). This completes the proof. 

Note that the proof of the theorem depends crucially on the
conditions~\eqref{ffe:FdotFs} and (\ref{ffe:FdotF}), which imply $\E_i
\B^i=0$ and $\B^2 > \E^2$.  The existence of a $u^\mu$ still holds, if
(\ref{ffe:FdotF}) is replaced by $\F^{\mu \nu} \F_{\mu \nu} \geq 0$
with $\F^{\mu \nu} \F_{\mu \nu}=0$, if and only if $\F^{\mu \nu}=0$,
in which case the condition $u_\nu \F^{\mu \nu}=0$ is automatically
satisfied.

\section{Redundancy of the energy equation in the $\ve{S}$-$\ve{\B}$ formulation of GRFFE}
\label{energy_redundant}

In this appendix we demonstrate that the energy
equation~(\ref{eq:S0dot}) is redundant because it can be derived from
Eqs.~(\ref{eq:divBeq0}), (\ref{eq:induction3}), (\ref{eq:Sidot}) and
(\ref{eq:E-SB}).

Equations~(\ref{eq:divBeq0}) and (\ref{eq:induction3}) are derived
from the Maxwell Eq. \eqref{nablastarFmunu} and
Eq.~(\ref{eq:E-SB}). It is straightforward to show that
Eqs.~(\ref{eq:divBeq0}), (\ref{eq:induction3}) and (\ref{eq:E-SB})
imply the Maxwell Eq. \eqref{nablastarFmunu}, which is also equivalent
to Eq.~(\ref{eq:maxwellabc}). Combining the left equation of 
Eq. \eqref{eq:maxwell} and Eq.~(\ref{eq:divTem})
yields 
\labeq{tempvas0}{
\nabla_\nu \tem^{\nu}{}_\mu=\F_{\nu \mu} \nabla_\alpha \F^{\nu
  \alpha}.}  
It can also be shown that Eq.~(\ref{eq:Sidot})
implies 
\labeq{}{
\nabla_\nu \tem^{\nu}{}_i=0.
}
Hence, Eqs.~(\ref{eq:divBeq0}),(\ref{eq:induction3}), (\ref{eq:Sidot}) and (\ref{eq:E-SB}) imply
\labeq{FmuinablaaFnua}{
\F_{\nu i} \nabla_\alpha \F^{\nu \alpha}=0.
}  
It remains to show Eq. \eqref{FmuinablaaFnua} implies $\F_{\nu 0}
\nabla_\alpha \F^{\nu \alpha}=0$. 

Introduce the new quantities
\labeq{zeta}{\zeta=-n_\nu \nabla_\alpha
\F^{\nu \alpha}}
 and
\labeq{Qmu}{Q^\mu = \gamma^\mu{}_\nu \nabla_\alpha \F^{\nu
  \alpha}.}
 Then Eqs. \eqref{zeta} and \eqref{Qmu} imply that 
\labeq{zetaetannuplQnu}{
\nabla_\alpha \F^{\nu \alpha}=\zeta n^\nu + Q^\nu.}
Hence Eq. \eqref{FmuinablaaFnua} via Eq. \eqref{zetaetannuplQnu} and Eqs.
\eqref{def:Emu}, \eqref{def:Bmu} yields
 \beq \zeta
\E_i + \epsilon_{ijk} Q^j \B^k = 0 ,
\label{eq:tmp0001}
\eeq
while
\labeq{tempvas}{
\begin{split}
  \F_{\nu 0} \nabla_\alpha \F^{\nu \alpha}=&\ \alpha (n^\mu \F_{\nu \mu} - n^i \F_{\nu i}) \nabla_\alpha \F^{\nu \alpha} \\
  = &\ \alpha n_\mu \F_\nu{}^\mu \nabla_\alpha \F^{\nu \alpha} \\ 
  = &\ -\alpha^2 \F_\nu{}^0 \nabla_\alpha \F^{\nu \alpha} = \alpha \E_i Q^i . 
\end{split}
}

Taking the cross product of Eq.~(\ref{eq:tmp0001}) with $\B^i$ gives 
\beq
  \zeta \epsilon^{ijk} \B_j \E_k + \B^2 Q^i - (Q^j \B_j) \B^i = 0.\label{tempvas2}
\eeq
Taking the dot product of Eq.~(\ref{tempvas2}) with $\E_i$ and using the degeneracy condition 
$\E_i \B^i=0$ [from Eq.~(\ref{eq:E-SB})] and $\B^2 \neq 0$ (otherwise the constraint $\B^2 > \E^2$ will be 
violated) gives 
\labeq{EdotQ}{\E_i Q^i=0.}
By virtue of Eq. \eqref{EdotQ} equation \eqref{tempvas}  implies
\labeq{}{
F_{\nu 0} \nabla_\alpha \F^{\nu \alpha}=0,
}
which via Eq. \eqref{tempvas0} implies 
\labeq{}{
\nabla_\nu \tem^{\nu}{}_0=0
} 
which equivalent to the energy equation~(\ref{eq:S0dot}). Thus, 
Eq.~(\ref{eq:S0dot}), as well as $\nabla_\nu \tem^{\mu \nu}=0$ and 
the condition $\F_{\nu \mu} \nabla_\alpha \F^{\nu \alpha}=0$, all
follow from Eqs.~(\ref{eq:divBeq0}), (\ref{eq:induction3}), (\ref{eq:Sidot}), (\ref{eq:E-SB}) 
plus the condition $\B^2 \neq 0$. Hence the 
energy equation~(\ref{eq:S0dot}) is redundant~\footnote{Note that we deliberately wrote
$\nabla_\alpha \F^{\nu \alpha}$ instead of $\J^\nu$, $\zeta$ instead of $\rho$, and $Q^\mu$ instead of 
$J^\mu$. This is to demonstrate that the Maxwell equation $\nabla_\nu \F^{\mu \nu} = \J^\mu$ is not 
needed to prove the redundancy of the energy equation.}.

\section{Evolution Equation for $\ve{C_{SB}}$}
\label{evolCsb}

In this appendix we derive the evolution equation for the constraint 
$C_{SB}=\B^i S_i$ using evolution equations~(\ref{eq:induction3}) 
and (\ref{eq:Sidot}). We demonstrate that the evolution equations preserve
this constraint, provided it is satisfied initially.

It is convenient to write the EM stress-energy tensor in the form 
\labeq{}{
  \bspl
  \tem^{\mu \nu} = &\ \frac{\B^2+\E^2}{2} (\gamma^{\mu \nu} + n^\mu n^\nu) - (\B^\mu \B^\nu + \E^\mu \E^\nu) 
\\ &\ + n^\mu \bS^\nu + n^\nu \bS^\mu ,
\end{split}
}
where $\bS^\mu = \gamma^{\mu \nu} S_\nu$. Recall that $\E^\mu$ is considered as 
a function of $S_i$ and $\B^i$ via 
\beq
  \E^\mu = \frac{\epsilon^{\mu \alpha \beta} \B_\alpha S_\beta}{\B^2} 
\eeq
and, hence, $\B_\mu \E^\mu = 0$, but $\B^\mu S_\mu = C_{SB}$ is not set to 0 in this analysis.
Define the purely spatial EM stress tensor according to: 
\beq
  \bar{T}_{\rm EM}^{\mu \nu} = \gamma^\mu{}_\alpha \gamma^\nu{}_\beta \tem^{\alpha \beta} 
= \frac{\B^2+\E^2}{2} \gamma^{\mu \nu} - (\B^\mu \B^\nu + \E^\mu \E^\nu) ,
\eeq
where the components of $\tem^{\mu \nu}$ are
\beq
  \tem^{00} = \frac{\B^2 + \E^2}{2\alpha^2} \ \ \ , \ \ \ 
\tem^{0j} = -\frac{\B^2 + \E^2}{2\alpha^2}\beta^j + \frac{\bS^j}{\alpha} ,
\eeq
\labeq{}{
  \bspl
  \tem^j{}_i =&\ \frac{\B^2 + \E^2}{2} \delta^j{}_i - (\B^j \B_i + \E^j \E_i) - \frac{\beta^j}{\alpha} S_i \\
= &\ \bar{T}_{\rm EM}^j{}_i - \frac{\beta^j}{\alpha} S_i ,
\end{split}
}
and where $\bar{T}_{\rm EM}^j{}_i \equiv \frac{\B^2 + \E^2}{2} \delta^j{}_i - (\B^j \B_i + \E^j \E_i)$. 
The following identities will be useful. For any purely {\it spatial} antisymmetric tensor $A^{ij}$ 
and purely {\it spatial} symmetric tensor $S^{ij}$, 
\labeq{}{
  \bspl
  D_j A^{ij} = &\ \frac{1}{\sg} \partial_j (\sg A^{ij}) , \\ 
  D_j S^j{}_i = &\ \frac{1}{\sg} \partial_j (\sg S^j{}_i) - \frac{1}{2} S^{jk} \partial_i \gamma_{jk} .
\end{split}
}

Now, it follows from Eqs.~(\ref{eq:induction3}) and (\ref{eq:Sidot}) that 
\beqn
  \p_t (\gamma C_{SB}) &=& \p_t (\tilde{\B}^i \tilde{S}_i) = \sg S_i \p_t \tilde{\B}^i + \sg \B^i \p_t \tilde{S}_i \cr
&=& -\sg S_i \p_j \left( \alpha \sg \frac{\B^i \bS^j-\B^j \bS^i}{\B^2} \right. \nonumber \\ 
&&  + \sg \beta^i \B^j - \sg \beta^j \B^i\bigg) \cr \nonumber  \\
&& -\sg \B^i \p_j (\alpha \sg \tem^j{}_i) + \frac{\alpha \gamma}{2} \B^i \tem^{\mu \nu} \p_i g_{\mu \nu} \cr
&=& X+Y , \label{A7}
\eeqn
where 
\beq
  X = -\alpha \gamma S_i D_j \left( \frac{\B^i \bS^j-\B^j \bS^i}{\B^2}\right) 
- \alpha \gamma \B^i D_j \bar{T}_{\rm EM}^j{}_i ,
\eeq
and 
\beqn
  Y &=& -\gamma S_i \frac{\B^i \bS^j-\B^j \bS^i}{\B^2} \p_j \alpha \cr 
 &&  -\sg S_i \p_j (\sg \beta^i \B^j -\sg \beta^j \B^i) +\sg \B^i \p_j (\sg \beta^j S_i) \cr
 &&- \gamma \B^i \tem^j{}_i \p_j \alpha 
+ \frac{\alpha \gamma}{2} \B^i (\tem^{\mu \nu} \p_i g_{\mu \nu} - \bar{T}_{\rm EM}^{jk} \p_i \gamma_{jk}) \cr \cr
&=& -\gamma \frac{\bS^j \partial_j \alpha}{\B^2} C_{SB} + \gamma \frac{\bS^2}{\B^2} \B^j \p_j \alpha 
  - \gamma S_i \beta^i C_{dB} - \gamma S_i \B^j \p_j \beta^i \cr
&& + \sg \p_j (\sg \beta^j) C_{SB}  + \gamma S_i \beta^j \p_j \B^i + \sg \p_j (\sg \beta^j) C_{SB} \cr
&&  + \gamma \B^i \beta^j \p_j S_i - \gamma \B^i \tem^j{}_i \p_j \alpha \cr \cr
&& + \frac{\alpha \gamma}{2} \B^i \bigg[ \tem^{00} \p_i g_{00} + 2 \tem^{0j} \p_i g_{0j} \cr
&& + \left( \frac{\B^2+\E^2}{2\alpha^2}\beta^j \beta^k - \frac{2 \beta^j \bS^k}{\alpha}\right) \p_i \gamma_{jk} 
\bigg] . \label{A9}
\eeqn

The motivation for the separation of $X$ and $Y$ is that in the flat
spacetime limit $Y=0$. We can now write some of the $Y$ terms as
\beq
  \frac{\bS^2}{\B^2} \B^j \p_j \alpha = \left(\E^2 +\frac{C_{SB}^2}{\B^4}\right) \B^i \p_i \alpha .\label{A10}
\eeq
\labeq{}{
  \bspl
  S_i \B^j \p_j \beta^i = &\ \B^j \p_j (S_i \beta^i) - \B^j \beta^i \p_j S_i  \\
= &\ \B^i \p_i (\bS^j \beta_j) - \B^i \beta^j \p_i S_j \\
= &\ \B^i \bS^j \p_i \beta_j + \B^i \beta_j \p_i \bS^j - \B^i \beta^j \p_i S_j .
\end{split}
}\\

\beq
  S_i \beta^j \p_j \B^i = \beta^i \p_i C_{SB} - \B^i \beta^j \p_j S_i .
\eeq
\beq
  \B^i \tem^j{}_i \p_j \alpha = \frac{\E^2-\B^2}{2} \B^i \p_i \alpha - \frac{\beta^i}{\alpha} \p_i \alpha C_{SB} .
\eeq

\labeq{}{
  \bspl
  \B^i \tem^{00} \p_i g_{00} = &\ \frac{\B^2+\E^2}{2\alpha^2}\B^i \p_i (-\alpha^2 + \beta^j \beta_j) \\
= &\ \frac{\B^2+\E^2}{2\alpha^2}\B^i (-2 \alpha \p_i \alpha + \beta^j \p_i \beta_j + \beta_j \p_i \beta^j) .
\end{split}
}

\beq
  \B^i \tem^{0j} \p_i g_{0j} = -\frac{\B^2+\E^2}{2\alpha^2} \B^i \beta^j \p_i \beta_j 
+ \frac{\B^i \bS^j}{\alpha} \p_i \beta_j .
\eeq
\labeq{}{
  \bspl
\B^i \beta^j \beta^k \p_i \gamma_{jk} 
= &\ \B^i [ \p_i (\beta^j \beta_j) - \gamma_{jk} \p_i (\beta^j \beta^k)] \\
= &\ \B^i (\beta^j \p_i \beta_j - \beta_j \p_i \beta^j) .
\end{split}
}
\labeq{}{
  \bspl
  \B^i \beta^j \bS^k \p_i \gamma_{jk} = &\ \B^i [ \p_i (\beta^j S_j) - \gamma_{jk} \p_i (\beta^j \bS^k)] \\
= &\ \B^i (\beta^j \p_i S_j - \beta_j \p_i \bS^j) .
\end{split} \label{A17}
}

Eqs. \eqref{A10}-\eqref{A17} imply
\begin{widetext}
\beqn
  \frac{\alpha \gamma}{2} \B^i \left[ \tem^{00} \p_i g_{00} + 2 \tem^{0j} \p_i g_{0j} 
+ \left( \frac{\B^2+\E^2}{2\alpha^2}\beta^j \beta^k - \frac{2 \beta^j \bS^k}{\alpha}\right) \p_i \gamma_{jk}
\right] && \cr \cr 
= -\gamma \frac{\B^2+\E^2}{2}\B^i \p_i \alpha + \gamma \B^i \bar{S}^j \p_i \beta_j 
- \gamma \B^i (\beta^j \p_i S_j - \beta_j \p_i \bar{S}^j ) . && \nonumber
\eeqn

Gathering all the terms gives 
\beqn
  Y &=& \p_i (\gamma \beta^i C_{SB}) + \gamma \left[ \p_i \beta^i - \left( 
\alpha \frac{\bS^i}{\B^2}-\beta^i \right) \frac{\p_i \alpha}{\alpha} \right] C_{SB} 
+ \gamma \frac{\B^i \p_i \alpha}{\B^4} C_{SB}^2 
- \gamma S_i \beta^i C_{dB} \cr \cr
&=& \p_i (\gamma \beta^i C_{SB}) + \gamma ( \p_i \beta^i - v^i a_i) C_{SB} 
+ \gamma \frac{\B^i \p_i \alpha}{\B^4} C_{SB}^2 - \gamma S_i \beta^i C_{dB} ,\label{A18}
\eeqn
\end{widetext}
where 
\beq
  a_i = n^\mu \nabla_\mu n_i = D_i (\ln \alpha) = \frac{\p_i \alpha}{\alpha} 
\eeq
is the 4-acceleration of $n^\mu$. 

The calculation of $X$ involves fewer algebraic operations: 
\beqn
  \B^i D_j \bar{T}_{\rm EM}^j{}_i &=& \B^i D_j \left[ \frac{\B^2+\E^2}{2}\delta^j{}_i 
- (\B^j \B_i + \E^j \E_i) \right] \cr
&=& \frac{\B^i}{2}D_i (\B^2+\E^2) - \B^i D_j (\B^j \B_i + \E^j \E_i) \cr 
&=& \B^i (\B^j D_i \B_j + \E^j D_i \E_j) - \B^i \B^j D_j \B_i \cr
&& - \B^2 C_{dB} - \B^i \E^j D_j \E_i \cr
&=& -\B^2 C_{dB} + \B^i \E^j ( D_i \E_j - D_j \E_i) \cr 
&=& -\B^2 C_{dB} + \B^i \E^j \delta^{lm}{}_{ij} D_l \E_m \cr 
&=& -\B^2 C_{dB} + (\epsilon^{lmk} D_l \E_m) (\epsilon_{ijk} \E^j \B^i) \cr 
&=& -\B^2 C_{dB} + (\epsilon^{lmk} D_l \E_m) \epsilon_{ijk} \epsilon^{jpn} \B^i \frac{\B_p S_n}{\B^2} \cr 
&=& -\B^2 C_{dB} + \frac{\epsilon^{lmk} D_l \E_m}{\B^2} \delta^{pn}{}_{ki} \B^i \B_p S_n \cr 
&=& -\B^2 C_{dB} + \frac{\epsilon^{lmk} D_l \E_m}{\B^2} \B_k C_{SB}  \cr
 && - S_k \epsilon^{lmk} D_l \E_m .\label{A20}
\eeqn

Now 
\beqn
  \epsilon^{lmk}  D_l \E_m &=& \epsilon^{lmk} D_l \left( \epsilon_{mij} \frac{\B^i \bS^j}{\B^2} \right) \cr \cr
&=& \delta^{kl}{}_{ij} D_l \left( \frac{\B^i \bS^j}{\B^2}\right) = D_l \left( \frac{\B^k \bS^l-\B^l \bS^k}{\B^2}\right) \cr \cr
&=& D_j \left( \frac{\B^k \bS^j-\B^j \bS^k}{\B^2}\right) .\label{A21}
\eeqn
By use of Eq. \eqref{A21}, Eq. \eqref{A20} yields 
\labeq{}{
  \bspl
  \B^i D_j \bar{T}_{\rm EM}^j{}_i = &\ -\B^2 C_{dB} + \B_i D_j \left( \frac{\B^i \bS^j-\B^j \bS^i}{\B^2}\right) C_{SB} \\
 &\ - S_i D_j \left( \frac{\B^i \bS^j-\B^j \bS^i}{\B^2}\right) 
  \end{split}
}
and 
\beq
  X = \alpha \gamma \B^2 C_{dB} - \alpha \sg \B_i \p_j \left( \sg 
\frac{\B^i \bS^j-\B^j \bS^i}{\B^2}\right) C_{SB} . \label{A23}
\eeq
Finally, using Eqs. \eqref{A18} and \eqref{A23} the evolution equation \eqref{A7} becomes
\beqn
  \p_t (\gamma C_{SB})  &=&  \p_i (\gamma \beta^i C_{SB}) + \gamma (\p_i \beta^i - v^i a_i) C_{SB} \cr\cr
 && - \alpha \sg \B_i \p_j \left( \sg \frac{\B^i \bS^j-\B^j \bS^i}{\B^2}\right) C_{SB} \cr 
 && + \gamma \frac{\B^i \p_i \alpha}{\B^4} C_{SB}^2 + \gamma (\alpha \B^2 - S_i \beta^i) C_{dB} .\qquad
\label{eq:CSBevol}
\eeqn
Hence if $C_{SB}=C_{dB}=0$ initially, the evolution equations preserve the constraints.

\section{An alternative derivation for the FFE current}
\label{FFE_current}

As in \S~\ref{sec:formalism1}, $J^i$ can be decomposed into a parallel
and a perpendicular components using Eqs.~(\ref{eq:Jidecomp}) and
(\ref{def:Jpara_Jperp}). The perpendicular component can be
determined, as in \S~\ref{sec:formalism1}, by taking the cross product
of Eq.~(\ref{ffe:FdotJ31}) with $\B^i$, resulting in the first
equality of Eq.~(\ref{eq:Jperpi}). It is straightforward to show that
the remaining piece $\rho = D_i \E^i$, which is not a priori guaranteed in the
$\ve{S}-\ve{\B}$ formulation, follows from the Maxwell equation
$\nabla_\nu \F^{\mu \nu} = \J^\mu$.
\beqn
  \rho &=& -n_\mu \J^\mu = -n_\mu \nabla_\nu \F^{\mu \nu} \cr 
&=& -n_\mu \nabla_\nu (n^\mu \E^\nu - n^\nu \E^\mu - \epsilon^{\mu \nu \alpha \beta} \B_\alpha n_\beta) \cr
&=& \nabla_\nu \E^\nu + n_\mu n^\nu \nabla_\nu \E^\mu 
+ \epsilon^{\mu \nu \alpha \beta} n_\mu \B_\alpha \nabla_\nu n_\beta \cr
&=& (\delta_\mu{}^\nu + n_\mu n^\nu) \nabla_\nu \E^\mu 
- \epsilon^{\mu \nu \alpha \beta} n_\mu \B_\alpha (K_{\nu \beta}+n_\nu a_\beta) \cr 
&=& \gamma_\mu{}^\nu \nabla_\nu \E^\mu \cr
&=& \gamma_\mu{}^\nu \delta^\mu{}_\alpha \nabla_\nu \E^\alpha \cr
&=& \gamma_\mu{}^\nu (\gamma^\mu{}_\alpha - n^\mu n_\alpha) \nabla_\nu \E^\alpha \cr
&=& \gamma_\mu{}^\nu \gamma^\mu{}_\alpha \nabla_\nu \E^\alpha \cr
&=& D_\mu \E^\mu \cr 
&=& D_i \E^i, \nonumber
\eeqn
which in terms of the $S_i$, $\B^i$ variables becomes
\labeq{}{
\rho =\epsilon^{ijk} D_i \left( \frac{\B_j S_k}{\B^2}\right). 
}

This takes care of the perpendicular component. The parallel component can be determined by 
computing the scalar $J_\parallel$:
\labeq{eq:Jpara1}{
\begin{split}
  J_\parallel =&\ \B_\mu \J^\mu = \B_\mu \nabla_\nu \F^{\mu \nu} \cr 
=&\ \B_\mu \nabla_\nu (n^\mu \E^\nu - n^\nu \E^\mu - \epsilon^{\mu \nu \alpha \beta} \B_\alpha n_\beta) \cr 
=&\ \B_\mu \E^\nu \nabla_\nu n^\mu - \B_\mu n^\nu \nabla_\nu \E^\mu 
- \epsilon^{\mu \nu \alpha \beta} \B_\mu n_\beta \nabla_\nu \B_\alpha .
\end{split}
}
The last term in this last equation can be simplified 
\beqn
  - \epsilon^{\mu \nu \alpha \beta} \B_\mu n_\beta \nabla_\nu \B_\alpha &=& 
n_\beta \B_\mu \epsilon^{\beta \mu \nu \alpha} \delta^\gamma{}_\nu 
\delta^\lambda{}_\alpha \nabla_\gamma \B_\lambda \cr 
&=& n_\beta \B_\mu \epsilon^{\beta \mu \nu \alpha} \gamma^\gamma{}_\nu 
\gamma^\lambda{}_\alpha \nabla_\gamma \B_\lambda \cr
&=& n_\beta \B_\mu \epsilon^{\beta \mu \nu \alpha} D_\nu \B_\alpha \cr 
&=& \epsilon^{ijk} \B_i D_j \B_k . \label{eq:id1}
\eeqn
The middle term in the last line of Eq.~(\ref{eq:Jpara1}) can be rewritten using the Maxwell 
equation $\nabla_\nu {}^*\F^{\mu \nu}=0$, which implies
\labeq{eq:id2}{
  \bspl
  0  = &\ \E_\mu \nabla_\nu {}^*\F^{\mu \nu}  = \E_\mu \nabla_\nu (-n^\mu \B^\nu + n^\nu \B^\mu 
- \epsilon^{\mu \nu \alpha \beta} \E_\alpha n_\beta) \cr 
=&\ -\E_\mu \B^\nu \nabla_\nu n^\mu + \E_\mu n^\nu \nabla_\nu \B^\mu 
- \epsilon^{\mu \nu \alpha \beta} n_\beta \E_\mu \nabla_\nu \E_\alpha \cr
=&\ -\E^\mu \B^\nu \nabla_\nu n_\mu - n^\nu \B^\mu \nabla_\nu \E_\mu 
+ n_\beta \epsilon^{\beta \mu \nu \alpha} \E_\mu \nabla_\nu \E_\alpha, \cr 
\end{split}
}
which yields
\labeq{}{
- \B_\mu n^\nu \nabla_\nu \E^\mu=\E^\mu \B^\nu \nabla_\nu n_\mu - \epsilon^{ijk} \E_i D_j \E_k , 
}
where the identity 
$n_\beta \epsilon^{\beta \mu \nu \alpha} \E_\mu \nabla_\nu \E_\alpha=\epsilon^{ijk} \E_i D_j \E_k$ 
[which can be proved in a similar way as in Eq.~(\ref{eq:id1})] has been used. 
Combining Eqs.~(\ref{eq:Jpara1})--(\ref{eq:id2}) gives 
\labeq{}{
  \bspl
  J_\parallel = &\ \epsilon^{ijk} (\B_i D_j \B_k - \E_i D_j \E_k) 
+ \E^\mu \B^\nu (\nabla_\nu n_\mu + \nabla_\mu n_\nu) \\
= &\ \epsilon^{ijk} (\B_i D_j \B_k - \E_i D_j \E_k) - 2 \E^i \E^j K_{ij} ,
\end{split}
}
which is the same as Eq.~(\ref{eq:Jpara}). Hence the current density $J^i$ is given by
Eq.~(\ref{ffe:Ji}), as expected.

\bibliography{paper}

\begin{thebibliography}{10}%
\makeatletter
\providecommand \@ifxundefined [1]{%
 \ifx #1\undefined \expandafter \@firstoftwo
 \else \expandafter \@secondoftwo
\fi
}%
\providecommand \@ifnum [1]{%
 \ifnum #1\expandafter \@firstoftwo
 \else \expandafter \@secondoftwo
\fi
}%
\providecommand \enquote [1]{``#1''}%
\providecommand \bibnamefont  [1]{#1}%
\providecommand \bibfnamefont [1]{#1}%
\providecommand \citenamefont [1]{#1}%
\providecommand\href[0]{\@sanitize\@href}%
\providecommand\@href[1]{\endgroup\@@startlink{#1}\endgroup\@@href}%
\providecommand\@@href[1]{#1\@@endlink}%
\providecommand \@sanitize [0]{\begingroup\catcode`\&12\catcode`\#12\relax}%
\@ifxundefined \pdfoutput {\@firstoftwo}{%
 \@ifnum{\z@=\pdfoutput}{\@firstoftwo}{\@secondoftwo}%
}{%
 \providecommand\@@startlink[1]{\leavevmode\special{html:<a href="#1">}}%
 \providecommand\@@endlink[0]{\special{html:</a>}}%
}{%
 \providecommand\@@startlink[1]{%
  \leavevmode
  \pdfstartlink
   attr{/Border[0 0 1 ]/H/I/C[0 1 1]}%
   user{/Subtype/Link/A<</Type/Action/S/URI/URI(#1)>>}%
  \relax
 }%
 \providecommand\@@endlink[0]{\pdfendlink}%
}%
\providecommand \url  [0]{\begingroup\@sanitize \@url }%
\providecommand \@url [1]{\endgroup\@href {#1}{\urlprefix}}%
\providecommand \urlprefix [0]{URL }%
\providecommand \Eprint[0]{\href }%
\@ifxundefined \urlstyle {%
  \providecommand \doi [1]{doi:\discretionary{}{}{}#1}%
}{%
  \providecommand \doi [0]{doi:\discretionary{}{}{}\begingroup
  \urlstyle{rm}\Url }%
}%
\providecommand \doibase [0]{http://dx.doi.org/}%
\providecommand \Doi[1]{\href{\doibase#1}}%
\providecommand \bibAnnote [3]{%
  \BibitemShut{#1}%
  \begin{quotation}\noindent
    \textsc{Key:}\ #2\\\textsc{Annotation:}\ #3%
  \end{quotation}%
}%
\providecommand \bibAnnoteFile [2]{%
  \IfFileExists{#2}{\bibAnnote {#1} {#2} {\input{#2}}}{}%
}%
\providecommand \typeout [0]{\immediate \write \m@ne }%
\providecommand \selectlanguage [0]{\@gobble}%
\providecommand \bibinfo [0]{\@secondoftwo}%
\providecommand \bibfield [0]{\@secondoftwo}%
\providecommand \translation [1]{[#1]}%
\providecommand \BibitemOpen[0]{}%
\providecommand \bibitemStop [0]{}%
\providecommand \bibitemNoStop [0]{.\EOS\space}%
\providecommand \EOS [0]{\spacefactor3000\relax}%
\providecommand \BibitemShut [1]{\csname bibitem#1\endcsname}%
\bibitem{Nissanke:2012dj}%
  \BibitemOpen
  \bibfield{author}{%
  \bibinfo {author} {\bibfnamefont{S.}~\bibnamefont{Nissanke}}, \bibinfo
  {author} {\bibfnamefont{M.}~\bibnamefont{Kasliwal}},\ and\ \bibinfo {author}
  {\bibfnamefont{A.}~\bibnamefont{Georgieva}}}%
   (\bibinfo {year} {2012}),\
  \Eprint{http://arxiv.org/abs/1210.6362}{arXiv:1210.6362 [astro-ph.HE]}%
  \bibAnnoteFile{NoStop}{Nissanke:2012dj}%
\bibitem{1998sese.conf..729R}%
  \BibitemOpen
  \bibinfo {author} {\bibfnamefont{S.}~\bibnamefont{{Rosswog}}}, \bibinfo
  {author} {\bibfnamefont{M.}~\bibnamefont{{Liebend{\"o} Rfer}}}, \bibinfo
  {author} {\bibfnamefont{F.-K.}\ \bibnamefont{{Thielemann}}},\ and\ \bibinfo
  {author} {\bibfnamefont{a.~.~P.}\
  \bibnamefont{l~=~{http://adsabs.harvard.edu/abs/1998sese.conf..729R}}}%
  \bibAnnoteFile{NoStop}{1998sese.conf..729R}%
\bibitem{Rosswog:1998hy}%
  \BibitemOpen
\bibfield{author}{%
    }%
  \bibfield{author}{%
  \bibinfo {author} {\bibfnamefont{S.}~\bibnamefont{Rosswog}}, \bibinfo
  {author} {\bibfnamefont{M.}~\bibnamefont{Liebendoerfer}}, \bibinfo {author}
  {\bibfnamefont{F.}~\bibnamefont{Thielemann}}, \bibinfo {author}
  {\bibfnamefont{M.}~\bibnamefont{Davies}}, \bibinfo {author}
  {\bibfnamefont{W.}~\bibnamefont{Benz}}, \emph{et~al.},\ }%
  \bibfield{journal}{%
  \bibinfo {journal} {Astron.Astrophys.}\ }%
  \textbf{\bibinfo {volume} {341}},\ \bibinfo {pages} {499} (\bibinfo {year}
  {1999}),\
  \Eprint{http://arxiv.org/abs/astro-ph/9811367}{arXiv:astro-ph/9811367
  [astro-ph]}%
  \bibAnnoteFile{NoStop}{Rosswog:1998hy}%
\bibitem{Li:1998bw}%
  \BibitemOpen
  \bibfield{author}{%
  \bibinfo {author} {\bibfnamefont{L.-X.}\ \bibnamefont{Li}}\ and\ \bibinfo
  {author} {\bibfnamefont{B.}~\bibnamefont{Paczynski}},\ }%
  \bibfield{journal}{%
  \Doi{10.1086/311680}{\bibinfo {journal} {Astrophys.J.}}\ }%
  \textbf{\bibinfo {volume} {507}},\ \bibinfo {pages} {L59} (\bibinfo {year}
  {1998}),\
  \Eprint{http://arxiv.org/abs/astro-ph/9807272}{arXiv:astro-ph/9807272
  [astro-ph]}%
  \bibAnnoteFile{NoStop}{Li:1998bw}%
\bibitem{Kulkarni:2005jw}%
  \BibitemOpen
  \bibfield{author}{%
  \bibinfo {author} {\bibfnamefont{S.}~\bibnamefont{Kulkarni}}}%
   (\bibinfo {year} {2005}),\
  \Eprint{http://arxiv.org/abs/astro-ph/0510256}{arXiv:astro-ph/0510256
  [astro-ph]}%
  \bibAnnoteFile{NoStop}{Kulkarni:2005jw}%
\bibitem{Metzger:2010sy}%
  \BibitemOpen
  \bibfield{author}{%
  \bibinfo {author} {\bibfnamefont{B.}~\bibnamefont{Metzger}}, \bibinfo
  {author} {\bibfnamefont{G.}~\bibnamefont{Martinez-Pinedo}}, \bibinfo {author}
  {\bibfnamefont{S.}~\bibnamefont{Darbha}}, \bibinfo {author}
  {\bibfnamefont{E.}~\bibnamefont{Quataert}}, \bibinfo {author}
  {\bibfnamefont{A.}~\bibnamefont{Arcones}}, \emph{et~al.}}%
   (\bibinfo {year} {2010}),\
  \Eprint{http://arxiv.org/abs/1001.5029}{arXiv:1001.5029 [astro-ph.HE]}%
  \bibAnnoteFile{NoStop}{Metzger:2010sy}%
\bibitem{Goriely:2011vg}%
  \BibitemOpen
  \bibfield{author}{%
  \bibinfo {author} {\bibfnamefont{S.}~\bibnamefont{Goriely}}, \bibinfo
  {author} {\bibfnamefont{A.}~\bibnamefont{Bauswein}},\ and\ \bibinfo {author}
  {\bibfnamefont{H.-T.}\ \bibnamefont{Janka}}}%
   (\bibinfo {year} {2011}),\
  \Eprint{http://arxiv.org/abs/1107.0899}{arXiv:1107.0899 [astro-ph.SR]}%
  \bibAnnoteFile{NoStop}{Goriely:2011vg}%
\bibitem{Hotokezaka:2012ze}%
  \BibitemOpen
  \bibfield{author}{%
  \bibinfo {author} {\bibfnamefont{K.}~\bibnamefont{Hotokezaka}}, \bibinfo
  {author} {\bibfnamefont{K.}~\bibnamefont{Kiuchi}}, \bibinfo {author}
  {\bibfnamefont{K.}~\bibnamefont{Kyutoku}}, \bibinfo {author}
  {\bibfnamefont{H.}~\bibnamefont{Okawa}}, \bibinfo {author}
  {\bibfnamefont{Y.-i.}\ \bibnamefont{Sekiguchi}}, \emph{et~al.},\ }%
  \bibfield{journal}{%
  \Doi{10.1103/PhysRevD.87.024001}{\bibinfo {journal} {Phys.Rev.}}\ }%
  \textbf{\bibinfo {volume} {D87}},\ \bibinfo {pages} {024001} (\bibinfo {year}
  {2013}),\ \Eprint{http://arxiv.org/abs/1212.0905}{arXiv:1212.0905
  [astro-ph.HE]}%
  \bibAnnoteFile{NoStop}{Hotokezaka:2012ze}%
\bibitem{Kyutoku:2012fv}%
  \BibitemOpen
  \bibfield{author}{%
  \bibinfo {author} {\bibfnamefont{K.}~\bibnamefont{Kyutoku}}, \bibinfo
  {author} {\bibfnamefont{K.}~\bibnamefont{Ioka}},\ and\ \bibinfo {author}
  {\bibfnamefont{M.}~\bibnamefont{Shibata}}}%
   (\bibinfo {year} {2012}),\
  \Eprint{http://arxiv.org/abs/1209.5747}{arXiv:1209.5747 [astro-ph.HE]}%
  \bibAnnoteFile{NoStop}{Kyutoku:2012fv}%
\bibitem{Rosswog:2013kqa}%
  \BibitemOpen
  \bibfield{author}{%
  \bibinfo {author} {\bibfnamefont{S.}~\bibnamefont{Rosswog}}, \bibinfo
  {author} {\bibfnamefont{O.}~\bibnamefont{Korobkin}}, \bibinfo {author}
  {\bibfnamefont{A.}~\bibnamefont{Arcones}},\ and\ \bibinfo {author}
  {\bibfnamefont{F.~K.}\ \bibnamefont{Thielemann}}}%
   (\bibinfo {year} {2013}),\
  \Eprint{http://arxiv.org/abs/1307.2939}{arXiv:1307.2939 [astro-ph.HE]}%
  \bibAnnoteFile{NoStop}{Rosswog:2013kqa}%
\bibitem{Grossman:2013lqa}%
  \BibitemOpen
  \bibfield{author}{%
  \bibinfo {author} {\bibfnamefont{D.}~\bibnamefont{Grossman}}, \bibinfo
  {author} {\bibfnamefont{O.}~\bibnamefont{Korobkin}}, \bibinfo {author}
  {\bibfnamefont{S.}~\bibnamefont{Rosswog}},\ and\ \bibinfo {author}
  {\bibfnamefont{T.}~\bibnamefont{Piran}}}%
   (\bibinfo {year} {2013}),\
  \Eprint{http://arxiv.org/abs/1307.2943}{arXiv:1307.2943 [astro-ph.HE]}%
  \bibAnnoteFile{NoStop}{Grossman:2013lqa}%
\bibitem{Kyutoku:2013wxa}%
  \BibitemOpen
  \bibfield{author}{%
  \bibinfo {author} {\bibfnamefont{K.}~\bibnamefont{Kyutoku}}, \bibinfo
  {author} {\bibfnamefont{K.}~\bibnamefont{Ioka}},\ and\ \bibinfo {author}
  {\bibfnamefont{M.}~\bibnamefont{Shibata}}}%
   (\bibinfo {year} {2013}),\
  \Eprint{http://arxiv.org/abs/1305.6309}{arXiv:1305.6309 [astro-ph.HE]}%
  \bibAnnoteFile{NoStop}{Kyutoku:2013wxa}%
\bibitem{Bauswein:2013yna}%
  \BibitemOpen
  \bibfield{author}{%
  \bibinfo {author} {\bibfnamefont{A.}~\bibnamefont{Bauswein}}, \bibinfo
  {author} {\bibfnamefont{S.}~\bibnamefont{Goriely}},\ and\ \bibinfo {author}
  {\bibfnamefont{H.-T.}\ \bibnamefont{Janka}},\ }%
  \bibfield{journal}{%
  \Doi{10.1088/0004-637X/773/1/78}{\bibinfo {journal} {Astrophys.J.}}\ }%
  \textbf{\bibinfo {volume} {773}},\ \bibinfo {pages} {78} (\bibinfo {year}
  {2013}),\ \Eprint{http://arxiv.org/abs/1302.6530}{arXiv:1302.6530
  [astro-ph.SR]}%
  \bibAnnoteFile{NoStop}{Bauswein:2013yna}%
\bibitem{UIUC_BHNS__BH_SPIN_PAPER}%
  \BibitemOpen
  \bibfield{author}{%
  \bibinfo {author} {\bibfnamefont{Z.~B.}\ \bibnamefont{{Etienne}}}, \bibinfo
  {author} {\bibfnamefont{Y.~T.}\ \bibnamefont{{Liu}}}, \bibinfo {author}
  {\bibfnamefont{S.~L.}\ \bibnamefont{{Shapiro}}},\ and\ \bibinfo {author}
  {\bibfnamefont{T.~W.}\ \bibnamefont{{Baumgarte}}},\ }%
  \bibfield{journal}{%
  \Doi{10.1103/PhysRevD.79.044024}{\bibinfo {journal} {\prd}}\ }%
  \textbf{\bibinfo {volume} {79}},\ \bibinfo {pages} {044024} (\bibinfo {month}
  {Feb.}\ \bibinfo {year} {2009})%
  \bibAnnoteFile{NoStop}{UIUC_BHNS__BH_SPIN_PAPER}%
\bibitem{Duez:2009yy}%
  \BibitemOpen
  \bibfield{author}{%
  \bibinfo {author} {\bibfnamefont{M.~D.}\ \bibnamefont{Duez}}, \bibinfo
  {author} {\bibfnamefont{F.}~\bibnamefont{Foucart}}, \bibinfo {author}
  {\bibfnamefont{L.~E.}\ \bibnamefont{Kidder}}, \bibinfo {author}
  {\bibfnamefont{C.~D.}\ \bibnamefont{Ott}},\ and\ \bibinfo {author}
  {\bibfnamefont{S.~A.}\ \bibnamefont{Teukolsky}},\ }%
  \bibfield{journal}{%
  \Doi{10.1088/0264-9381/27/11/114106}{\bibinfo {journal} {Class.Quant.Grav.}}\
  }%
  \textbf{\bibinfo {volume} {27}},\ \bibinfo {pages} {114106} (\bibinfo {year}
  {2010}),\ \Eprint{http://arxiv.org/abs/0912.3528}{arXiv:0912.3528
  [astro-ph.HE]}%
  \bibAnnoteFile{NoStop}{Duez:2009yy}%
\bibitem{Shibata:2009cn}%
  \BibitemOpen
  \bibfield{author}{%
  \bibinfo {author} {\bibfnamefont{M.}~\bibnamefont{Shibata}}, \bibinfo
  {author} {\bibfnamefont{K.}~\bibnamefont{Kyutoku}}, \bibinfo {author}
  {\bibfnamefont{T.}~\bibnamefont{Yamamoto}},\ and\ \bibinfo {author}
  {\bibfnamefont{K.}~\bibnamefont{Taniguchi}},\ }%
  \bibfield{journal}{%
  \Doi{10.1103/PhysRevD.85.127502, 10.1103/PhysRevD.79.044030}{\bibinfo
  {journal} {Phys.Rev.}}\ }%
  \textbf{\bibinfo {volume} {D79}},\ \bibinfo {pages} {044030} (\bibinfo {year}
  {2009}),\ \Eprint{http://arxiv.org/abs/0902.0416}{arXiv:0902.0416 [gr-qc]}%
  \bibAnnoteFile{NoStop}{Shibata:2009cn}%
\bibitem{Foucart:2010eq}%
  \BibitemOpen
  \bibfield{author}{%
  \bibinfo {author} {\bibfnamefont{F.}~\bibnamefont{Foucart}}, \bibinfo
  {author} {\bibfnamefont{M.~D.}\ \bibnamefont{Duez}}, \bibinfo {author}
  {\bibfnamefont{L.~E.}\ \bibnamefont{Kidder}},\ and\ \bibinfo {author}
  {\bibfnamefont{S.~A.}\ \bibnamefont{Teukolsky}},\ }%
  \bibfield{journal}{%
  \Doi{10.1103/PhysRevD.83.024005}{\bibinfo {journal} {Phys.Rev.}}\ }%
  \textbf{\bibinfo {volume} {D83}},\ \bibinfo {pages} {024005} (\bibinfo {year}
  {2011}),\ \Eprint{http://arxiv.org/abs/1007.4203}{arXiv:1007.4203
  [astro-ph.HE]}%
  \bibAnnoteFile{NoStop}{Foucart:2010eq}%
\bibitem{Kyutoku:2011vz}%
  \BibitemOpen
  \bibfield{author}{%
  \bibinfo {author} {\bibfnamefont{K.}~\bibnamefont{Kyutoku}}, \bibinfo
  {author} {\bibfnamefont{H.}~\bibnamefont{Okawa}}, \bibinfo {author}
  {\bibfnamefont{M.}~\bibnamefont{Shibata}},\ and\ \bibinfo {author}
  {\bibfnamefont{K.}~\bibnamefont{Taniguchi}},\ }%
  \bibfield{journal}{%
  \Doi{10.1103/PhysRevD.84.064018}{\bibinfo {journal} {Phys.Rev.}}\ }%
  \textbf{\bibinfo {volume} {D84}},\ \bibinfo {pages} {064018} (\bibinfo {year}
  {2011}),\ \Eprint{http://arxiv.org/abs/1108.1189}{arXiv:1108.1189
  [astro-ph.HE]}%
  \bibAnnoteFile{NoStop}{Kyutoku:2011vz}%
\bibitem{2011ApJ...737L...5S}%
  \BibitemOpen
  \bibfield{author}{%
  \bibinfo {author} {\bibfnamefont{B.~C.}\ \bibnamefont{{Stephens}}}, \bibinfo
  {author} {\bibfnamefont{W.~E.}\ \bibnamefont{{East}}},\ and\ \bibinfo
  {author} {\bibfnamefont{F.}~\bibnamefont{{Pretorius}}},\ }%
  \bibfield{journal}{%
  \Doi{10.1088/2041-8205/737/1/L5}{\bibinfo {journal} {\apjl}}\ }%
  \textbf{\bibinfo {volume} {737}},\ \bibinfo {eid} {L5} (\bibinfo {month}
  {Aug.}\ \bibinfo {year} {2011}),\
  \Eprint{http://arxiv.org/abs/1105.3175}{arXiv:1105.3175 [astro-ph.HE]}%
  \bibAnnoteFile{NoStop}{2011ApJ...737L...5S}%
\bibitem{Lackey:2011vz}%
  \BibitemOpen
  \bibfield{author}{%
  \bibinfo {author} {\bibfnamefont{B.~D.}\ \bibnamefont{Lackey}}, \bibinfo
  {author} {\bibfnamefont{K.}~\bibnamefont{Kyutoku}}, \bibinfo {author}
  {\bibfnamefont{M.}~\bibnamefont{Shibata}}, \bibinfo {author}
  {\bibfnamefont{P.~R.}\ \bibnamefont{Brady}},\ and\ \bibinfo {author}
  {\bibfnamefont{J.~L.}\ \bibnamefont{Friedman}},\ }%
  \bibfield{journal}{%
  \Doi{10.1103/PhysRevD.85.044061}{\bibinfo {journal} {Phys.Rev.}}\ }%
  \textbf{\bibinfo {volume} {D85}},\ \bibinfo {pages} {044061} (\bibinfo {year}
  {2012}),\ \Eprint{http://arxiv.org/abs/1109.3402}{arXiv:1109.3402
  [astro-ph.HE]}%
  \bibAnnoteFile{NoStop}{Lackey:2011vz}%
\bibitem{ShibataBHNSreview}%
  \BibitemOpen
  \bibfield{author}{%
  \bibinfo {author} {\bibfnamefont{M.}~\bibnamefont{Shibata}}\ and\ \bibinfo
  {author} {\bibfnamefont{K.}~\bibnamefont{Taniguchi}},\ }%
  \bibfield{journal}{%
  \bibinfo {journal} {Living Reviews in Relativity}\ }%
  \textbf{\bibinfo {volume} {14}} (\bibinfo {year} {2011}),\ \doi{\bibinfo
  {doi} {10.12942/lrr-2011-6}},\ \url{http://www.livingreviews.org/lrr-2011-6}%
  \bibAnnoteFile{NoStop}{ShibataBHNSreview}%
\bibitem{Foucart:2012vn}%
  \BibitemOpen
  \bibfield{author}{%
  \bibinfo {author} {\bibfnamefont{F.}~\bibnamefont{Foucart}}, \bibinfo
  {author} {\bibfnamefont{M.~B.}\ \bibnamefont{Deaton}}, \bibinfo {author}
  {\bibfnamefont{M.~D.}\ \bibnamefont{Duez}}, \bibinfo {author}
  {\bibfnamefont{L.~E.}\ \bibnamefont{Kidder}}, \bibinfo {author}
  {\bibfnamefont{I.}~\bibnamefont{MacDonald}}, \emph{et~al.}}%
   (\bibinfo {year} {2012}),\
  \Eprint{http://arxiv.org/abs/1212.4810}{arXiv:1212.4810 [gr-qc]}%
  \bibAnnoteFile{NoStop}{Foucart:2012vn}%
\bibitem{2012PhRvD..85l4009E}%
  \BibitemOpen
  \bibfield{author}{%
  \bibinfo {author} {\bibfnamefont{W.~E.}\ \bibnamefont{{East}}}, \bibinfo
  {author} {\bibfnamefont{F.}~\bibnamefont{{Pretorius}}},\ and\ \bibinfo
  {author} {\bibfnamefont{B.~C.}\ \bibnamefont{{Stephens}}},\ }%
  \bibfield{journal}{%
  \Doi{10.1103/PhysRevD.85.124009}{\bibinfo {journal} {\prd}}\ }%
  \textbf{\bibinfo {volume} {85}},\ \bibinfo {eid} {124009} (\bibinfo {month}
  {Jun.}\ \bibinfo {year} {2012}),\
  \Eprint{http://arxiv.org/abs/1111.3055}{arXiv:1111.3055 [astro-ph.HE]}%
  \bibAnnoteFile{NoStop}{2012PhRvD..85l4009E}%
\bibitem{Lovelace:2013vma}%
  \BibitemOpen
  \bibfield{author}{%
  \bibinfo {author} {\bibfnamefont{G.}~\bibnamefont{Lovelace}}, \bibinfo
  {author} {\bibfnamefont{M.~D.}\ \bibnamefont{Duez}}, \bibinfo {author}
  {\bibfnamefont{F.}~\bibnamefont{Foucart}}, \bibinfo {author}
  {\bibfnamefont{L.~E.}\ \bibnamefont{Kidder}}, \bibinfo {author}
  {\bibfnamefont{H.~P.}\ \bibnamefont{Pfeiffer}}, \emph{et~al.}}%
   (\bibinfo {year} {2013}),\
  \Eprint{http://arxiv.org/abs/1302.6297}{arXiv:1302.6297 [gr-qc]}%
  \bibAnnoteFile{NoStop}{Lovelace:2013vma}%
\bibitem{Foucart:2013psa}%
  \BibitemOpen
  \bibfield{author}{%
  \bibinfo {author} {\bibfnamefont{F.}~\bibnamefont{Foucart}}, \bibinfo
  {author} {\bibfnamefont{L.}~\bibnamefont{Buchman}}, \bibinfo {author}
  {\bibfnamefont{M.~D.}\ \bibnamefont{Duez}}, \bibinfo {author}
  {\bibfnamefont{M.}~\bibnamefont{Grudich}}, \bibinfo {author}
  {\bibfnamefont{L.~E.}\ \bibnamefont{Kidder}}, \emph{et~al.}}%
   (\bibinfo {year} {2013}),\
  \Eprint{http://arxiv.org/abs/1307.7685}{arXiv:1307.7685 [gr-qc]}%
  \bibAnnoteFile{NoStop}{Foucart:2013psa}%
\bibitem{2002PThPh.107..265S}%
  \BibitemOpen
  \bibfield{author}{%
  \bibinfo {author} {\bibfnamefont{M.}~\bibnamefont{{Shibata}}}\ and\ \bibinfo
  {author} {\bibfnamefont{K.}~\bibnamefont{{Ury{\= u}}}},\ }%
  \bibfield{journal}{%
  \Doi{10.1143/PTP.107.265}{\bibinfo {journal} {Progress of Theoretical
  Physics}}\ }%
  \textbf{\bibinfo {volume} {107}},\ \bibinfo {pages} {265} (\bibinfo {month}
  {Feb.}\ \bibinfo {year} {2002}),\
  \Eprint{http://arxiv.org/abs/arXiv:gr-qc/0203037}{arXiv:gr-qc/0203037}%
  \bibAnnoteFile{NoStop}{2002PThPh.107..265S}%
\bibitem{Shibata:2003ga}%
  \BibitemOpen
  \bibfield{author}{%
  \bibinfo {author} {\bibfnamefont{M.}~\bibnamefont{Shibata}}, \bibinfo
  {author} {\bibfnamefont{K.}~\bibnamefont{Taniguchi}},\ and\ \bibinfo {author}
  {\bibfnamefont{K.}~\bibnamefont{Uryu}},\ }%
  \bibfield{journal}{%
  \Doi{10.1103/PhysRevD.68.084020}{\bibinfo {journal} {Phys.Rev.}}\ }%
  \textbf{\bibinfo {volume} {D68}},\ \bibinfo {pages} {084020} (\bibinfo {year}
  {2003}),\ \Eprint{http://arxiv.org/abs/gr-qc/0310030}{arXiv:gr-qc/0310030
  [gr-qc]}%
  \bibAnnoteFile{NoStop}{Shibata:2003ga}%
\bibitem{Shibata:2006nm}%
  \BibitemOpen
  \bibfield{author}{%
  \bibinfo {author} {\bibfnamefont{M.}~\bibnamefont{Shibata}}\ and\ \bibinfo
  {author} {\bibfnamefont{K.}~\bibnamefont{Taniguchi}},\ }%
  \bibfield{journal}{%
  \Doi{10.1103/PhysRevD.73.064027}{\bibinfo {journal} {Phys.Rev.}}\ }%
  \textbf{\bibinfo {volume} {D73}},\ \bibinfo {pages} {064027} (\bibinfo {year}
  {2006}),\
  \Eprint{http://arxiv.org/abs/astro-ph/0603145}{arXiv:astro-ph/0603145
  [astro-ph]}%
  \bibAnnoteFile{NoStop}{Shibata:2006nm}%
\bibitem{Anderson:2007kz}%
  \BibitemOpen
  \bibfield{author}{%
  \bibinfo {author} {\bibfnamefont{M.}~\bibnamefont{Anderson}}, \bibinfo
  {author} {\bibfnamefont{E.~W.}\ \bibnamefont{Hirschmann}}, \bibinfo {author}
  {\bibfnamefont{L.}~\bibnamefont{Lehner}}, \bibinfo {author}
  {\bibfnamefont{S.~L.}\ \bibnamefont{Liebling}}, \bibinfo {author}
  {\bibfnamefont{P.~M.}\ \bibnamefont{Motl}}, \emph{et~al.},\ }%
  \bibfield{journal}{%
  \Doi{10.1103/PhysRevD.77.024006}{\bibinfo {journal} {Phys.Rev.}}\ }%
  \textbf{\bibinfo {volume} {D77}},\ \bibinfo {pages} {024006} (\bibinfo {year}
  {2008}),\ \Eprint{http://arxiv.org/abs/0708.2720}{arXiv:0708.2720 [gr-qc]}%
  \bibAnnoteFile{NoStop}{Anderson:2007kz}%
\bibitem{2008PhRvD..78h4033B}%
  \BibitemOpen
  \bibfield{author}{%
  \bibinfo {author} {\bibfnamefont{L.}~\bibnamefont{{Baiotti}}}, \bibinfo
  {author} {\bibfnamefont{B.}~\bibnamefont{{Giacomazzo}}},\ and\ \bibinfo
  {author} {\bibfnamefont{L.}~\bibnamefont{{Rezzolla}}},\ }%
  \bibfield{journal}{%
  \Doi{10.1103/PhysRevD.78.084033}{\bibinfo {journal} {\prd}}\ }%
  \textbf{\bibinfo {volume} {78}},\ \bibinfo {eid} {084033} (\bibinfo {month}
  {Oct.}\ \bibinfo {year} {2008}),\
  \Eprint{http://arxiv.org/abs/0804.0594}{arXiv:0804.0594 [gr-qc]}%
  \bibAnnoteFile{NoStop}{2008PhRvD..78h4033B}%
\bibitem{2009CQGra..26k4005B}%
  \BibitemOpen
  \bibfield{author}{%
  \bibinfo {author} {\bibfnamefont{L.}~\bibnamefont{{Baiotti}}}, \bibinfo
  {author} {\bibfnamefont{B.}~\bibnamefont{{Giacomazzo}}},\ and\ \bibinfo
  {author} {\bibfnamefont{L.}~\bibnamefont{{Rezzolla}}},\ }%
  \bibfield{journal}{%
  \Doi{10.1088/0264-9381/26/11/114005}{\bibinfo {journal} {Classical and
  Quantum Gravity}}\ }%
  \textbf{\bibinfo {volume} {26}},\ \bibinfo {eid} {114005} (\bibinfo {month}
  {Jun.}\ \bibinfo {year} {2009}),\
  \Eprint{http://arxiv.org/abs/0901.4955}{arXiv:0901.4955 [gr-qc]}%
  \bibAnnoteFile{NoStop}{2009CQGra..26k4005B}%
\bibitem{Kiuchi:2010ze}%
  \BibitemOpen
  \bibfield{author}{%
  \bibinfo {author} {\bibfnamefont{K.}~\bibnamefont{Kiuchi}}, \bibinfo {author}
  {\bibfnamefont{Y.}~\bibnamefont{Sekiguchi}}, \bibinfo {author}
  {\bibfnamefont{M.}~\bibnamefont{Shibata}},\ and\ \bibinfo {author}
  {\bibfnamefont{K.}~\bibnamefont{Taniguchi}},\ }%
  \bibfield{journal}{%
  \Doi{10.1103/PhysRevLett.104.141101}{\bibinfo {journal} {Phys.Rev.Lett.}}\ }%
  \textbf{\bibinfo {volume} {104}},\ \bibinfo {pages} {141101} (\bibinfo {year}
  {2010}),\ \Eprint{http://arxiv.org/abs/1002.2689}{arXiv:1002.2689
  [astro-ph.HE]}%
  \bibAnnoteFile{NoStop}{Kiuchi:2010ze}%
\bibitem{2012PhRvD..85j4030B}%
  \BibitemOpen
  \bibfield{author}{%
  \bibinfo {author} {\bibfnamefont{S.}~\bibnamefont{{Bernuzzi}}}, \bibinfo
  {author} {\bibfnamefont{M.}~\bibnamefont{{Thierfelder}}},\ and\ \bibinfo
  {author} {\bibfnamefont{B.}~\bibnamefont{{Br{\"u}gmann}}},\ }%
  \bibfield{journal}{%
  \Doi{10.1103/PhysRevD.85.104030}{\bibinfo {journal} {\prd}}\ }%
  \textbf{\bibinfo {volume} {85}},\ \bibinfo {eid} {104030} (\bibinfo {month}
  {May}\ \bibinfo {year} {2012}),\
  \Eprint{http://arxiv.org/abs/1109.3611}{arXiv:1109.3611 [gr-qc]}%
  \bibAnnoteFile{NoStop}{2012PhRvD..85j4030B}%
\bibitem{FaberNSNSreview}%
  \BibitemOpen
  \bibfield{author}{%
  \bibinfo {author} {\bibfnamefont{J.~A.}\ \bibnamefont{Faber}}\ and\ \bibinfo
  {author} {\bibfnamefont{F.~A.}\ \bibnamefont{Rasio}},\ }%
  \bibfield{journal}{%
  \bibinfo {journal} {Living Reviews in Relativity}\ }%
  \textbf{\bibinfo {volume} {15}} (\bibinfo {year} {2012}),\ \doi{\bibinfo
  {doi} {10.12942/lrr-2012-8}},\ \url{http://www.livingreviews.org/lrr-2012-8}%
  \bibAnnoteFile{NoStop}{FaberNSNSreview}%
\bibitem{Sekiguchi:2011zd}%
  \BibitemOpen
  \bibfield{author}{%
  \bibinfo {author} {\bibfnamefont{Y.}~\bibnamefont{Sekiguchi}}, \bibinfo
  {author} {\bibfnamefont{K.}~\bibnamefont{Kiuchi}}, \bibinfo {author}
  {\bibfnamefont{K.}~\bibnamefont{Kyutoku}},\ and\ \bibinfo {author}
  {\bibfnamefont{M.}~\bibnamefont{Shibata}},\ }%
  \bibfield{journal}{%
  \Doi{10.1103/PhysRevLett.107.051102}{\bibinfo {journal} {Phys.Rev.Lett.}}\ }%
  \textbf{\bibinfo {volume} {107}},\ \bibinfo {pages} {051102} (\bibinfo {year}
  {2011}),\ \Eprint{http://arxiv.org/abs/1105.2125}{arXiv:1105.2125 [gr-qc]}%
  \bibAnnoteFile{NoStop}{Sekiguchi:2011zd}%
\bibitem{Paschalidis:2012ff}%
  \BibitemOpen
  \bibfield{author}{%
  \bibinfo {author} {\bibfnamefont{V.}~\bibnamefont{Paschalidis}}, \bibinfo
  {author} {\bibfnamefont{Z.~B.}\ \bibnamefont{Etienne}},\ and\ \bibinfo
  {author} {\bibfnamefont{S.~L.}\ \bibnamefont{Shapiro}},\ }%
  \bibfield{journal}{%
  \Doi{10.1103/PhysRevD.86.064032}{\bibinfo {journal} {Phys.Rev.}}\ }%
  \textbf{\bibinfo {volume} {D86}},\ \bibinfo {pages} {064032} (\bibinfo {year}
  {2012}),\ \Eprint{http://arxiv.org/abs/1208.5487}{arXiv:1208.5487
  [astro-ph.HE]}%
  \bibAnnoteFile{NoStop}{Paschalidis:2012ff}%
\bibitem{Bauswein:2013jpa}%
  \BibitemOpen
  \bibfield{author}{%
  \bibinfo {author} {\bibfnamefont{A.}~\bibnamefont{Bauswein}}, \bibinfo
  {author} {\bibfnamefont{T.}~\bibnamefont{Baumgarte}},\ and\ \bibinfo {author}
  {\bibfnamefont{H.~T.}\ \bibnamefont{Janka}}}%
   (\bibinfo {year} {2013}),\
  \Eprint{http://arxiv.org/abs/1307.5191}{arXiv:1307.5191 [astro-ph.SR]}%
  \bibAnnoteFile{NoStop}{Bauswein:2013jpa}%
\bibitem{Chawla:2010sw}%
  \BibitemOpen
  \bibfield{author}{%
  \bibinfo {author} {\bibfnamefont{S.}~\bibnamefont{Chawla}}, \bibinfo {author}
  {\bibfnamefont{M.}~\bibnamefont{Anderson}}, \bibinfo {author}
  {\bibfnamefont{M.}~\bibnamefont{Besselman}}, \bibinfo {author}
  {\bibfnamefont{L.}~\bibnamefont{Lehner}}, \bibinfo {author}
  {\bibfnamefont{S.~L.}\ \bibnamefont{Liebling}}, \emph{et~al.},\ }%
  \bibfield{journal}{%
  \Doi{10.1103/PhysRevLett.105.111101}{\bibinfo {journal} {Phys.Rev.Lett.}}\ }%
  \textbf{\bibinfo {volume} {105}},\ \bibinfo {pages} {111101} (\bibinfo {year}
  {2010})%
  \bibAnnoteFile{NoStop}{Chawla:2010sw}%
\bibitem{UIUC_MAGNETIZED_BHNS_PAPER1}%
  \BibitemOpen
  \bibfield{author}{%
  \bibinfo {author} {\bibfnamefont{Z.~B.}\ \bibnamefont{{Etienne}}}, \bibinfo
  {author} {\bibfnamefont{Y.~T.}\ \bibnamefont{{Liu}}}, \bibinfo {author}
  {\bibfnamefont{V.}~\bibnamefont{{Paschalidis}}},\ and\ \bibinfo {author}
  {\bibfnamefont{S.~L.}\ \bibnamefont{{Shapiro}}},\ }%
  \bibfield{journal}{%
  \Doi{10.1103/PhysRevD.85.064029}{\bibinfo {journal} {\prd}}\ }%
  \textbf{\bibinfo {volume} {85}},\ \bibinfo {eid} {064029} (\bibinfo {month}
  {Mar.}\ \bibinfo {year} {2012})%
  \bibAnnoteFile{NoStop}{UIUC_MAGNETIZED_BHNS_PAPER1}%
\bibitem{UIUC_MAGNETIZED_BHNS_PAPER2}%
  \BibitemOpen
  \bibfield{author}{%
  \bibinfo {author} {\bibfnamefont{Z.~B.}\ \bibnamefont{{Etienne}}}, \bibinfo
  {author} {\bibfnamefont{V.}~\bibnamefont{{Paschalidis}}},\ and\ \bibinfo
  {author} {\bibfnamefont{S.~L.}\ \bibnamefont{{Shapiro}}},\ }%
  \bibfield{journal}{%
  \Doi{10.1103/PhysRevD.86.084026}{\bibinfo {journal} {\prd}}\ }%
  \textbf{\bibinfo {volume} {86}},\ \bibinfo {eid} {084026} (\bibinfo {month}
  {Oct.}\ \bibinfo {year} {2012}),\
  \Eprint{http://arxiv.org/abs/1209.1632}{arXiv:1209.1632 [astro-ph.HE]}%
  \bibAnnoteFile{NoStop}{UIUC_MAGNETIZED_BHNS_PAPER2}%
\bibitem{2008PhRvL.100s1101A}%
  \BibitemOpen
  \bibinfo {author} {\bibfnamefont{M.}~\bibnamefont{{Anderson}}}, \bibinfo
  {author} {\bibfnamefont{E.~W.}\ \bibnamefont{{Hirschmann}}}, \bibinfo
  {author} {\bibfnamefont{L.}~\bibnamefont{{Lehner}}},\ and\ \bibinfo {author}
  {\bibnamefont{Motl}}%
  \bibAnnoteFile{NoStop}{2008PhRvL.100s1101A}%
\bibitem{Liu:2008xy}%
  \BibitemOpen
\bibfield{author}{%
    }%
  \bibfield{author}{%
  \bibinfo {author} {\bibfnamefont{Y.~T.}\ \bibnamefont{Liu}}, \bibinfo
  {author} {\bibfnamefont{S.~L.}\ \bibnamefont{Shapiro}}, \bibinfo {author}
  {\bibfnamefont{Z.~B.}\ \bibnamefont{Etienne}},\ and\ \bibinfo {author}
  {\bibfnamefont{K.}~\bibnamefont{Taniguchi}},\ }%
  \bibfield{journal}{%
  \Doi{10.1103/PhysRevD.78.024012}{\bibinfo {journal} {Phys.Rev.}}\ }%
  \textbf{\bibinfo {volume} {D78}},\ \bibinfo {pages} {024012} (\bibinfo {year}
  {2008}),\ \Eprint{http://arxiv.org/abs/0803.4193}{arXiv:0803.4193
  [astro-ph]}%
  \bibAnnoteFile{NoStop}{Liu:2008xy}%
\bibitem{2011PhRvD..83d4014G}%
  \BibitemOpen
  \bibfield{author}{%
  \bibinfo {author} {\bibfnamefont{B.}~\bibnamefont{{Giacomazzo}}}, \bibinfo
  {author} {\bibfnamefont{L.}~\bibnamefont{{Rezzolla}}},\ and\ \bibinfo
  {author} {\bibfnamefont{L.}~\bibnamefont{{Baiotti}}},\ }%
  \bibfield{journal}{%
  \Doi{10.1103/PhysRevD.83.044014}{\bibinfo {journal} {\prd}}\ }%
  \textbf{\bibinfo {volume} {83}},\ \bibinfo {eid} {044014} (\bibinfo {month}
  {Feb.}\ \bibinfo {year} {2011}),\
  \Eprint{http://arxiv.org/abs/1009.2468}{arXiv:1009.2468 [gr-qc]}%
  \bibAnnoteFile{NoStop}{2011PhRvD..83d4014G}%
\bibitem{Rezzolla:2011da}%
  \BibitemOpen
  \bibfield{author}{%
  \bibinfo {author} {\bibfnamefont{L.}~\bibnamefont{Rezzolla}}, \bibinfo
  {author} {\bibfnamefont{B.}~\bibnamefont{Giacomazzo}}, \bibinfo {author}
  {\bibfnamefont{L.}~\bibnamefont{Baiotti}}, \bibinfo {author}
  {\bibfnamefont{J.}~\bibnamefont{Granot}}, \bibinfo {author}
  {\bibfnamefont{C.}~\bibnamefont{Kouveliotou}}, \emph{et~al.},\ }%
  \bibfield{journal}{%
  \bibinfo {journal} {Astrophys.J.}\ }%
  \textbf{\bibinfo {volume} {732}},\ \bibinfo {pages} {L6} (\bibinfo {year}
  {2011}),\ \Eprint{http://arxiv.org/abs/1101.4298}{arXiv:1101.4298
  [astro-ph.HE]}%
  \bibAnnoteFile{NoStop}{Rezzolla:2011da}%
\bibitem{Goldreich:1969sb}%
  \BibitemOpen
  \bibfield{author}{%
  \bibinfo {author} {\bibfnamefont{P.}~\bibnamefont{Goldreich}}\ and\ \bibinfo
  {author} {\bibfnamefont{W.~H.}\ \bibnamefont{Julian}},\ }%
  \bibfield{journal}{%
  \bibinfo {journal} {Astrophys.J.}\ }%
  \textbf{\bibinfo {volume} {157}},\ \bibinfo {pages} {869} (\bibinfo {year}
  {1969})%
  \bibAnnoteFile{NoStop}{Goldreich:1969sb}%
\bibitem{Hansen:2000am}%
  \BibitemOpen
  \bibfield{author}{%
  \bibinfo {author} {\bibfnamefont{B.~M.}\ \bibnamefont{Hansen}}\ and\ \bibinfo
  {author} {\bibfnamefont{M.}~\bibnamefont{Lyutikov}},\ }%
  \bibfield{journal}{%
  \Doi{10.1046/j.1365-8711.2001.04103.x}{\bibinfo {journal}
  {Mon.Not.Roy.Astron.Soc.}}\ }%
  \textbf{\bibinfo {volume} {322}},\ \bibinfo {pages} {695} (\bibinfo {year}
  {2001}),\
  \Eprint{http://arxiv.org/abs/astro-ph/0003218}{arXiv:astro-ph/0003218
  [astro-ph]}%
  \bibAnnoteFile{NoStop}{Hansen:2000am}%
\bibitem{McWilliams:2011zi}%
  \BibitemOpen
  \bibfield{author}{%
  \bibinfo {author} {\bibfnamefont{S.~T.}\ \bibnamefont{McWilliams}}\ and\
  \bibinfo {author} {\bibfnamefont{J.}~\bibnamefont{Levin}},\ }%
  \bibfield{journal}{%
  \Doi{10.1088/0004-637X/742/2/90}{\bibinfo {journal} {Astrophys.J.}}\ }%
  \textbf{\bibinfo {volume} {742}},\ \bibinfo {pages} {90} (\bibinfo {year}
  {2011}),\ \Eprint{http://arxiv.org/abs/1101.1969}{arXiv:1101.1969
  [astro-ph.HE]}%
  \bibAnnoteFile{NoStop}{McWilliams:2011zi}%
\bibitem{Lyutikov:2011tq}%
  \BibitemOpen
  \bibfield{author}{%
  \bibinfo {author} {\bibfnamefont{M.}~\bibnamefont{Lyutikov}},\ }%
  \bibfield{journal}{%
  \Doi{10.1103/PhysRevD.83.124035}{\bibinfo {journal} {Phys.Rev.}}\ }%
  \textbf{\bibinfo {volume} {D83}},\ \bibinfo {pages} {124035} (\bibinfo {year}
  {2011}),\ \Eprint{http://arxiv.org/abs/1104.1091}{arXiv:1104.1091
  [astro-ph.HE]}%
  \bibAnnoteFile{NoStop}{Lyutikov:2011tq}%
\bibitem{Piro:2012rq}%
  \BibitemOpen
  \bibfield{author}{%
  \bibinfo {author} {\bibfnamefont{A.~L.}\ \bibnamefont{Piro}},\ }%
  \bibfield{journal}{%
  \Doi{10.1088/0004-637X/755/1/80}{\bibinfo {journal} {Astrophys.J.}}\ }%
  \textbf{\bibinfo {volume} {755}},\ \bibinfo {pages} {80} (\bibinfo {year}
  {2012}),\ \Eprint{http://arxiv.org/abs/1205.6482}{arXiv:1205.6482
  [astro-ph.HE]}%
  \bibAnnoteFile{NoStop}{Piro:2012rq}%
\bibitem{Lai:2012qe}%
  \BibitemOpen
  \bibfield{author}{%
  \bibinfo {author} {\bibfnamefont{D.}~\bibnamefont{Lai}}}%
   (\bibinfo {year} {2012}),\ \doi{\bibinfo {doi}
  {10.1088/2041-8205/757/1/L3}},\
  \Eprint{http://arxiv.org/abs/1206.3723}{arXiv:1206.3723 [astro-ph.HE]}%
  \bibAnnoteFile{NoStop}{Lai:2012qe}%
\bibitem{D'Orazio:2013kgo}%
  \BibitemOpen
  \bibfield{author}{%
  \bibinfo {author} {\bibfnamefont{D.~J.}\ \bibnamefont{D'Orazio}}\ and\
  \bibinfo {author} {\bibfnamefont{J.}~\bibnamefont{Levin}}}%
   (\bibinfo {year} {2013}),\
  \Eprint{http://arxiv.org/abs/1302.3885}{arXiv:1302.3885 [astro-ph.HE]}%
  \bibAnnoteFile{NoStop}{D'Orazio:2013kgo}%
\bibitem{UI1969}%
  \BibitemOpen
  \bibfield{author}{%
  \bibinfo {author} {\bibfnamefont{P.}~\bibnamefont{{Goldreich}}}\ and\
  \bibinfo {author} {\bibfnamefont{D.}~\bibnamefont{{Lynden-Bell}}},\ }%
  \bibfield{journal}{%
  \Doi{10.1086/149947}{\bibinfo {journal} {\apj}}\ }%
  \textbf{\bibinfo {volume} {156}},\ \bibinfo {pages} {59} (\bibinfo {month}
  {Apr.}\ \bibinfo {year} {1969})%
  \bibAnnoteFile{NoStop}{UI1969}%
\bibitem{McL2011}%
  \BibitemOpen
  \bibfield{author}{%
  \bibinfo {author} {\bibfnamefont{S.~T.}\ \bibnamefont{{McWilliams}}}\ and\
  \bibinfo {author} {\bibfnamefont{J.}~\bibnamefont{{Levin}}},\ }%
  \bibfield{journal}{%
  \Doi{10.1088/0004-637X/742/2/90}{\bibinfo {journal} {\apj}}\ }%
  \textbf{\bibinfo {volume} {742}},\ \bibinfo {eid} {90} (\bibinfo {month}
  {Dec.}\ \bibinfo {year} {2011}),\
  \Eprint{http://arxiv.org/abs/1101.1969}{arXiv:1101.1969 [astro-ph.HE]}%
  \bibAnnoteFile{NoStop}{McL2011}%
\bibitem{Ioka:2000yb}%
  \BibitemOpen
  \bibfield{author}{%
  \bibinfo {author} {\bibfnamefont{K.}~\bibnamefont{Ioka}}\ and\ \bibinfo
  {author} {\bibfnamefont{K.}~\bibnamefont{Taniguchi}},\ }%
  \bibfield{journal}{%
  \bibinfo {journal} {Astrophys.J.}}%
   (\bibinfo {year} {2000}),\
  \Eprint{http://arxiv.org/abs/astro-ph/0001218}{arXiv:astro-ph/0001218
  [astro-ph]}%
  \bibAnnoteFile{NoStop}{Ioka:2000yb}%
\bibitem{Bucciantini:2012sm}%
  \BibitemOpen
  \bibfield{author}{%
  \bibinfo {author} {\bibfnamefont{N.}~\bibnamefont{Bucciantini}}\ and\
  \bibinfo {author} {\bibfnamefont{L.}~\bibnamefont{Del~Zanna}}}%
   (\bibinfo {year} {2012}),\
  \Eprint{http://arxiv.org/abs/1205.2951}{arXiv:1205.2951 [astro-ph.HE]}%
  \bibAnnoteFile{NoStop}{Bucciantini:2012sm}%
\bibitem{Dionysopoulou:2012zv}%
  \BibitemOpen
  \bibfield{author}{%
  \bibinfo {author} {\bibfnamefont{K.}~\bibnamefont{Dionysopoulou}}, \bibinfo
  {author} {\bibfnamefont{D.}~\bibnamefont{Alic}}, \bibinfo {author}
  {\bibfnamefont{C.}~\bibnamefont{Palenzuela}}, \bibinfo {author}
  {\bibfnamefont{L.}~\bibnamefont{Rezzolla}},\ and\ \bibinfo {author}
  {\bibfnamefont{B.}~\bibnamefont{Giacomazzo}}}%
   (\bibinfo {year} {2012}),\
  \Eprint{http://arxiv.org/abs/1208.3487}{arXiv:1208.3487 [gr-qc]}%
  \bibAnnoteFile{NoStop}{Dionysopoulou:2012zv}%
\bibitem{Palenzuela:2012my}%
  \BibitemOpen
  \bibfield{author}{%
  \bibinfo {author} {\bibfnamefont{C.}~\bibnamefont{Palenzuela}},\ }%
  \bibfield{journal}{%
  \bibinfo {journal} {Mon. Not. R. Aston. Soc.}\ }%
  \textbf{\bibinfo {volume} {431}},\ \bibinfo {pages} {, 1853} (\bibinfo {year}
  {2}),\ \Eprint{http://arxiv.org/abs/1212.0130}{arXiv:1212.0130
  [astro-ph.HE]}%
  \bibAnnoteFile{NoStop}{Palenzuela:2012my}%
\bibitem{Lehner:2011aa}%
  \BibitemOpen
  \bibfield{author}{%
  \bibinfo {author} {\bibfnamefont{L.}~\bibnamefont{Lehner}}, \bibinfo {author}
  {\bibfnamefont{C.}~\bibnamefont{Palenzuela}}, \bibinfo {author}
  {\bibfnamefont{S.~L.}\ \bibnamefont{Liebling}}, \bibinfo {author}
  {\bibfnamefont{C.}~\bibnamefont{Thompson}},\ and\ \bibinfo {author}
  {\bibfnamefont{C.}~\bibnamefont{Hanna}},\ }%
  \bibfield{journal}{%
  \Doi{10.1103/PhysRevD.86.104035}{\bibinfo {journal} {Phys.Rev.}}\ }%
  \textbf{\bibinfo {volume} {D86}},\ \bibinfo {pages} {104035} (\bibinfo {year}
  {2012}),\ \Eprint{http://arxiv.org/abs/1112.2622}{arXiv:1112.2622
  [astro-ph.HE]}%
  \bibAnnoteFile{NoStop}{Lehner:2011aa}%
\bibitem{Paschalidis:2013jsa}%
  \BibitemOpen
  \bibfield{author}{%
  \bibinfo {author} {\bibfnamefont{V.}~\bibnamefont{Paschalidis}}, \bibinfo
  {author} {\bibfnamefont{Z.~B.}\ \bibnamefont{Etienne}},\ and\ \bibinfo
  {author} {\bibfnamefont{S.~L.}\ \bibnamefont{Shapiro}},\ }%
  \bibfield{journal}{%
  \Doi{10.1103/PhysRevD.88.021504}{\bibinfo {journal} {Phys. Rev. D 88,}}\ }%
  \textbf{\bibinfo {volume} {021504}},\ \bibinfo {pages} {(2013)} (\bibinfo
  {year} {R}),\ \Eprint{http://arxiv.org/abs/1304.1805}{arXiv:1304.1805
  [astro-ph.HE]}%
  \bibAnnoteFile{NoStop}{Paschalidis:2013jsa}%
\bibitem{Palenzuela:2013hu}%
  \BibitemOpen
  \bibfield{author}{%
  \bibinfo {author} {\bibfnamefont{C.}~\bibnamefont{Palenzuela}}, \bibinfo
  {author} {\bibfnamefont{L.}~\bibnamefont{Lehner}}, \bibinfo {author}
  {\bibfnamefont{M.}~\bibnamefont{Ponce}}, \bibinfo {author}
  {\bibfnamefont{S.~L.}\ \bibnamefont{Liebling}}, \bibinfo {author}
  {\bibfnamefont{M.}~\bibnamefont{Anderson}}, \emph{et~al.}}%
   (\bibinfo {year} {2013}),\
  \Eprint{http://arxiv.org/abs/1301.7074}{arXiv:1301.7074 [gr-qc]}%
  \bibAnnoteFile{NoStop}{Palenzuela:2013hu}%
\bibitem{Palenzuela:2013kra}%
  \BibitemOpen
  \bibfield{author}{%
  \bibinfo {author} {\bibfnamefont{C.}~\bibnamefont{Palenzuela}}, \bibinfo
  {author} {\bibfnamefont{L.}~\bibnamefont{Lehner}}, \bibinfo {author}
  {\bibfnamefont{S.~L.}\ \bibnamefont{Liebling}}, \bibinfo {author}
  {\bibfnamefont{M.}~\bibnamefont{Ponce}}, \bibinfo {author}
  {\bibfnamefont{M.}~\bibnamefont{Anderson}}, \emph{et~al.}}%
   (\bibinfo {year} {2013}),\
  \Eprint{http://arxiv.org/abs/1307.7372}{arXiv:1307.7372 [gr-qc]}%
  \bibAnnoteFile{NoStop}{Palenzuela:2013kra}%
\bibitem{Contopoulos:1999ga}%
  \BibitemOpen
  \bibfield{author}{%
  \bibinfo {author} {\bibfnamefont{I.}~\bibnamefont{Contopoulos}}, \bibinfo
  {author} {\bibfnamefont{D.}~\bibnamefont{Kazanas}},\ and\ \bibinfo {author}
  {\bibfnamefont{C.}~\bibnamefont{Fendt}},\ }%
  \bibfield{journal}{%
  \Doi{10.1086/306652}{\bibinfo {journal} {Astrophys.J.}}\ }%
  \textbf{\bibinfo {volume} {511}},\ \bibinfo {pages} {351} (\bibinfo {year}
  {1999}),\
  \Eprint{http://arxiv.org/abs/astro-ph/9903049}{arXiv:astro-ph/9903049
  [astro-ph]}%
  \bibAnnoteFile{NoStop}{Contopoulos:1999ga}%
\bibitem{Komissarov:2005xc}%
  \BibitemOpen
  \bibfield{author}{%
  \bibinfo {author} {\bibfnamefont{S.}~\bibnamefont{Komissarov}},\ }%
  \bibfield{journal}{%
  \Doi{10.1111/j.1365-2966.2005.09932.x}{\bibinfo {journal}
  {Mon.Not.Roy.Astron.Soc.}}\ }%
  \textbf{\bibinfo {volume} {367}},\ \bibinfo {pages} {19} (\bibinfo {year}
  {2006}),\
  \Eprint{http://arxiv.org/abs/astro-ph/0510310}{arXiv:astro-ph/0510310
  [astro-ph]}%
  \bibAnnoteFile{NoStop}{Komissarov:2005xc}%
\bibitem{McKinney:2006sd}%
  \BibitemOpen
  \bibfield{author}{%
  \bibinfo {author} {\bibfnamefont{J.~C.}\ \bibnamefont{McKinney}},\ }%
  \bibfield{journal}{%
  \Doi{10.1111/j.1745-3933.2006.00150.x}{\bibinfo {journal}
  {Mon.Not.Roy.Astron.Soc.Lett.}}\ }%
  \textbf{\bibinfo {volume} {368}},\ \bibinfo {pages} {L30} (\bibinfo {year}
  {2006}),\
  \Eprint{http://arxiv.org/abs/astro-ph/0601411}{arXiv:astro-ph/0601411
  [astro-ph]}%
  \bibAnnoteFile{NoStop}{McKinney:2006sd}%
\bibitem{Spitkovsky:2006np}%
  \BibitemOpen
  \bibfield{author}{%
  \bibinfo {author} {\bibfnamefont{A.}~\bibnamefont{Spitkovsky}},\ }%
  \bibfield{journal}{%
  \Doi{10.1086/507518}{\bibinfo {journal} {Astrophys.J.}}\ }%
  \textbf{\bibinfo {volume} {648}},\ \bibinfo {pages} {L51} (\bibinfo {year}
  {2006}),\
  \Eprint{http://arxiv.org/abs/astro-ph/0603147}{arXiv:astro-ph/0603147
  [astro-ph]}%
  \bibAnnoteFile{NoStop}{Spitkovsky:2006np}%
\bibitem{BSBook}%
  \BibitemOpen
  \bibfield{author}{%
  \bibinfo {author} {\bibfnamefont{T.~W.}\ \bibnamefont{{Baumgarte}}}\ and\
  \bibinfo {author} {\bibfnamefont{S.~L.}\ \bibnamefont{{Shapiro}}},\ }%
  \emph{\bibinfo {title} {{Numerical Relativity: Solving Einstein's Equations
  on the Computer}}}\ (\bibinfo {publisher} {Cambridge University Press},\
  \bibinfo {year} {2010})%
  \bibAnnoteFile{NoStop}{BSBook}%
\bibitem{bs03}%
  \BibitemOpen
  \bibfield{author}{%
  \bibinfo {author} {\bibfnamefont{T.~W.}\ \bibnamefont{{Baumgarte}}}\ and\
  \bibinfo {author} {\bibfnamefont{S.~L.}\ \bibnamefont{{Shapiro}}},\ }%
  \bibfield{journal}{%
  \Doi{10.1086/346103}{\bibinfo {journal} {\apj}}\ }%
  \textbf{\bibinfo {volume} {585}},\ \bibinfo {pages} {921} (\bibinfo {month}
  {Mar.}\ \bibinfo {year} {2003}),\
  \Eprint{http://arxiv.org/abs/arXiv:astro-ph/0211340}{arXiv:astro-ph/0211340}%
  \bibAnnoteFile{NoStop}{bs03}%
\bibitem{Duez:2005sf}%
  \BibitemOpen
  \bibfield{author}{%
  \bibinfo {author} {\bibfnamefont{M.~D.}\ \bibnamefont{Duez}}, \bibinfo
  {author} {\bibfnamefont{Y.~T.}\ \bibnamefont{Liu}}, \bibinfo {author}
  {\bibfnamefont{S.~L.}\ \bibnamefont{Shapiro}},\ and\ \bibinfo {author}
  {\bibfnamefont{B.~C.}\ \bibnamefont{Stephens}},\ }%
  \bibfield{journal}{%
  \Doi{10.1103/PhysRevD.72.024028}{\bibinfo {journal} {Phys.Rev.}}\ }%
  \textbf{\bibinfo {volume} {D72}},\ \bibinfo {pages} {024028} (\bibinfo {year}
  {2005}),\
  \Eprint{http://arxiv.org/abs/astro-ph/0503420}{arXiv:astro-ph/0503420
  [astro-ph]}%
  \bibAnnoteFile{NoStop}{Duez:2005sf}%
\bibitem{Etienne:2010ui}%
  \BibitemOpen
  \bibfield{author}{%
  \bibinfo {author} {\bibfnamefont{Z.~B.}\ \bibnamefont{Etienne}}, \bibinfo
  {author} {\bibfnamefont{Y.~T.}\ \bibnamefont{Liu}},\ and\ \bibinfo {author}
  {\bibfnamefont{S.~L.}\ \bibnamefont{Shapiro}},\ }%
  \bibfield{journal}{%
  \Doi{10.1103/PhysRevD.82.084031}{\bibinfo {journal} {Phys.Rev.}}\ }%
  \textbf{\bibinfo {volume} {D82}},\ \bibinfo {pages} {084031} (\bibinfo {year}
  {2010}),\ \Eprint{http://arxiv.org/abs/1007.2848}{arXiv:1007.2848
  [astro-ph.HE]}%
  \bibAnnoteFile{NoStop}{Etienne:2010ui}%
\bibitem{UIUCEMGAUGEPAPER}%
  \BibitemOpen
  \bibfield{author}{%
  \bibinfo {author} {\bibfnamefont{Z.~B.}\ \bibnamefont{{Etienne}}}, \bibinfo
  {author} {\bibfnamefont{V.}~\bibnamefont{{Paschalidis}}}, \bibinfo {author}
  {\bibfnamefont{Y.~T.}\ \bibnamefont{{Liu}}},\ and\ \bibinfo {author}
  {\bibfnamefont{S.~L.}\ \bibnamefont{{Shapiro}}},\ }%
  \bibfield{journal}{%
  \Doi{10.1103/PhysRevD.85.024013}{\bibinfo {journal} {\prd}}\ }%
  \textbf{\bibinfo {volume} {85}},\ \bibinfo {eid} {024013} (\bibinfo {month}
  {Jan.}\ \bibinfo {year} {2012})%
  \bibAnnoteFile{NoStop}{UIUCEMGAUGEPAPER}%
\bibitem{k04}%
  \BibitemOpen
  \bibfield{author}{%
  \bibinfo {author} {\bibfnamefont{S.~S.}\ \bibnamefont{{Komissarov}}},\ }%
  \bibfield{journal}{%
  \Doi{10.1111/j.1365-2966.2004.07598.x}{\bibinfo {journal} {\mnras}}\ }%
  \textbf{\bibinfo {volume} {350}},\ \bibinfo {pages} {427} (\bibinfo {month}
  {May}\ \bibinfo {year} {2004}),\
  \Eprint{http://arxiv.org/abs/arXiv:astro-ph/0402403}{arXiv:astro-ph/0402403}%
  \bibAnnoteFile{NoStop}{k04}%
\bibitem{Komissarov:2002my}%
  \BibitemOpen
  \bibfield{author}{%
  \bibinfo {author} {\bibfnamefont{S.}~\bibnamefont{Komissarov}},\ }%
  \bibfield{journal}{%
  \Doi{10.1046/j.1365-8711.2002.05313.x}{\bibinfo {journal}
  {Mon.Not.Roy.Astron.Soc.}}\ }%
  \textbf{\bibinfo {volume} {336}},\ \bibinfo {pages} {759} (\bibinfo {year}
  {2002}),\
  \Eprint{http://arxiv.org/abs/astro-ph/0202447}{arXiv:astro-ph/0202447
  [astro-ph]}%
  \bibAnnoteFile{NoStop}{Komissarov:2002my}%
\bibitem{m06}%
  \BibitemOpen
  \bibfield{author}{%
  \bibinfo {author} {\bibfnamefont{J.~C.}\ \bibnamefont{{McKinney}}},\ }%
  \bibfield{journal}{%
  \Doi{10.1111/j.1365-2966.2006.10087.x}{\bibinfo {journal} {\mnras}}\ }%
  \textbf{\bibinfo {volume} {367}},\ \bibinfo {pages} {1797} (\bibinfo {month}
  {Apr.}\ \bibinfo {year} {2006}),\
  \Eprint{http://arxiv.org/abs/arXiv:astro-ph/0601410}{arXiv:astro-ph/0601410}%
  \bibAnnoteFile{NoStop}{m06}%
\bibitem{bz77}%
  \BibitemOpen
  \bibfield{author}{%
  \bibinfo {author} {\bibfnamefont{R.~D.}\ \bibnamefont{{Blandford}}}\ and\
  \bibinfo {author} {\bibfnamefont{R.~L.}\ \bibnamefont{{Znajek}}},\ }%
  \bibfield{journal}{%
  \bibinfo {journal} {\mnras}\ }%
  \textbf{\bibinfo {volume} {179}},\ \bibinfo {pages} {433} (\bibinfo {month}
  {May}\ \bibinfo {year} {1977})%
  \bibAnnoteFile{NoStop}{bz77}%
\bibitem{k11}%
  \BibitemOpen
  \bibfield{author}{%
  \bibinfo {author} {\bibfnamefont{S.~S.}\ \bibnamefont{{Komissarov}}},\ }%
  \bibfield{journal}{%
  \Doi{10.1111/j.1745-3933.2011.01150.x}{\bibinfo {journal} {\mnras}}\ }%
  \textbf{\bibinfo {volume} {418}},\ \bibinfo {pages} {L94} (\bibinfo {month}
  {Nov.}\ \bibinfo {year} {2011}),\
  \Eprint{http://arxiv.org/abs/1108.3511}{arXiv:1108.3511 [astro-ph.HE]}%
  \bibAnnoteFile{NoStop}{k11}%
\bibitem{w74}%
  \BibitemOpen
  \bibfield{author}{%
  \bibinfo {author} {\bibfnamefont{R.~M.}\ \bibnamefont{{Wald}}},\ }%
  \bibfield{journal}{%
  \Doi{10.1103/PhysRevD.10.1680}{\bibinfo {journal} {\prd}}\ }%
  \textbf{\bibinfo {volume} {10}},\ \bibinfo {pages} {1680} (\bibinfo {month}
  {Sep.}\ \bibinfo {year} {1974})%
  \bibAnnoteFile{NoStop}{w74}%
\bibitem{k02}%
  \BibitemOpen
  \bibfield{author}{%
  \bibinfo {author} {\bibfnamefont{S.~S.}\ \bibnamefont{{Komissarov}}},\ }%
  \bibfield{journal}{%
  \Doi{10.1046/j.1365-8711.2002.05313.x}{\bibinfo {journal} {\mnras}}\ }%
  \textbf{\bibinfo {volume} {336}},\ \bibinfo {pages} {759} (\bibinfo {month}
  {Nov.}\ \bibinfo {year} {2002}),\
  \Eprint{http://arxiv.org/abs/arXiv:astro-ph/0202447}{arXiv:astro-ph/0202447}%
  \bibAnnoteFile{NoStop}{k02}%
\bibitem{1982MNRAS.198..339T}%
  \BibitemOpen
  \bibfield{author}{%
  \bibinfo {author} {\bibfnamefont{K.~S.}\ \bibnamefont{{Thorne}}}\ and\
  \bibinfo {author} {\bibfnamefont{D.}~\bibnamefont{{MacDonald}}},\ }%
  \bibfield{journal}{%
  \bibinfo {journal} {\mnras}\ }%
  \textbf{\bibinfo {volume} {198}},\ \bibinfo {pages} {339} (\bibinfo {month}
  {Jan.}\ \bibinfo {year} {1982})%
  \bibAnnoteFile{NoStop}{1982MNRAS.198..339T}%
\bibitem{g99}%
  \BibitemOpen
  \bibfield{author}{%
  \bibinfo {author} {\bibfnamefont{A.}~\bibnamefont{{Gruzinov}}},\ }%
  \bibfield{journal}{%
  \bibinfo {journal} {ArXiv Astrophysics e-prints}}%
   (\bibinfo {month} {Feb.}\ \bibinfo {year} {1999}),\
  \Eprint{http://arxiv.org/abs/arXiv:astro-ph/9902288}{arXiv:astro-ph/9902288}%
  \bibAnnoteFile{NoStop}{g99}%
\bibitem{Pfeiffer:2013wza}%
  \BibitemOpen
  \bibfield{author}{%
  \bibinfo {author} {\bibfnamefont{H.~P.}\ \bibnamefont{Pfeiffer}}\ and\
  \bibinfo {author} {\bibfnamefont{A.~I.}\ \bibnamefont{MacFadyen}}}%
   (\bibinfo {year} {2013}),\
  \Eprint{http://arxiv.org/abs/1307.7782}{arXiv:1307.7782 [gr-qc]}%
  \bibAnnoteFile{NoStop}{Pfeiffer:2013wza}%
\bibitem{Tóth2000605}%
  \BibitemOpen
  \bibfield{author}{%
  \bibinfo {author} {\bibfnamefont{G.}~\bibnamefont{Tóth}},\ }%
  \bibfield{journal}{%
  \Doi{http://dx.doi.org/10.1006/jcph.2000.6519}{\bibinfo {journal} {Journal of
  Computational Physics}}\ }%
  \textbf{\bibinfo {volume} {161}},\ \bibinfo {pages} {605 } (\bibinfo {year}
  {2000}),\ ISSN \bibinfo {issn} {0021-9991},\
  \url{http://www.sciencedirect.com/science/article/pii/S0021999100965197}%
  \bibAnnoteFile{NoStop}{Tóth2000605}%
\bibitem{1988ApJ...332..659E}%
  \BibitemOpen
  \bibfield{author}{%
  \bibinfo {author} {\bibfnamefont{C.~R.}\ \bibnamefont{{Evans}}}\ and\
  \bibinfo {author} {\bibfnamefont{J.~F.}\ \bibnamefont{{Hawley}}},\ }%
  \bibfield{journal}{%
  \Doi{10.1086/166684}{\bibinfo {journal} {\apj}}\ }%
  \textbf{\bibinfo {volume} {332}},\ \bibinfo {pages} {659} (\bibinfo {month}
  {Sep.}\ \bibinfo {year} {1988})%
  \bibAnnoteFile{NoStop}{1988ApJ...332..659E}%
\bibitem{Dedner2002645}%
  \BibitemOpen
  \bibfield{author}{%
  \bibinfo {author} {\bibfnamefont{A.}~\bibnamefont{Dedner}}, \bibinfo {author}
  {\bibfnamefont{F.}~\bibnamefont{Kemm}}, \bibinfo {author}
  {\bibfnamefont{D.}~\bibnamefont{Kröner}}, \bibinfo {author}
  {\bibfnamefont{C.-D.}\ \bibnamefont{Munz}}, \bibinfo {author}
  {\bibfnamefont{T.}~\bibnamefont{Schnitzer}},\ and\ \bibinfo {author}
  {\bibfnamefont{M.}~\bibnamefont{Wesenberg}},\ }%
  \bibfield{journal}{%
  \Doi{http://dx.doi.org/10.1006/jcph.2001.6961}{\bibinfo {journal} {Journal of
  Computational Physics}}\ }%
  \textbf{\bibinfo {volume} {175}},\ \bibinfo {pages} {645 } (\bibinfo {year}
  {2002}),\ ISSN \bibinfo {issn} {0021-9991},\
  \url{http://www.sciencedirect.com/science/article/pii/S002199910196961X}%
  \bibAnnoteFile{NoStop}{Dedner2002645}%
\bibitem{Palenzuela:2010xn}%
  \BibitemOpen
  \bibfield{author}{%
  \bibinfo {author} {\bibfnamefont{C.}~\bibnamefont{Palenzuela}}, \bibinfo
  {author} {\bibfnamefont{T.}~\bibnamefont{Garrett}}, \bibinfo {author}
  {\bibfnamefont{L.}~\bibnamefont{Lehner}},\ and\ \bibinfo {author}
  {\bibfnamefont{S.~L.}\ \bibnamefont{Liebling}},\ }%
  \bibfield{journal}{%
  \Doi{10.1103/PhysRevD.82.044045}{\bibinfo {journal} {Phys.Rev.}}\ }%
  \textbf{\bibinfo {volume} {D82}},\ \bibinfo {pages} {044045} (\bibinfo {year}
  {2010}),\ \Eprint{http://arxiv.org/abs/1007.1198}{arXiv:1007.1198 [gr-qc]}%
  \bibAnnoteFile{NoStop}{Palenzuela:2010xn}%
\bibitem{Alic:2012df}%
  \BibitemOpen
  \bibfield{author}{%
  \bibinfo {author} {\bibfnamefont{D.}~\bibnamefont{Alic}}, \bibinfo {author}
  {\bibfnamefont{P.}~\bibnamefont{Mosta}}, \bibinfo {author}
  {\bibfnamefont{L.}~\bibnamefont{Rezzolla}}, \bibinfo {author}
  {\bibfnamefont{O.}~\bibnamefont{Zanotti}},\ and\ \bibinfo {author}
  {\bibfnamefont{J.~L.}\ \bibnamefont{Jaramillo}},\ }%
  \bibfield{journal}{%
  \Doi{10.1088/0004-637X/754/1/36}{\bibinfo {journal} {Astrophys.J.}}\ }%
  \textbf{\bibinfo {volume} {754}},\ \bibinfo {pages} {36} (\bibinfo {year}
  {2012}),\ \Eprint{http://arxiv.org/abs/1204.2226}{arXiv:1204.2226 [gr-qc]}%
  \bibAnnoteFile{NoStop}{Alic:2012df}%
\bibitem{Farris:2012ux}%
  \BibitemOpen
  \bibfield{author}{%
  \bibinfo {author} {\bibfnamefont{B.~D.}\ \bibnamefont{Farris}}, \bibinfo
  {author} {\bibfnamefont{R.}~\bibnamefont{Gold}}, \bibinfo {author}
  {\bibfnamefont{V.}~\bibnamefont{Paschalidis}}, \bibinfo {author}
  {\bibfnamefont{Z.~B.}\ \bibnamefont{Etienne}},\ and\ \bibinfo {author}
  {\bibfnamefont{S.~L.}\ \bibnamefont{Shapiro}},\ }%
  \bibfield{journal}{%
  \Doi{10.1103/PhysRevLett.109.221102}{\bibinfo {journal} {Phys.Rev.Lett.}}\ }%
  \textbf{\bibinfo {volume} {109}},\ \bibinfo {pages} {221102} (\bibinfo {year}
  {2012}),\ \Eprint{http://arxiv.org/abs/1207.3354}{arXiv:1207.3354
  [astro-ph.HE]}%
  \bibAnnoteFile{NoStop}{Farris:2012ux}%
\bibitem{PPM}%
  \BibitemOpen
  \bibfield{author}{%
  \bibinfo {author} {\bibfnamefont{P.}~\bibnamefont{{Colella}}}\ and\ \bibinfo
  {author} {\bibfnamefont{P.~R.}\ \bibnamefont{{Woodward}}},\ }%
  \bibfield{journal}{%
  \Doi{10.1016/0021-9991(84)90143-8}{\bibinfo {journal} {Journal of
  Computational Physics}}\ }%
  \textbf{\bibinfo {volume} {54}},\ \bibinfo {pages} {174} (\bibinfo {month}
  {Sep.}\ \bibinfo {year} {1984})%
  \bibAnnoteFile{NoStop}{PPM}%
\bibitem{HLL}%
  \BibitemOpen
  \bibfield{author}{%
  \bibinfo {author} {\bibfnamefont{A.}~\bibnamefont{{Harten}}}, \bibinfo
  {author} {\bibfnamefont{P.}~\bibnamefont{{Lax}}},\ and\ \bibinfo {author}
  {\bibfnamefont{B.}~\bibnamefont{{van Leer}}},\ }%
  \bibfield{journal}{%
  \bibinfo {journal} {SIAM Rev.}\ }%
  \textbf{\bibinfo {volume} {25}},\ \bibinfo {pages} {35} (\bibinfo {year}
  {1983})%
  \bibAnnoteFile{NoStop}{HLL}%
\bibitem{Note1}%
  \BibitemOpen
  \bibinfo {note} {Note that~\cite {k02,k04} give the initial data for
  $\protect \mathcal B^i$ and $\protect \mathcal E^i$. However, since the FFE
  equations are invariant if $\protect \mathcal B^i$ and $\protect \mathcal
  E^i$ are multiplied by a constant factor, initial data with $B^i$ and $E^i$
  having the same values as the ones with $\protect \mathcal B^i$ and $\protect
  \mathcal E^i$ are equally valid and the subsequent evolution will be exactly
  the same as the old set of initial data after multiplying an appropriate
  factor. Therefore, the initial data listed in this note are not multiplied by
  the factor $\protect \sqrt {4\pi }$.}%
  \bibAnnoteFile{Stop}{Note1}%
\bibitem{mg04}%
  \BibitemOpen
  \bibfield{author}{%
  \bibinfo {author} {\bibfnamefont{J.~C.}\ \bibnamefont{{McKinney}}}\ and\
  \bibinfo {author} {\bibfnamefont{C.~F.}\ \bibnamefont{{Gammie}}},\ }%
  \bibfield{journal}{%
  \Doi{10.1086/422244}{\bibinfo {journal} {\apj}}\ }%
  \textbf{\bibinfo {volume} {611}},\ \bibinfo {pages} {977} (\bibinfo {month}
  {Aug.}\ \bibinfo {year} {2004}),\
  \Eprint{http://arxiv.org/abs/arXiv:astro-ph/0404512}{arXiv:astro-ph/0404512}%
  \bibAnnoteFile{NoStop}{mg04}%
\bibitem{BriansLatest2012}%
  \BibitemOpen
  \bibfield{author}{%
  \bibinfo {author} {\bibfnamefont{B.~D.}\ \bibnamefont{{Farris}}}, \bibinfo
  {author} {\bibfnamefont{R.}~\bibnamefont{{Gold}}}, \bibinfo {author}
  {\bibfnamefont{V.}~\bibnamefont{{Paschalidis}}}, \bibinfo {author}
  {\bibfnamefont{Z.~B.}\ \bibnamefont{{Etienne}}},\ and\ \bibinfo {author}
  {\bibfnamefont{S.~L.}\ \bibnamefont{{Shapiro}}},\ }%
  \bibfield{journal}{%
  \bibinfo {journal} {ArXiv e-prints}}%
   (\bibinfo {month} {Jul.}\ \bibinfo {year} {2012}),\
  \Eprint{http://arxiv.org/abs/1207.3354}{arXiv:1207.3354 [astro-ph.HE]}%
  \bibAnnoteFile{NoStop}{BriansLatest2012}%
\bibitem{Note2}%
  \BibitemOpen
  \bibinfo {note} {Note that we deliberately wrote $\nabla _\alpha \protect
  \mathcal F^{\nu \alpha }$ instead of $\protect \mathcal J^\nu $, $\zeta $
  instead of $\rho $, and $Q^\mu $ instead of $J^\mu $. This is to demonstrate
  that the Maxwell equation $\nabla _\nu \protect \mathcal F^{\mu \nu } =
  \protect \mathcal J^\mu $ is not needed to prove the redundancy of the energy
  equation.}%
  \bibAnnoteFile{Stop}{Note2}%
\end{thebibliography}%

\end{document}